\documentclass[structabstract]{aa} 

\usepackage{natbib}
\bibpunct{(}{)}{;}{a}{}{,} 

\usepackage{color}
\usepackage{amsmath,amssymb,rotating}      
\usepackage{graphicx}
\usepackage[linktocpage=true,unicode=true]{hyperref}
\usepackage{hypcap}
\usepackage{multirow}

\newcommand{\citeeq}[1]{Eq.~(\ref{#1})}

\newcommand{\citesec}[1]{Sect.~\ref{#1}}
\newcommand{\citesecs}[1]{Sects.~\ref{#1}}
\newcommand{\citeapp}[1]{Appendix~\ref{#1}}
\newcommand{\citefig}[1]{Fig.~\ref{#1}}
\newcommand{\citetab}[1]{Table~\ref{#1}}

\newcommand{\beq}{\begin{equation}}
\newcommand{\eeq}{\end{equation}}
\newcommand{\beal}{\begin{align}}
\newcommand{\eeqal}{\end{align}}
\newcommand{\ben}{\begin{eqnarray}}
\newcommand{\een}{\end{eqnarray}}
\newcommand{\nn}{\nonumber}
\newcommand{\bi}{\begin{itemize}}
\newcommand{\ei}{\end{itemize}}

\newcommand{\ie}{\textit{i.e.}}
\newcommand{\eg}{\textit{e.g.}}

\newcommand{\propmin}{\mbox{\em min}}
\newcommand{\propmed}{\mbox{\em med}}
\newcommand{\propmax}{\mbox{\em max}}

\newcommand{\greenf}{\mbox{${\cal G}$}}
\newcommand{\tgammae}{\mbox{$\widetilde{\Gamma}_E$}}
\newcommand{\etilde}{\mbox{$\widetilde{E}_\star$}}
\newcommand{\Msol}{\mbox{$M_{\odot}$}}
\newcommand{\Rsol}{\mbox{$R_{\odot}$}}

\newcommand{\mymean}[1]{\mbox{$\langle #1 \rangle$}}

\newcommand{\lsim}{\mathrel{\mathop{\kern 0pt \rlap
  {\raise.2ex\hbox{$<$}}}
  \lower.9ex\hbox{\kern-.190em $\sim$}}}
\newcommand{\gsim}{\mathrel{\mathop{\kern 0pt \rlap
  {\raise.2ex\hbox{$>$}}}
  \lower.9ex\hbox{\kern-.190em $\sim$}}}

\graphicspath{{./NewFigs/}}
\begin{document}

\title{Galactic electrons and positrons at the Earth:\\
  new estimate of the primary and secondary fluxes}
\author{ 
  T. Delahaye\inst{1,2} \and 
  J. Lavalle\inst{2} \and 
  R. Lineros\inst{2} \and 
  F. Donato\inst{2} \and 
  N. Fornengo\inst{2} } 

\offprints{\tt\\ 
  delahaye@lapp.in2p3.fr \\ 
  lavalle@to.infn.it \\ 
  lineros@to.infn.it}

\institute{
  LAPTH, Universit\'e de Savoie \& CNRS, 
  BP 110, F-74941 Annecy-le-Vieux Cedex --- France.
  \and 
  Dipartimento di Fisica Teorica, 
  Universit\`a di Torino \& INFN - Sezione di Torino, 
  Via Giuria 1, I-10122 Torino --- Italia.}

\date{Received / Accepted}


\abstract
%
{
  The so-called excess of cosmic ray (CR) positrons observed by the PAMELA 
  satellite up to 100 GeV has led to many interpretation attempts, from 
  standard astrophysics to a possible exotic contribution from 
  dark matter annihilation or decay. The Fermi data subsequently obtained 
  about CR electrons and positrons in the range 0.02-1 TeV, and HESS data above 
  1 TeV have provided additional information about the leptonic content of 
  local Galactic CRs.
}
%
{
 We analyse predictions of the CR lepton fluxes at the Earth of both secondary 
 and primary origins, evaluate the theoretical uncertainties, and determine 
 their level of consistency with respect to the available data.
}
%
{
  For propagation, we use a relativistic treatment of the energy losses 
  for which we provide useful parameterizations. We compute the secondary 
  components by improving on the method that we derived earlier for positrons. 
  For primaries, we estimate the contributions from astrophysical sources 
  (supernova remnants and pulsars) by considering all known local objects 
  within 2 kpc and a smooth distribution beyond.
}
%
{
  We find that the electron flux in the energy range 5-30 GeV is well 
  reproduced by a smooth distant distribution of sources with index $\gamma
  \sim 2.3-2.4$, while local sources dominate the flux at higher energy. For 
  positrons, local pulsars have an important effect above 5-10 GeV. 
  Uncertainties affecting the source modeling and propagation are degenerate 
  and each translates into about one order of magnitude error in terms of local 
  flux. The spectral shape at high energy is weakly correlated with the 
  spectral indices of local sources, but more strongly with the hierarchy in 
  their distance, age and power. Despite the large theoretical errors that we 
  describe, our global and self-consistent analysis can explain all available 
  data without over-tuning the parameters, and therefore without the need to 
  consider any exotic physics.
}
%
{
  Though a \emph{standard paradigm} of Galactic CRs is well established, our 
  results show that we can hardly talk about any \emph{standard model} of CR 
  leptons, because of the very large theoretical uncertainties. Our analysis 
  provides details about the impact of these uncertainties, thereby sketching a 
  roadmap for future improvements.
}

\titlerunning{Galactic electrons and positrons at the Earth}
\authorrunning{Delahaye et al.}  

\maketitle

\begin{flushleft}
  Preprint DFTT-51/2009 and LAPTH-1339/09
\end{flushleft}

\section{Introduction}
\label{sec:intro}

Cosmic ray (CR) electrons and positrons\footnotemark~constitute $\sim 1$\% of 
the CR budget at the Earth in the GeV-TeV energy range, and provide interesting
means of probing the acceleration processes in CR sources, propagation 
phenomenology, and the interstellar environment itself, complementary to 
protons~\citep[\eg,][]{1987PhR...154....1B}. At energies $\gtrsim$ 100 GeV, 
their observed properties are mostly set by the very local environment. Their 
typical propagation scale is indeed limited to the kpc scale because of the very
efficient electromagnetic energy losses caused by Compton scattering with 
the interstellar radiation fields (ISRF), the cosmic microwave background 
(CMB), and the magnetic field~\citep{1965PhRv..137.1306J,1970RvMP...42..237B}. 
High energy CR electrons are produced directly by well-known astrophysical CR 
accelerators such as supernova remnants (SNRs) and pulsars, in which case they 
are referred to as \emph{primary} electrons. They can also be created by 
secondary processes, mostly nuclear interactions of cosmic protons and light 
nuclei with the interstellar medium (ISM) gas concentrated in the Galactic disk 
(spallation), in which case they are referred to as \emph{secondary} electrons. 
Because they have been assumed to hardly be produced in astrophysical sources, 
positrons have been proposed as potential tracers of new physics, in particular 
the annihilation or decay of dark matter~\citep{1984PhRvL..53..624S}. Although 
the main theoretical ideas regarding the origin and propagation of cosmic 
electrons were formalized a long time ago in the seminal monograph 
of~\cite{1964ocr..book.....G}, their ongoing measurements are still far from 
being completely understood.

\footnotetext{Hereafter, \emph{electrons} will denote both electrons and 
positrons, unless specified.}

The observation by the PAMELA satellite~\citep{2009Natur.458..607A} of 
a rising positron fraction up to $\sim$100 GeV has triggered a considerable 
amount of interpretation attempts. Estimates of the cosmic electron and 
positron fluxes were first calculated in detail in~\cite{1998ApJ...493..694M}, 
where only secondaries were considered for positrons, that fail to match the 
PAMELA data. We derived novel predictions of the secondary positron flux at the 
Earth, with a particular focus on sizing the theoretical errors caused by 
uncertainties in spallation cross-sections, in the modeling of the progenitor 
interstellar (IS) CR flux, in the characterization of the energy losses, and in 
the propagation parameters~\citep{2009A&A...501..821D}. Although the overall 
theoretical uncertainty is about one order of magnitude, we have still shown 
that a rising positron fraction was not expected unless a very soft electron 
spectrum was considered. Even in that case, however, we have also illustrated 
how difficult it was to accommodate a good fit to the PAMELA data in spectral 
shape as well as in amplitude. This soft electron spectrum is at the lowest 
statistical edge of the current electron cosmic ray data below 30 GeV. Likewise,
it is not supported by the unprecedented measurements performed with the Fermi 
satellite between 20 GeV and 1 TeV of CR electrons plus 
positrons~\citep{2009PhRvL.102r1101A}, which sets the true denominator of the 
positron fraction. At this stage, separate data of positrons and electrons 
would be of particular interest and would provide stronger grounds to any 
interpretation attempt, but, unfortunately, are not yet available. Nevertheless,
from both predictions of the secondary positron flux and the current data, it 
appears unlikely that this increase observed in the positron fraction is purely 
of secondary origin. Therefore, this positron excess points towards the 
existence of primary sources of positrons in the neighborhood. Note finally 
that the cut-off in the electron flux observed by HESS around 3 TeV provides 
interesting and complementary information~\citep{2008PhRvL.101z1104A}.

It has long been demonstrated that astrophysical sources may supply this 
extra-yield of cosmic positrons. For instance, as discussed
by~\cite{1989ApJ...342..807B}~\citep[see also~\eg][]{1995A&A...294L..41A,1996ApJ...459L..83C,2001A&A...368.1063Z}, pulsars could provide sizable contributions 
to the positron flux from pair conversions of $\gamma$-rays in the strong 
magnetic fields that they host. This has been recently revisited by several 
authors~\citep[\eg][]{2009JCAP...01..025H,2009PhRvL.103e1101Y,2008arXiv0812.4457P,2009PhRvD..80f3005M}, who have drawn similar conclusions. Another class of 
contributions invokes spallation processes with the ISM gas during the 
acceleration stage of cosmic rays inside SNRs that had not been considered 
before 
\citep{2003A&A...410..189B,2009PhRvL.103e1104B,2009PhRvL.103h1103B,2009PhRvL.103h1104M,2009PhRvD..80l3017A}. 
This hypothesis leads to the additional production of antiprotons or heavier 
secondary nuclei, providing interesting counterparts that should be observed in 
the near future. Finally, using a more refined spatial 
distribution of sources and interstellar gas might also lead to a rising 
positron fraction in the PAMELA energy range~\citep{2009PhRvL.103k1302S}.

The large amount of standard, but still different, astrophysical interpretations
of the observed positron fraction is noteworthy and points chiefly towards 
significant lacks in our understanding of the cosmic electron production in 
sources and their subsequent propagation in the Galaxy. This also points towards
a {\em standard model of Galactic cosmic rays} still being far from complete, in
spite of the progresses achieved so far in the description of cosmic ray 
sources, propagation and interaction with the ISM and ISRF. Because high energy 
electrons have a propagation horizon much smaller than light cosmic ray nuclei, 
they offer an interesting means to improve the overall phenomenological 
modeling, the Galactic environment being indeed much more tightly constrained 
locally. We nevertheless emphasize the robustness of the {\em standard paradigm 
of cosmic ray physics} as formalized in the seminal book of 
\citet{1964ocr..book.....G}: distinguishing {\em standard model} from 
{\em standard paradigm} appears to us important to avoid considering 
departures from peculiar observational data as deep failures of generic 
astrophysical explanations.

Our purpose is to develop novel calculations of the electron and positron 
fluxes to assess the relative roles of the different primaries and 
secondaries in the positron fraction. This is somehow a continuation of the 
study that we performed on secondary positrons~\citep{2009A&A...501..821D}. We 
treat all of these components in a self-consistent framework that includes 
\eg\ improved propagation modeling (with a full relativistic treatment of the 
energy losses) as well as constrained properties of local sources, including 
both SNRs and pulsars. In addition to improving and clarifying the 
interpretation of the PAMELA data from standard astrophysical processes, this
study also helps us to verify whether the cosmic positron spectrum can provide 
interesting perspectives in the search for new physics. A particularly important
issue is whether positrons injected from dark matter annihilation could be 
differentiated from all other astrophysical contributions. Dark matter could 
indeed in some cases manifests itself in this channel~\citep[\eg][]{1998PhRvD..59b3511B,2004PhRvD..70k5004H,2007A&A...462..827L,2007PhRvD..75f3506A,2008PhRvD..78j3520B,2008NuPhB.800..204C,2008PhRvD..77f3527D,2009arXiv0908.0195P,2009arXiv0912.4421C}, 
and the discovery of an exotic contribution to the positron budget would be a 
spectacular result. Any such result would, however, have to rely on solid 
grounds, in particular a good understanding of the astrophysical contributions. 
We show that the theoretical uncertainties are very large, and discuss in 
detail the relative impact of each ingredient. This variance in the predictions 
is quite bad news for exotic searches, because it indicates that the background 
is poorly constrained. Despite these uncertainties, we show that our 
calculations, involving pure astrophysical processes in a self-consistent 
framework, can very well explain the whole set of available data on CR leptons, 
without over-tuning the parameters, and therefore without any need of exotic 
physics.

The outline of the paper is the following. In \citesec{sec:propag}, we  
describe in detail our propagation model, with a particular focus on the 
relativistic treatment of the energy losses. In Sect.~\ref{sec:secondaries}, 
we revisit the predictions of the local secondary electron and positron 
fluxes and discuss the effects of our improved propagation model compared to 
the results we derived in~\citet{2009A&A...501..821D}. In 
\citesec{sec:primary_el}, we compute the primary electron component by 
considering a smooth distribution of SNRs beyond $\sim$ 2 kpc from the Earth, 
and by determining the contribution of each known SNR within this distance; we 
also discuss in detail the impact of the source modeling. In 
\citesec{sec:primary_pos}, we briefly revisit the primary positrons that pulsars
could generate by using the same approach as for electrons. We finally compare 
our results with all available data on CR leptons in \citesec{sec:disc}, before 
concluding in \citesec{sec:concl}.

\section{Propagation of electrons and positrons}
\label{sec:propag}

\subsection{General aspects}
\label{subsec:propag}

CR propagation in the Galaxy involves quite complex processes. The spatial 
diffusion is caused by convection upwards and downwards from the Galactic disk 
and by the erratic bouncing of CRs off moving magnetic inhomogeneities, 
which also induces a diffusion in momentum space, more precisely diffusive
reacceleration. Energy losses along the CR journey have additional effects 
on the diffusion in momentum space. Nuclei can also experience nuclear 
interactions (spallation); this is of course irrelevant to electron 
propagation, but spallation will still be considered as the source
of secondaries. The propagation zone spreads beyond the disk, and 
is very often modeled as a cylindrical slab of radius 
$R\simeq R_{\rm disk} \simeq 20$ kpc, and a vertical half-height of 
$L \simeq 1-15$ kpc. Astrophysical sources of CRs and the ISM gas are 
mostly located within the disk, which has a vertical extent of $z_{\rm disk} 
\simeq 0.1$ kpc. More details on propagation phenomenology can be found in 
\eg\ \cite{berezinsky_book_90},~\cite{1994hea..book.....L} 
and~\cite{2007ARNPS..57..285S}.

Throughout this paper, we discuss high energy electrons particularly
of energies above $\sim$ 10 GeV, for which the effects of solar 
modulation are much weaker. We demonstrated in \cite{2009A&A...501..821D}
that convection and reacceleration could be neglected above a few GeV,
so that the propagation of electrons can be expressed in terms of the usual 
current conservation equation $\widehat{\cal D} {\cal N} = 
{\cal Q}(E,\vec{x},t)$, where the transport operator $\widehat{\cal D}$ can be 
expanded as
\ben
\partial_t {\cal N} - \vec{\nabla}\cdot 
\left\{ K(E)  \vec{\nabla}{\cal N} \right\} + 
\partial_E \left\{ \frac{dE}{dt}{\cal N} \right\} = {\cal Q}(E,\vec{x},t)\;.
\label{eq:prop}
\een
The electron number density per unit of energy is denoted 
${\cal N}={\cal N}(E,\vec{x},t)\equiv dn/dE$, $K(E)$ is the 
energy-dependent diffusion coefficient assumed isotropic and homogeneous, 
$dE/dt$ is the energy-loss term and ${\cal Q}$ is the source term. As 
mentioned above, we neglected convection and reacceleration.

The above equation can be solved numerically, \eg\ by means of the public 
code \verb|GALPROP|~\citep{1998ApJ...509..212S}, which treats CR nuclei and 
electrons in the same framework.
However, most of the studies using this code do not usually correlate the 
features of protons at sources with those of electrons
\citep[\eg][]{1998ApJ...493..694M,2000ApJ...537..763S}, which alleviates 
the relevance of treating nuclei and electrons in the same global numerical 
framework. In that case, one can always tune the source modeling differently 
for each of these species to accommodate the observational constraints.

We adopted instead a semi-analytical propagation modeling, which is designed to 
survey a wider parameter space and both clarifies and simplifies the 
discussion on theoretical uncertainties. Analytical steady-state solutions to 
\citeeq{eq:prop}, in terms of Green functions, can be found in \eg\
\cite{1974Ap&SS..29..305B},~\cite{berezinsky_book_90},
~\cite{1998PhRvD..59b3511B},~\cite{2007A&A...462..827L}, 
or~\cite{2008PhRvD..77f3527D}, in the non-relativistic Thomson approximation
of the inverse Compton energy losses. We improve this model by including a full 
relativistic calculation of the energy losses (see 
\citesec{subsec:eloss}) and the time-dependent solution to 
\citeeq{eq:prop}, which has to be used when dealing with local sources (see 
\citesec{subsec:time}). The propagation parameters are constrained as usual, by 
means of the ratio of secondary to primary stable nuclei, except for the 
energy-loss parameters, which are constrained from the description of the local 
ISRF and magnetic field. This latter point is discussed in detail in 
\citesec{subsec:prop_par}.

For the sake of completeness, we briefly recall the Green functions that 
are steady-state solutions to \citeeq{eq:prop}, disregarding the energy-loss 
features for the moment. Assuming that spatial diffusion and energy losses are 
isotropic and homogeneous, it is an academic exercise to derive the steady-state
Green function in an infinite 3D space, which obeys $\widehat{\cal D}_{\bar{t}}
\,\greenf = \delta^3(\vec{x}-\vec{x}_s)\delta(E-E_s)$, \ie,
\ben
\greenf(\vec{x},E\leftarrow \vec{x}_s,E_s) = 
\frac{1}{b(E)\,(\pi\,\lambda^2)^{\frac{3}{2}} } \cdot 
\exp\left\{ \frac{(\vec{x}_s - \vec{x})^2}{\lambda^2} \right\}\;,
\label{eq:3D}
\een
where the subscript $s$  flags quantities at source ($E_s\geq E$), and we 
define the energy-loss rate and the diffusion scale to be
\ben
b(E)\equiv -\frac{dE}{dt} \;;\;\; 
\lambda^2 \equiv 4 \int_{E}^{E_s} dE' \frac{K(E')}{b(E')}\;.
\label{eq:def_lambda}
\een
The propagation scale $\lambda$ characterizes the CR electron horizon and 
depends on energy in terms of the ratio of the diffusion coefficient to the 
energy-loss rate. If these are both described by power laws, \eg,
$K(E)\propto E^\delta$ and $b(E) \propto E^\alpha$, then 
$\greenf \propto E^{\frac{\alpha}{2}-\frac{3}{2}(\delta+1)}$; this is of 
importance when discussing the primary and secondary contributions later on.
For definiteness, we define
\ben
K(E) &\equiv& \beta\, K_0\,\left(\frac{\cal R}{1\,{\rm GV}}\right)^\delta
\simeq K_0 \,\epsilon^\delta\nn\\
b(E) &\equiv& b_0 \, \epsilon^\alpha = \frac{E_0}{\tau_l}\, \epsilon^\alpha
\;\;{\rm with}\;\;\epsilon \equiv \frac{E}{E_0 = 1\,{\rm GeV}}\;,
\label{eq:k_and_b}
\een
where $K_0$ and $b_0$ are the normalizations of the diffusion coefficient and 
the energy-loss rate, respectively, that carry the appropriate dimensions, and
$\tau_l$ is the characteristic energy-loss time.

Because of the finite spatial extent of the diffusion slab, boundary conditions 
must be taken into account when the propagation scale is on the order of the 
vertical or radial boundaries. At the Earth location, which we fix to be 
$(x_\odot,y_\odot,z_\odot) = (8,0,0)$ kpc throughout the paper, the radial 
boundary is irrelevant while $(R-r_\odot)\gtrsim L$, which is almost always the 
case for reasonable values of $L$ and $R$, constrained by observations
\citep[\eg][]{1998ApJ...509..212S,2001ApJ...555..585M}. Therefore, we briefly 
review the solutions accounting for the vertical boundary condition only. In 
that case, one can split the general Green function into two terms, one radial 
and the other vertical, such as $\greenf = (\greenf_r\times\greenf_z)/
b(E)$. The radial term is merely the infinite 2D solution
\ben
\greenf_r (\vec{r},E\leftarrow \vec{r}_s,E_s) = 
\frac{1}{\pi \,\lambda^2}
\exp\left\{ -\frac{(\vec{r}-\vec{r}_s)^2}{\lambda^2} \right\}\;,
\een
where $\vec{r}$ is the projection of the electron position in the $z=0$ plane, 
and the subscript $s$ refers to the source. The vertical solution can be 
determined by different methods. On small propagation scales, more precisely 
for $\lambda<L$, one can use the image method~\citep[\eg][]{1979ApJ...228..297C,1998PhRvD..59b3511B}
\ben
\greenf_z (z,E\leftarrow z_s,E_s) = \sum_{n=-\infty}^{+\infty}
\frac{(-1)^n}{\sqrt{\pi} \lambda} 
\exp\left\{ -\frac{(z-z_{s,n})^2}{\lambda^2}\right\} \;,
\een
where $z_{s,n} \equiv 2\,n\,L + (-1)^nz_s$. On larger propagation scales, \ie~ 
$\lambda\gtrsim L$, the basis defined by the Helmholtz eigen-functions allows a 
better numerical convergence~\citep{2007A&A...462..827L}. In that case, we have 
instead
\ben
\greenf_z (z,E\leftarrow z_s,E_s) &=& \frac{1}{L}\sum_{n=1}^{+\infty} 
\Bigg\{  e^{-\left[\frac{k_n\lambda}{2}\right]^2} \phi_n(z)\phi_n(z_s) \nn \\
&& + e^{-\left[\frac{k_n'\lambda}{2}\right]^2} \phi_n'(z)\phi_n'(z_s)\Bigg\}\;,
\een
where the pair and odd eigen-modes and eigen-functions read, respectively
\ben
k_n = (n-1/2) \pi/L\;&;& \;\;\; k_n' = n\pi/L\;;\nn\\
\phi_n(z) = \sin\left( k_n(L-|z|)\right)\;&;& \;\;\; 
\phi_n'(z) = \sin\left( k_n'(L-z)\right)\;.
\een
The radial boundary condition becomes relevant when 
$(R-r_\odot)\sim L \sim \lambda $. We accounted for it using the image method 
for the radial component, or, since the smooth source term exhibits a 
cylindrical symmetry, by expanding the solution in terms of Bessel series~\citep[see \eg\ ][]{1974Ap&SS..29..305B,berezinsky_book_90,2008PhRvD..77f3527D}. The 
radial boundary condition is, however, mostly irrelevant in the following, since
we will mainly consider electron energies $\gtrsim 10$ GeV, for which the 
propagation scale is no more than a few kpc.


\subsection{Time-dependent solution}
\label{subsec:time}

The steady-state solutions derived above are safe approximations for a 
continuous injection of CRs in the ISM, as in the case of secondaries. In 
opposition, primary CRs are released after violent and localized events such as 
supernova explosions, the remnants and sometimes pulsars of which are assumed to
be the most common Galactic CR accelerators. Since the supernova explosion rate 
$\Gamma_\star$ is most likely a few per century, the CR injection rate could 
exhibit significant local variations over the CR lifetime (confinement time, or 
energy-loss time, depending on the species) provided this latter is much longer 
than the individual source lifetime. Since electrons lose energy very 
efficiently, there is a spatial scale (an energy scale, equivalently), below 
(above) which these local variations will have a significant effect on the local
electron density. To roughly estimate this scale, one can compare the energy 
loss rate $b(E)$ with the {\em local} injection rate. Assuming that source 
events are all identical and homogeneously distributed in an infinitely thin 
disk of radius $R = 20$ kpc, local fluctuations are expected to be smoothed 
when integrated over an electron horizon $\lambda$ such that 
$\Gamma_\star(\lambda/R)^2 \gg b(E)/E$. Using 
$K_0\approx 0.01\,{\rm kpc^2/Myr}$, $b(E)\approx ({\rm GeV/Myr}/315)\epsilon^2$ 
and $\Gamma_\star \approx 1/100\,{\rm yr}$, we find that $E\ll 80$ GeV, which 
means that local fluctuations of the flux are probably important above a few 
tens of GeV. A similar reasoning was presented a few decades ago by 
\citet{1970ApJ...162L.181S}. We recall that a significant number of SNRs and 
pulsars are actually observed within a few kpc of the Earth. Therefore, current 
multi-wavelength measurements may help us to feature them as electron sources, 
and thereby predict the local electron density.

To estimate the contribution of local transient sources, we need to 
solve the full time-dependent transport equation given in~\citeeq{eq:prop}, and 
we demonstrate that the method used for the steady-state case can also be used, 
though partly, for the transient case. The time-dependent Green function, 
$\greenf_{t}$, is defined in terms of the transport operator, asking that 
$\widehat{\cal D}\,\greenf_{t}=\delta^3(\vec{x}-\vec{x}_s)\delta(E-E_s)
\delta(t-t_s)$. To solve this equation, we generally work in Fourier space
\citep[\eg~][]{1964ocr..book.....G,berezinsky_book_90,1995PhRvD..52.3265A,2004ApJ...601..340K,2004PhRvD..70b3512B}, using
\ben
\greenf_{t}(t,E,\vec{x}) &=&  
\frac{1}{\left(2 \pi\right)^{2}} \iint d^3 k \, d\omega  \nn\\ 
& & \times \exp\left\{ i \big(\vec{k} \cdot \vec{x} + \omega t \big)\right\}\, 
\phi(\omega,E,k) \, .
\label{eq:expan_time_green}
\een
In Fourier space, we derive the ordinary differential equation for $E$, for 
each pair $(\omega,k)$
\begin{align}
\left\{i \omega + k^2 K(E) \right\} \phi -  \partial_E & 
\Big(b(E) \phi \Big) = \, \delta(E - E_s) \\ 
 & \times\frac{1}{(2\pi)^{2}} 
\exp\left\{ -i (\vec{k}\cdot\vec{x}_s+ \omega t_s) \right\} \, ,\nn
\end{align}
which is solved by the function
\begin{align}
  \phi(\omega,E,k) = & \, \frac{1}{b(E)} \frac{1}{(2\pi)^2} 
  \exp\left\{ -\frac{1}{4} k^2 \lambda^2 -i \vec{k}\cdot\vec{x}_s\right\} \\
  & \times \exp\left\{ -i \omega(t_s + \Delta\tau) \right\} \, .\nn
\end{align}
This solution is only valid for $E \le E_s$ because it describes 
processes ruled by energy losses. It contains the propagation scale $\lambda$ 
previously defined in \citeeq{eq:def_lambda} and the {\em loss time} 
defined as
\ben
\Delta\tau(E, E_s) \equiv \int_{E}^{E_s} \frac{dE'}{b(E')} \;.
\een
This loss time corresponds to the average time during which the energy of a 
particle decreases from $E_s$ to $E$ because of losses. The inverse Fourier 
transformation is straightforward from \citeeq{eq:expan_time_green}, and we 
eventually obtain
\ben
\label{eq:comp_green}
\greenf_{t}(t,E,\vec{x}\leftarrow t_s,E_s,\vec{x}_s) = 
\frac{\delta(\Delta t - \Delta\tau)}{b(E)}
\frac{\exp\left\{- \frac{(\vec{x}-\vec{x}_s)^2}{\lambda^2}\right\}}
     {(\pi \lambda^2)^{3/2}}\; , \nn\\ 
\een
where $\Delta t=t-t_s$. We recognize the product of the steady-state solution 
and a delta function mixing real time and loss time. As in the steady-state 
case, we can account for the vertical boundary condition by expanding this 3D 
solution by means of the image method or on the basis of Helmholtz 
eigen-functions. The final result can therefore be expressed in terms of the 
full steady-state solution
\ben
\greenf_{t}(t,E,\vec{x}\leftarrow t_s,E_s,\vec{x}_s) = 
 \delta(\Delta t - \Delta\tau) \,
\greenf (E,\vec{x}\leftarrow E_s,\vec{x}_s)\;. \nn\\
\een
An alternative interpretation of the time dependence comes up when the temporal 
delta function is converted into an energy delta function, which has been shown 
to be appropriate for bursting sources for which $\Delta t$ is fixed. In this 
case, the Green function is instead given by
\ben
\greenf_{t}(t,E,\vec{x}\leftarrow t_s,E_s,\vec{x}_s) &=& 
\delta(E_s - E^{\star}) \, b(E^{\star})\nn\\ \,
&& \times \greenf (E,\vec{x}\leftarrow E_s,\vec{x}_s)\;,
\een
where the energy $E^{\star}$ satisfies
\ben
\Delta\tau(E,E^{\star}) = \Delta t \, .
\een
Thus, $E^{\star}$ corresponds to the injection energy needed to observe a 
particle with energy $E$ after a time $\Delta t= t - t_s$. Although there is no 
analytical solution to this equation in the full relativistic treatment of 
the energy losses (see \citesec{subsec:eloss}), we can still derive it in the 
Thomson approximation
\ben
\label{eq:estar}
E^{\star} &\overset{\rm Th.}{\underset{\rm approx.}{=}} &
\frac{E}{1-E/E_{\rm max}^{\rm Th}} \\
\;\;{\rm with }\;\; 
E_{\rm max}^{\rm Th}&\equiv& \left[ b_0\,\Delta t \right]^{-1}
=\frac{\tau_l}{\Delta t}E_0\;,\nn
\een
where we used the energy-loss term from \citeeq{eq:k_and_b}. We see that when 
the energy-loss timescale $\tau_l\gg \Delta t$, we have $E^{\star}\approx E$. We
also see that a maximal energy is set by the ratio $\tau_l / \Delta t$: in the 
Thomson approximation, a particle injected with energy $\ge E_{\rm max}^{\rm Th}$
will have already lost all its energy by $\Delta t$. We emphasize that 
$E_{\rm max}\ge E_{\rm max}^{\rm Th}$ in the general relativistic case (see 
\citesec{subsec:eloss}).

Note that an additional consequence of this energy $E^{\star}$ is that the 
propagation scale $\lambda$ is no longer set by energy losses but instead by the
injection time $\Delta t$. In the simplified case of a constant diffusion 
coefficient $K$, we would indeed have found that $\lambda^2 = 4\,K\,\Delta t$. 
Of course, the energy dependence of the diffusion coefficient slightly modifies 
this relation, but this remark will further help us to make a rough prediction 
about the observed spectrum for a bursting source 
(see \citesec{subsec:spectral_indices}).

Finally, we underline that solutions to the time-dependent transport equation do
not always provide causality, which is important to avoid incorrect predictions 
when varying the source age and distance. To ensure causality at zeroth order 
and for the sake of definitiveness, we use
\ben
\greenf_{t}(t,E,\vec{x}\leftarrow t_s,E_s,\vec{x}_s) &=& 
\theta(c\Delta t - ||\vec{x}-\vec{x}_s||)\,
\delta(E_s - E^{\star}) \nn\\ \,
&& \times  b(E^{\star})\, \greenf (E,\vec{x}\leftarrow E_s,\vec{x}_s)\;,
\label{eq:time_dep_prop}
\een
as our time-dependent propagator. A more accurate causal solution would need
more specific methods inferred from~\eg~detailed studies of relativistic 
heat conduction.

\subsection{Approximated links between propagation models and observed spectra}
\label{subsec:spectral_indices}

To anticipate the discussion about the observed versus predicted spectra primary 
and secondary electrons, it is useful to show how observed indices can be 
formally linked to the propagation ingredients. We now establish 
approximate relations between the observed spectral index $\widetilde{\gamma}$, 
the source spectral index $\gamma$, and the propagation parameters. In the 
most general case, the interstellar (IS) flux at the Earth, \ie~without 
accounting for solar modulation, is expressed as
\begin{align}
\label{eq:flux_def}
\phi_\odot ( & E) = \frac{\beta \, c}{4\,\pi} \\
\times & \iiint dt_sdE_s d^3\vec{x}_s \, 
\greenf (E,\vec{x}_\odot\leftarrow t_s,E_s,\vec{x}_s)\,
{\cal Q}(t_s,E_s,\vec{x}_s)\nn\;.
\end{align}
We first discuss the steady-state case, before continuing to the case of 
time-dependent sources.

The energy dependence arising in the electron propagator (see 
\citesec{subsec:propag}) comes from spatial diffusion and energy losses. At 
high energy, one can assume that the propagation scale is short enough to 
allow us to neglect the vertical boundary condition, so that one can use the 3D 
propagator to predict the electron flux on Earth, given a source 
${\cal Q}(E,\vec{x})$. Since we consider a short propagation scale, and since 
sources are located in the Galactic disk, we can assume a source term that is 
homogeneously distributed in the disk. This is a very good approximation for 
secondaries (see \citesec{sec:secondaries}), and fair enough for primaries (see 
\citesec{subsec:smooth_snrs}). Likewise, we consider that the source 
spectrum is a mere power law of index $\gamma$, so that the source term can
be written as
${\cal Q}(E,\vec{x}) = 2\,h\,{\cal Q}_0\,\delta(z)\,\epsilon^{-\gamma}$, where 
$h$ is the half-height of the disk and $\epsilon$, which is defined in 
\citeeq{eq:k_and_b}, is the dimensionless energy parameter. Given this source 
term, the flux on Earth is given by
\ben
\phi_\odot ( E) &\simeq& 
\frac{o\, c \,h}{2\,\pi^{3/2}}\,\frac{{\cal Q}_0}{\sqrt{K_0/\tau_l}} 
\epsilon^{-\widetilde{\gamma}}\;,
\label{eq:disk_approx}
\een
where, $o= \sqrt{\alpha-\delta-1}/(\gamma-1)= {\cal O}(1)$, and we have used 
the 3D propagator defined in \citeeq{eq:3D}, the energy dependence of which is 
fully determined from Eqs.~(\ref{eq:def_lambda}) and (\ref{eq:k_and_b}). 
Accordingly, the spectral index $\widetilde{\gamma}$ after propagation reads
\ben
\widetilde{\gamma} = \gamma + \frac{1}{2}(\alpha+\delta-1)\;.
\label{eq:gamma_obs_disk}
\een
As discussed in \citesec{subsec:eloss}, the energy-loss rate is dominated by 
inverse Compton and synchrotron processes. In the non-relativistic Thomson 
approximation, we have $\alpha = 2$, leading to $\widetilde{\gamma} = \gamma + 
\frac{1}{2}(1+\delta)$. From this basic calculation, it is easy to derive rough 
values for $\gamma$ and $\delta$ consistent with any spectral index 
$\widetilde{\gamma}$ measured on Earth. For instance, $\widetilde{\gamma}\approx
3$ translates into a source index $\gamma$ in the range $[2.1,2.35]$ for 
$\delta\in [0.3,0.8]$. Although very useful to first order, this crude spectral 
analysis is only valid for a smooth and flat distribution of source(s), and 
significantly differs when local discrete effects are taken into consideration. 
Implementing full relativistic losses induces $\alpha = \alpha_{\rm eff} 
\lesssim 2$, which implies a harder $\widetilde{\gamma}$. This will be delved 
into in more detail in \citesec{sec:primary_el}.

Finally, we extract the observed spectral index $\widetilde{\gamma}_\star$ for a
single event-like source, which differs slightly from the above calculation. 
The source term can be expressed as ${\cal Q}_\star = Q_{\star,0}
\delta(|\vec{x}_s|-d)\,\delta(t_s-t_\star)\,\epsilon^{-\gamma}$ --- we discuss 
this injection spectrum in more detail in \citesec{subsec:local_snrs}. Assuming 
further that the source is located within the propagation horizon $d\ll \lambda$
and that a burst occurs at a time much earlier than the energy-loss timescale 
$t_\star \ll \tau_l$, we readily find that
\ben
\phi_\odot ( E) = \frac{\beta\, c}{4\pi }\, 
\frac{b(E^\star)}{b(E)} \,
\frac{Q_{\star,0}\,\epsilon_\star^{-\gamma}}{(\pi\lambda^2)^{3/2}}
\simeq \frac{c}{4\pi } \,
\frac{Q_{\star,0}\,\epsilon^{-\widetilde{\gamma}_\star}}
     {(4 \, \pi \, K_0\,t_\star)^{3/2}}
\;,
\label{eq:approx_point_flux}
\een
where
\ben
\widetilde{\gamma}_\star = \gamma +\frac{3}{2}\delta\;.
\label{eq:gamma_obs_single}
\een
Here, we have considered that the propagation scale is no longer fixed by energy
losses, since $t_\star \ll \tau_l$, but instead by $t_\star$ (see the discussion 
at the end of \citesec{subsec:time}). In this case, since $E_\star 
\sim E$, the spectral index is not directly affected by the energy losses.


\subsection{Full relativistic energy losses}
\label{subsec:eloss}

In the GeV-TeV energy range, electrons lose their energy by electromagnetic
interactions with the ISRF (inverse Compton scattering) and the magnetic 
field (synchrotron emission), while Bremsstrahlung, ionization and Coulomb 
interactions with the ISM are negligible. Most studies have used the Thomson 
approximation to account for inverse Compton losses, which is valid for 
an electron Lorentz factor $\gamma_e \lesssim m_e c^2/E_{\rm ph}$, where 
$E_{\rm ph}$ is the photon energy \citep[\eg][]{1998ApJ...493..694M,2009A&A...501..821D}. This translates into a maximal electron energy 
of $\sim 1.11\times 10^6$ GeV for interactions with CMB ($E_{\rm ph}\simeq 
2.35\times 10^{-4}$ eV), and of $7.58\times 10^4/8.66\times 10^2$ GeV for IR / 
starlight radiation, respectively (with $E_{\rm ph, IR/\star} 
\simeq 3.45\times 10^{-3}/0.3$ eV). From those numbers, it is clear that the 
Thomson approximation is no longer valid for energies at Earth above a 
few tens of GeV, for which a full relativistic description of the term 
$dE/dt$ in \citeeq{eq:prop} is consequently necessary. Few other studies 
have implemented this relativistic 
treatment~\citep[\eg][]{2004ApJ...601..340K,2009arXiv0908.2183S}.

The calculation of inverse Compton scattering of electrons with photons in the 
relativistic regime was derived in the astrophysical context by
\cite{1965PhRv..137.1306J}. It was subsequently extensively 
revisited and complemented by~\cite{1970RvMP...42..237B}. In the following, we 
rely on the latter reference to derive our relativistic version of the inverse 
Compton energy losses, to which we refer the reader for more details.

We consider relativistic electrons propagating in an isotropic and homogeneous 
gas of photons, which, moreover, exhibits a black-body energy distribution. 
The relevance of these assumptions will be discussed in 
\citesec{subsec:prop_par}. The electron energy-loss rate can be expressed in 
terms of the energies $\epsilon$ and $\epsilon_1$ of a photon before and after 
the collision, respectively, as
\ben
- \frac{dE}{dt} = \int d\epsilon \int d\epsilon_1 (\epsilon_1 - \epsilon)
\frac{dN_{\rm coll}}{dt\,d\epsilon\, d\epsilon_1}\;.
\een
The collision rate is given by
\ben
\frac{dN_{\rm coll}}{dt\,d\epsilon \, d\epsilon_1} &=& 
\frac{3\, \sigma_{\rm T}\, c}{4\,\gamma_e^2} 
\frac{dn(\epsilon)/d\epsilon}{\epsilon} \times\nn\\  
\Bigg\{ 1 & + & 2q \left( \ln q - q + \frac{1}{2}\right) +
\frac{(1-q)}{2} \frac{(\Gamma q)^2)}{(1+\Gamma q)} \Bigg\}\;,
\een
where $dn(\epsilon)/d\epsilon$ is the initial photon density in the energy range
$d\epsilon$, which, for black-body radiation has the form (including the 
two polarization states)
\ben
\frac{dn}{d\epsilon} = 2 \times \frac{4\pi \epsilon^2}{(2\pi\hbar c)^3}
\left( e^{\epsilon/(k_b T)} - 1\right)^{-1}\;,
\een
and
\ben
q \equiv \frac{\hat{\epsilon}_1}{\Gamma(1-\hat{\epsilon}_1)} \;,\;\;
\hat{\epsilon}_1 \equiv \frac{\epsilon_1}{\gamma_e m c^2}\;,\;\;
\Gamma \equiv  \frac{4 \gamma_e \epsilon}{m c^2}\;.
\een
From kinematics, the range for $\hat{\epsilon}_1$ is readily found to be 
$\lbrack \hat{\epsilon}, \frac{\Gamma}{(1+\Gamma)} \rbrack$, which translates 
into $\lbrack \frac{1}{4\gamma_e^2} ,1\rbrack$ for $q$. It is convenient to 
rewrite the energy loss rate in terms of an integral over $q$
\ben
-\frac{dE}{dt} &=& \int d\epsilon \int dq \,
\frac{\Gamma^2 (\gamma_e m c^2)^2 }{(1 + \Gamma q)^2} \, 
\left\{ \frac{q}{(1+\Gamma q)}-\frac{1}{4\gamma_e^2} \right\} \nn\\
&&\times\, \frac{dN_{\rm coll}}{dt\,d\epsilon \, d\epsilon_1}\;,
\een
where the integral over $q$ is found to be analytical, so that one can 
easily check the full numerical calculation.

We define a dimensionless parameter that characterizes the relevant regime 
to be used for the energy loss rate
\ben
\alpha \equiv \frac{\gamma_e\, (k_b T_0)}{m_ec^2} \;,
\een
where $T_0$ is the mean temperature of the radiation field.

The non-relativistic Thomson limit is recovered for inverse Compton processes 
within a black-body radiation field, using $\Gamma\ll 1$ or equivalently 
$\alpha \ll 1$
\ben
-\frac{dE}{dt} = \frac{4}{3} \sigma_{\rm T}\, c\, U_{\rm rad} \,\gamma_e^2\;,
\label{eq:thomson}
\een
where $U_{\rm rad} = \int d\epsilon \, \epsilon\, dn/d\epsilon$, whereas 
the Klein-Nishina regime applies for $\alpha \gg 1$
\ben
-\frac{dE}{dt} = \frac{\sigma_{\rm T}}{16} \frac{(m_ec\,k_bT_0)^2}{\hbar^3}
\left\{ \ln\frac{4\gamma_e \,k_b T_0}{m_e c^2} - 1.9805 \right\}
\label{eq:kn}
\een
In \citefig{fig:rel_eloss}, we compare both regimes with the full calculation. 
From our numerical results, we derived a parameterization that is valid for any 
black-body radiation field, given by
\ben
- \frac{dE}{dt} =
\begin{cases}
  {\rm Thomson }  \;\;\;\;\;\;\;\;\;\;\;\;\;\;\;\;\;\;\;\;\;\;\;\;\;\;\;\;\;\;\;\;\;\;\;\;\;  {\rm for} \;{\cal C}_{\rm n-r}\\
  \frac{E^2 (k_b T_0)^4}{\alpha}
  \exp \left\{  \sum_{i=0} c_i \, \left( \ln \alpha \right )^i \right\}
  \;\;  {\rm for} \;{\cal C}_{\rm int}\\
       {\rm Klein-Nishina} \;\;\;\;\;\;\;\;\;\;\;\;\;\;\;\;\;\;\;\;\;\;\;\;\;\;\; {\rm for}\;{\cal C}_{\rm u-r}
\end{cases},
\label{eq:eloss}
\een
where the conditions ${\cal C}$ read
\ben
{\cal C}_{\rm n-r} & : & \alpha < 3.8 \times 10^{-4}\;,\nn \\
{\cal C}_{\rm int} & : & 3.8\times 10^{-4}\leq \alpha \leq 1.8\times 10^{3}\;,
\nn \\
{\cal C}_{\rm u-r} & : & \alpha > 1.8 \times 10^{3} \;.
\een
The fitting formula associated with the intermediate regime provided in 
\citeeq{eq:eloss} may be used with the parameters
\ben
c_i &=& \Big\{ 74.77,-0.1953,-9.97\times 10^{-2},\nn\\
&& 4.352\times 10^{-3},3.546\times 10^{-4},-3.01\times 10^{-5}\Big\}\;.
\een
An additional smooth interpolation between these three regimes might 
improve the calculation by avoiding tiny gaps at connections, 
which could arise \eg\ from very small numerical differences in the unit 
conversions or constants used above. This parameterization is valid for any 
black-body distribution of photons. If one considers a black-body distribution, 
the absolute energy density $U_{\rm rad}$ of which differs from that generally 
derived $U_{\rm rad}^{\rm bb}$, then one can simply renormalize \citeeq{eq:eloss}
by a factor $U_{\rm rad}/U_{\rm rad}^{\rm bb}$ to obtain the correct energy-loss 
rate.

\begin{figure}[t]
\begin{center}
\includegraphics[width=\columnwidth, clip]{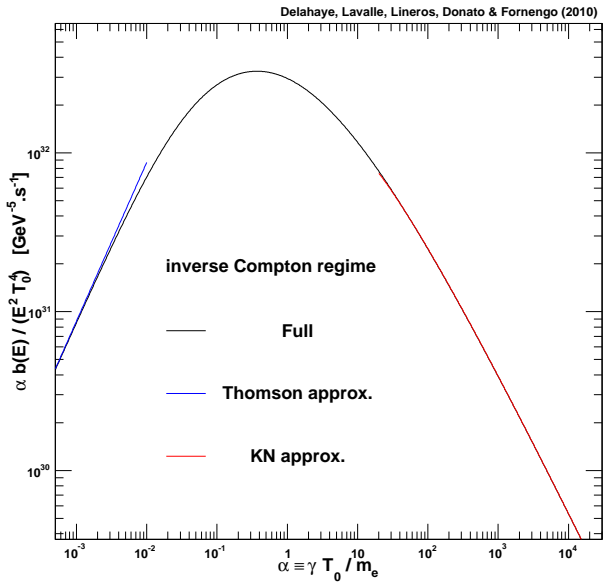}
\caption{\small Comparison between the different relevant regimes of the 
inverse Compton energy loss for any black-body radiation field.}
\label{fig:rel_eloss}
\end{center}
\end{figure}

In the following, we use \citeeq{eq:eloss} to describe the energy loss
rates associated with Compton processes.


\subsection{Review of the propagation parameters}
\label{subsec:prop_par}

In addition to energy losses, five parameters regulate the diffusion properties 
of Galactic CRs: $K_0$ and $\delta$ defining the diffusion coefficient 
(see Eq.~\ref{eq:k_and_b}), the half-thickness of the diffusion zone $L$, 
the convective wind velocity $V_c$, and the Alfv\`en speed of magnetic field 
inhomogeneities $V_a$, responsible for reacceleration. It was shown in 
\cite{2009A&A...501..821D} that the last two effects can be neglected above 
a few GeV. These parameters were self-consistently constrained in 
\citet{2001ApJ...555..585M} with ratios of secondary to primary nuclei 
--- mostly boron to carbon B/C 
\citep[see also][for a more recent analysis]{2010A&A...516A..66P}. 
In the following, we use the available parameter space provided by these 
authors. Nevertheless, as useful beacons for bracketting the theoretical
uncertainties, we also use the {\em min}, {\em med}, and {\em max} 
subsets of propagation parameters, which were derived in 
\citet{2004PhRvD..69f3501D} and called so after the hierarchy found on the 
primary antiproton fluxes, for sources spread all over the diffusion zone 
(not only in the disk). These models are described in~\citetab{tab:prop}.

For the normalization of the diffusion coefficient $K_0$, we note that $B/C$ 
measurements actually constrain $K_0/L$, not $K_0$ alone 
\citep{2001ApJ...555..585M}. Moreover, using radioactive species does not
allow yet to fully break this degeneracy~\citep{2002A&A...381..539D}. This 
explains why the {\em min} ({\em max}) configuration, which has a small (large) 
$L$, is associated with a small (large) value of $K_0$. The spectral index 
of the diffusion coefficient $\delta$ decreases from {\em min} to 
{\em max}, which is important to the spectral analysis of the electron flux.

\begin{table}[t]
\begin{center}
\begin{tabular}{|c||c|c|c|c|c|c|}
\hline
Model  & $\delta$ & $K_0$ & $L$ & $V_{c}$ & $V_{a}$ \\
       &          & [kpc$^2$/Myr] & [kpc] & [km/s] & [km/s] \\
\hline
\propmin  & 0.85 &  0.0016 & 1  & 13.5 &  22.4 \\
\textcolor{blue}{\propmed}  & 0.70 &  0.0112 & 4  & 12   &  52.9 \\
\propmax  & 0.46 &  0.0765 & 15 &  5   & 117.6 \\
\hline
\end{tabular}
\vskip 0.25cm
\caption{\small Beacon sets of diffusion parameters derived in 
  \cite{2004PhRvD..69f3501D} compatible with the B/C analysis performed in 
  \citet{2001ApJ...555..585M}. The \propmed~setup will be our default model.}
\label{tab:prop}
\end{center}
\end{table}

Tighter constraints are expected to be possible with future PAMELA data, and 
hopefully with AMS2~\citep{battiston_07}. The current uncertainty in those 
parameters leads to large theoretical errors in secondary positrons
\citep{2009A&A...501..821D} and therefore electrons, as reviewed in 
\citesec{sec:secondaries}. The error in astrophysical primaries is assessed in 
\citesec{subsec:el_pred}

In contrast to stable nuclei in the GeV-TeV energy range, electrons are strongly
affected by energy losses, which have a significant effect on their transport. 
It was shown in \citet{2009A&A...501..821D} that inverse Compton and 
synchrotron processes dominate in this energy domain. Therefore, it is crucial 
to constrain the ISRF --- including the CMB, and radiation from dust and stars 
--- and the magnetic field as accurately as possible, within the horizon of 
GeV-TeV electrons, \ie~ ${\cal O}(1\,{\rm kpc})$.

In \citesec{subsec:eloss}, we developed a method to calculate the inverse 
Compton energy losses in a fully relativistic formalism, provided the target 
radiation fields can be described in terms of black-body distributions. This is 
obviously the case for the CMB, the temperature of which was recently 
re-estimated in~\citet{2009arXiv0911.1955F} to be 2.7260$\pm$0.0013~K. However, 
it is well known that the ISRF is not simply Planckian radiation, since it 
consists of many different components --- IR radiation from dust, optical and 
UV radiation from stars, diffuse X-ray emission, {\it etc}. --- with different 
spatial distributions. Since by using CMB only we estimate the electron 
propagation scale to be $\lesssim$ 2 kpc for electron energies above 10 GeV, we 
disregard the spatial dependence of the ISRF, and only consider local averages.

\begin{figure*}[t]
\begin{center}
\includegraphics[width=0.65\columnwidth,clip]{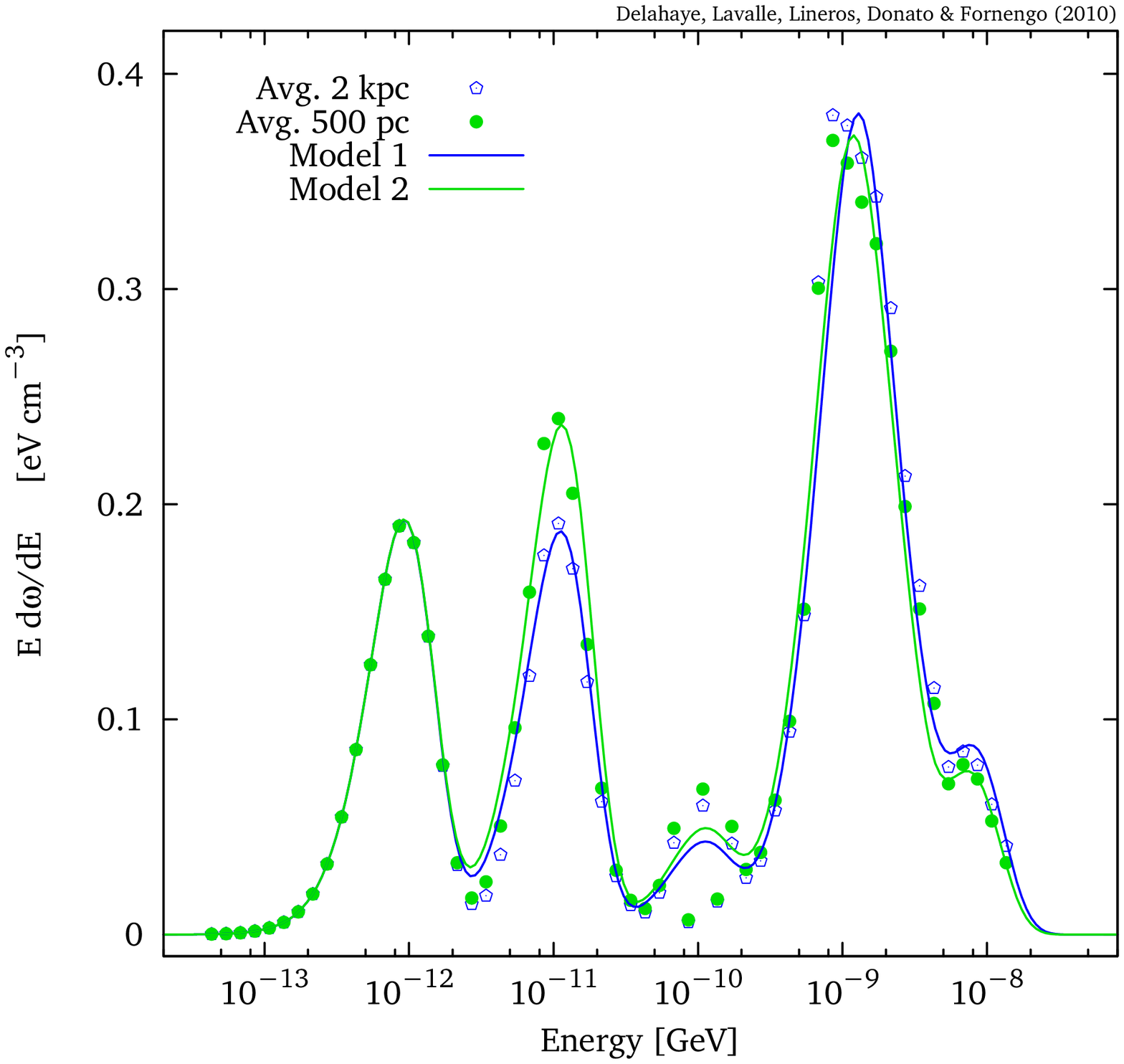}
\includegraphics[width=0.65\columnwidth,clip]{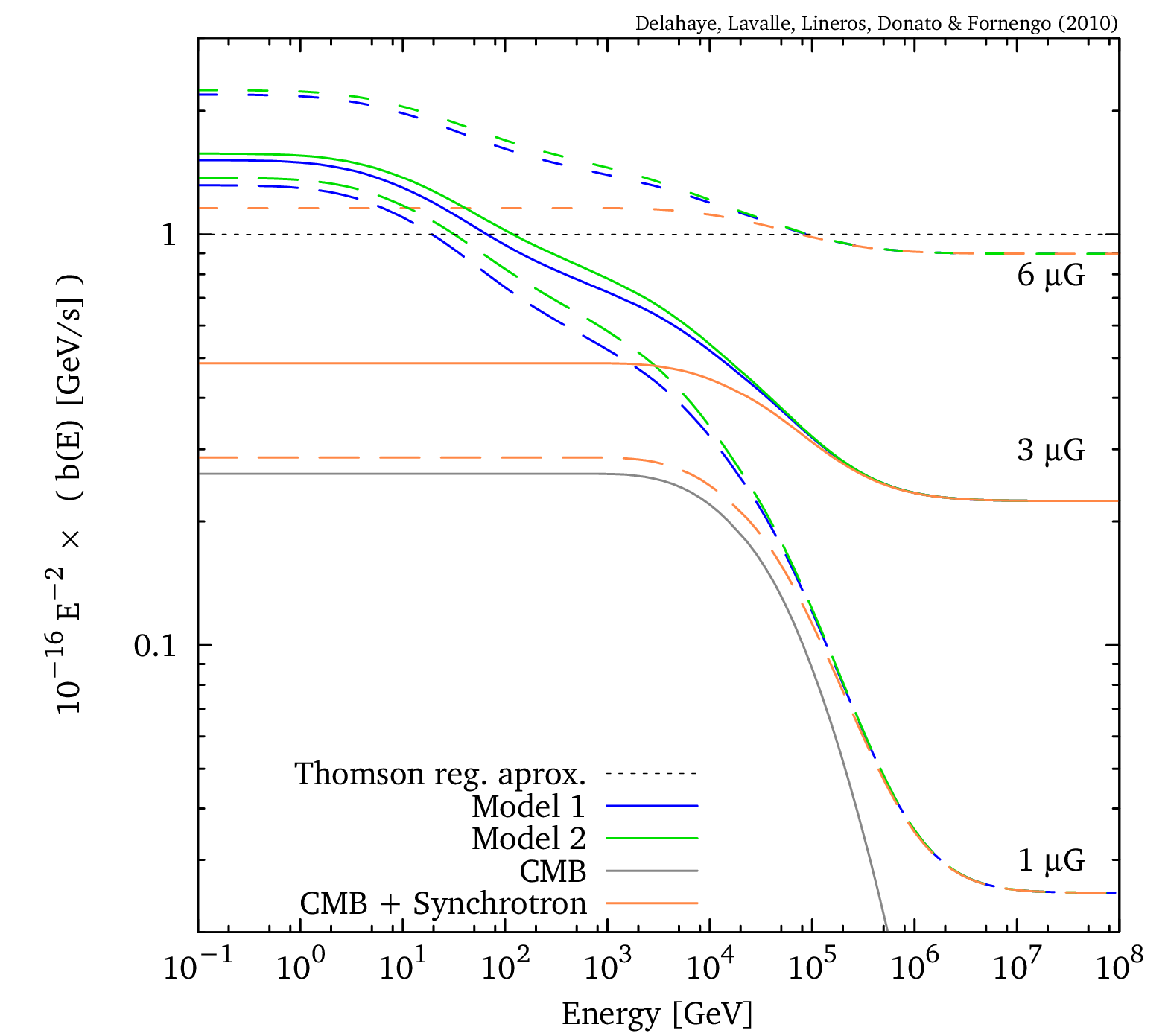}
\includegraphics[width=0.65\columnwidth,clip]{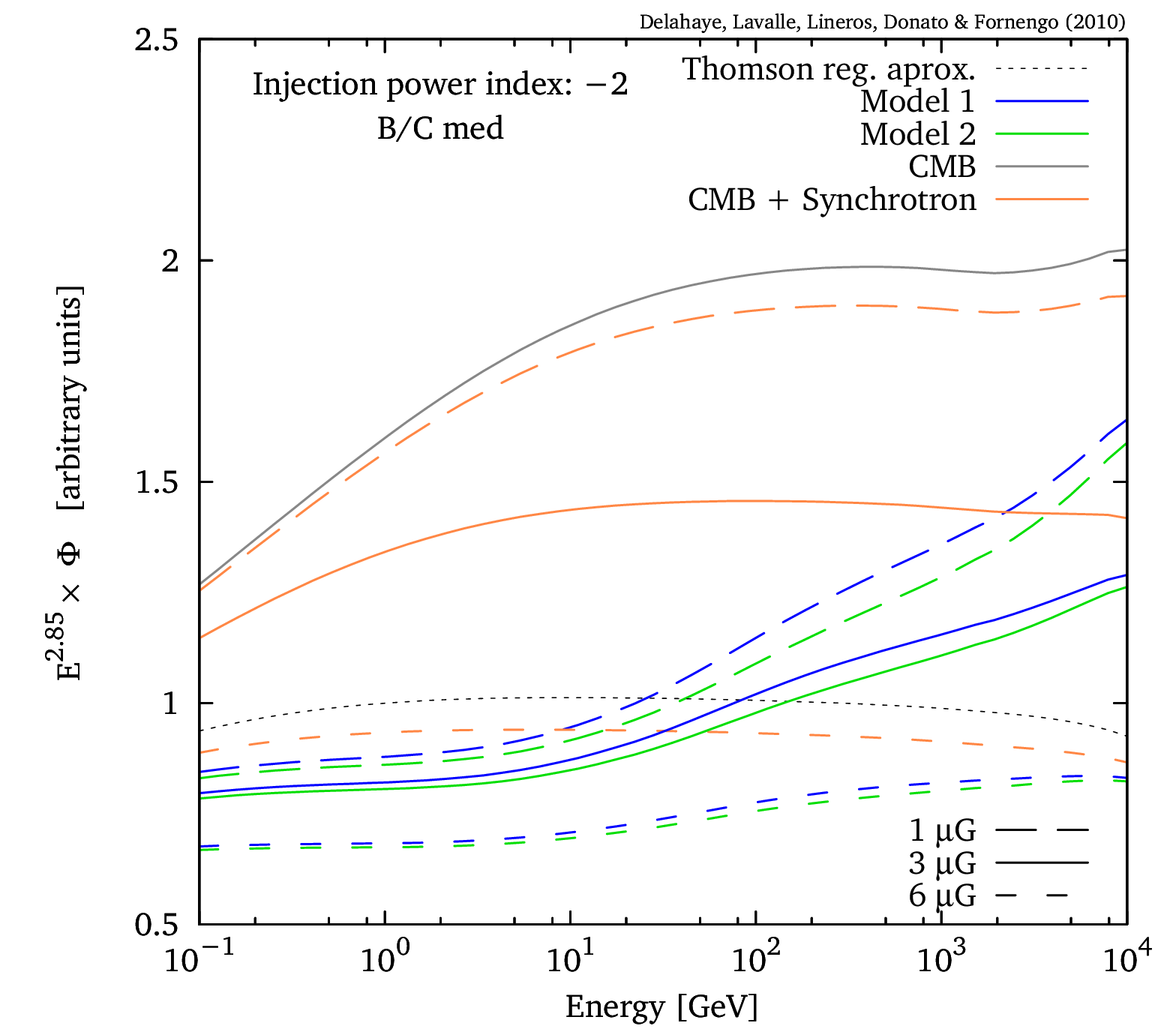}
\caption{\small Left: Energy density distribution of the ISRF averaged in 2 
  boxes of different volumes, where all components appear (data taken from the 
  analysis by~\citet{2008ApJ...682..400P}). Our models 1 and 2, using both 
  black-bodies for all components, are reported against the data.
  Middle: Corresponding energy loss rate. Right: Toy electron fluxes associated
  with the previous energy loss configurations, assuming
  an injection spectrum $\propto E^{-2}$.}
\label{fig:isrf}
\end{center}
\end{figure*}

In the left panel of \citefig{fig:isrf}, we report the ISRF data that we 
extracted from the analysis of \citet{2008ApJ...682..400P}, and averaged in 
cylinders of radius and half-height of $500$ pc (model 2, M2) and $2$ kpc 
(model 1, M1) about the Earth, on top of which we show that a sum of black-body 
distributions can provide a reasonable fit. These two models, defined with a set
of components characterized by their temperatures and energy densities, are 
summarized in \citetab{tab:e_losses}. 
They can be used to estimate the theoretical error coming from uncertainties in 
the characterization of the ISRF. We may assume that these uncertainties reflect
those affecting the data that we used, though error bars are not available. 
Note that the parameterizations appearing in \citetab{tab:e_losses} are not 
designed to reflect the true radiative physics operating in the ISM, which is 
beyond the scope of this paper. Nevertheless we observe an interplay between the
IR and UV components, depending on the averaging volume: taking a smaller 
volume results in a larger (smaller) IR (UV) contribution due to the efficient 
UV-absorption and IR-emission properties of the dust, which is mostly 
concentrated in the disk.

\begin{table}[t]
\begin{center}
\begin{tabular}{|c|c||c|c|}
\hline
& & $T_0$[K] & U$_{\text{rad}}$ [$10^{-11}$ GeV.cm$^{-3}$]\\
\hline
& CMB & 2.726 & Planckian (b-b)\\
\hline
\multirow{4}{*}{\rotatebox{90}{\textcolor{blue}{M1}}}& 
IR & $33.07$ & $25.4 $ ($4.5\,10^{-5}\times $ b-b) \\
\cline{2-4}
& Stellar & $313.32$ & $5.47$ ($1.2\,10^{-9}\times $ b-b)\\
\cline{2-4}
& \multirow{3}{*}{UV} & $3,249.3$ & $37$ ($7.03\,10^{-13}\times $ b-b)\\
& & $6,150.4$ & $22.9$ ($3.39\,10^{-14}\times $ b-b)\\
& & $23,209.0$ & $11.89$ ($8.67\,10^{-17}\times $ b-b)\\
\hline
\multirow{6}{*}{\rotatebox{90}{M2}} & IR & $33.653$ & 
$32.12$ ($5.3\,10^{-5}\times $ b-b) \\
\cline{2-4}
& Stellar & $313.32$ & $6.2$ ($1.36\, 10^{-9}\times $ b-b)\\
\cline{2-4}
& \multirow{3}{*}{UV} & $2, 901.13$ & $33.76$ ($1.01\,10^{-12}\times $ b-b)\\
& & $5,570.1$ & $25.93$  ($5.7\, 10^{-14}\times $ b-b) \\
& & $22,048.56$ & $10.26$ ($ 9.2\, 10^{-17}\times $ b-b) \\
\hline
\end{tabular}
\vskip 0.25cm
\caption{\small Parameters used to fit the local ISRF with black-bodies. They 
  correspond to fits performed on the data extracted from the analysis of 
  \citet{2008ApJ...682..400P} and averaged over cylinders of radius and 
  half-height of 2 kpc (model M1) and 0.5 kpc (model M2) about the Earth. We 
  choose M1 as our default ISRF model. In the last column, 
  (\mbox{$a\times $}b-b) indicates that the component can be obtained from a 
  standard black-body spectrum renormalized by a factor of \mbox{$a$}.}
\label{tab:e_losses}
\end{center}
\end{table}

The synchrotron emission can also be expressed as an inverse Compton scattering 
on a black-body distribution of virtual photons from the magnetic field. In 
this case, the characteristic energy of the radiation field is given by the 
cyclotron frequency
\begin{equation}
E_{B} = h\,\nu_c = \frac{h\,e\,B}{2\,\pi\,m} = 
1.16\times 10^{-14}\left[\frac{B}{1\,\mu{\rm G}}\right]\,{\rm eV}\;,
\end{equation} 
where $m$ is the electron mass, $k_B$ is Boltzmann's constant, and $B$ is the 
value of the Galactic magnetic field. It is clear that the condition 
$\gamma \,E_B\ll m\,c^2$ is fulfilled for the whole electron energy range
considered in this paper, so that the Thomson approximation is fully valid.
We estimate the local magnetic field relevant to the synchrotron losses 
to  $B_{\rm sync}\approx 1 \mu$G \citep[see \eg\ ][]{2001RvMP...73.1031F}, for
which the corresponding energy density derived from classical electrodynamics 
is $U_B = B_{\rm sync}^2/(2\mu_0)$.

The synchrotron energy losses do not depend on the mean value of the 
magnetic field \mymean{B}, but on the mean value of the {\em squared} 
field \mymean{B^2}. Although the mean value $\mymean{B} = B_r$,
namely the regular component of the magnetic field, the irregular component 
implies that $\mymean{B^2} > \mymean{B}^2$.

\citet{2010MNRAS.401.1013J} provide constraints on the different components of 
the magnetic field inside the disk. There are actually three components: the 
regular component $B_r$ and two irregular components (one aligned with the 
regular one $B_a$, the other completely isotropic $B_i$). The relevant value we 
need for the synchrotron losses is
\ben
B_{\rm sync} = \sqrt{ \mymean{B_r}^2 + \sigma^2 }\;,
\een
where it is easy to show that the variance is given by 
$\sigma^2 = \mymean{B_a^2} + \mymean{B_i^2}$. (Note that 
$\mymean{B_r^2} = \mymean{B_r}^2$).

From the results obtained by \citet{2010MNRAS.401.1013J}, we find that
\ben
\mymean{B_r} &=& 1 - 3 \, \mu{\rm G}\;, \\
\sqrt{ \mymean{B_i^2} } &=& 2.1 - 4.2 \, \mu {\rm G} \;\;\; 
[{\rm interarm - ridge}] \;,\nn \\
\sqrt{ \mymean{B_a^2} } &=& 0.0 - 3.3 \, \mu{\rm G} \;\;\; 
[{\rm interarm - ridge}] \;,\nn
\een
which translates into the range 
\ben
2.32\, \mu{\rm G} < B_{\rm sync} < 6.13\, \mu{\rm G}.
\label{eq:b_range}
\een
Nevertheless, we should not neglect the vertical dependence of the magnetic 
field, which is usually found to be exponential, with a typical scale of $\sim$ 
1 kpc. We emphasize that this scale is generally obtained from Faraday 
rotation measures as well as from observations of the all-sky polarized 
synchrotron emission, which translate into magnetic field intensity
only after deconvolution of the thermal and non-thermal electron density.
This density is usually grossly modeled by assuming the sum of Boltzmann and
single-index power-law spectra, and a $z$-exponential spatial dependence,
which is itself motivated from radio observations of the same synchrotron
emission \citep[see][for a recent analysis]{2009JCAP...07..021J}. Therefore,
these estimates might be affected by potentially large systematic errors.

We assume for simplicity that all components are constant in the disk 
(justified at the kpc scale around the Earth) and have the same vertical 
behavior, so that
\ben
B_{\rm sync} (r,z) = B_0 \exp\left\{ -\frac{|z|}{z_0}\right\}\;. \nn
\een
If we average $B_{\rm sync}^2$ inside a spherical volume of radius 2 kpc, 
which corresponds to the typical propagation scale for electrons, we find 
that $\sqrt{\mymean{B_{\rm sync}^2}} = 0.05\,(0.37) \times B_0$ assuming 
that $z_0 = 0.1\,(1.0)\,$ kpc. Given the range in \citeeq{eq:b_range}, we 
obtain
\ben
0.11\, \mu{\rm G} < & \mymean{B_{\rm sync}} & < 0.30\, \mu{\rm G} 
\;\;\;[z_0 = 0.1 \,{\rm kpc}] \;,\\
0.87\, \mu {\rm G} < &\mymean{B_{\rm sync}} & < 2.29\, \mu {\rm G} 
\;\;\; [z_0 = 1.0\,{\rm  kpc}]\;. \nn
\een
Using $z_0 = 0.1$ kpc is an extreme assumption, which is certainly not 
realistic. Nevertheless, assuming values in the range 1-3 $\mu$G seems 
reasonable.

Therefore, although there are uncertainties in the local value of the magnetic 
field, we suppose $B=1\,\mu$G in the following, so that the ISRF model M1 
complemented with the corresponding synchrotron losses leads to $\tau_l = 7.5
\times 10^{15}$ s in the Thomson approximation. The overall energy-loss rate 
$b(E)/E^2$ is plotted in the middle panel of Fig.~\ref{fig:isrf}, where it is 
shown to differ from the Thomson approximation with $\tau_l = {\rm cst} = 
10^{16}$ s very often used in the literature, and which appears as the dashed 
straight line. In particular, we observe a cascade transition due to 
Klein-Nishina effects, where it appears that the loss rate index $\alpha$ 
defined in \citeeq{eq:k_and_b} decreases step by step from 2, its Thomson 
value: at about 1 GeV, relativistic corrections become sizable for interactions 
with the main UV component which is felt less and less by electrons; then, 
across the range 10-100 GeV the IR component gradually loses its braking 
potential, and finally, above 10 TeV, interactions with CMB also cease. The 
value of the magnetic field sets the minimal value of the energy loss rate at 
higher energies. Since this latter is proportional to $B^2$, varying $B$ from 
1 to 3 $\mu$G translates into $\sim 1$ additional order of magnitude in the 
energy-loss rate at high energy, as also depicted in the middle panel of 
Fig.~\ref{fig:isrf}; for completeness, we also display the case of taking 6 
$\mu$G. Note that considering CMB only provides a robust estimate of the 
minimal energy-loss rate, which converts into a maximal flux by virtue of 
\citeeq{eq:disk_approx}; adding the synchrotron losses would instead define a 
next-to-minimal model for the energy losses.

In the right panel of Fig.~\ref{fig:isrf}, we quantify the impact of using 
different energy-loss models to derive IS flux predictions, for which we adopt 
the {\em med} propagation setup and a template injection spectrum $\propto 
E^{-2}$ homogeneously distributed in a thin disk. The dotted curve corresponds 
to the Thomson approximation with $\tau_l = 10^{16}$s, where we recover a flux 
with predicted index $\widetilde{\gamma} = \gamma + (\delta+1)/2 = 2.85$, as
predicted from \citeeq{eq:gamma_obs_disk} with $\alpha=2$. The higher 
curve is the flux obtained with the minimal case for the energy-loss rate 
(the minimal $\tau_l$), \ie~considering the CMB only, which provides the 
maximal flux. Indeed, in the Thomson approximation, the flux scales like 
$\sim \sqrt{\tau_l}=1/\sqrt{b_0}$, as seen from \citeeq{eq:disk_approx}. The 
index reaches a plateau around $\widetilde{\gamma} \approx 2.85$ in the 
range 10-1000 GeV, and then substantially hardens above 1 TeV because of 
relativistic effects. The next-to-minimal case exhibits the same feature, 
though the amplitude is slightly reduced, as expected. Finally, we report the 
flux associated with our complete models M1 and M2, both associated with a 
magnetic field of 1 $\mu$G (short dashed curves), 3 $\mu$G (solid curves), and 
6 $\mu$G (long dashed curves). We remark that the naive prediction of 
$\widetilde{\gamma}$ in the Thomson regime does not hold anymore, since the 
energy dependence of the energy loss $\alpha$ is no longer equal to 2, and the 
observed spectral index is significantly harder. Indeed, we have to consider 
instead an effective value $\alpha_{\rm eff}(E)\lesssim 2$ to account for 
relativistic effects. Taking a larger value of the magnetic field slightly 
softens the index and decreases the amplitude, as expected.

From this analysis of the local energy losses, we can estimate that the 
related uncertainties translate into a factor of $\lesssim 2$ in terms of IS 
flux amplitude ($\phi \propto \sqrt{\tau_l}$), and, above 10 GeV, 
$\pm 0.1$ in terms of spectral index (see the left panel of \citefig{fig:isrf}).
Note, however, that this crude spectral analysis is valid for a smooth 
distribution of sources only, \ie~for secondaries. We see in 
\citesec{subsec:local_snrs} that considering discrete nearby sources of 
primaries strongly modifies this simplistic view.

\section{Secondary CR electrons and positrons}
\label{sec:secondaries}

We performed an exhaustive study of the secondary positron flux
in~\cite{2009A&A...501..821D}, which is qualitatively fully valid for electrons 
and to which we refer the reader for more details.

Secondary electrons originate from the spallation of hadronic cosmic ray 
species (mainly protons and $\alpha$ particles) in the interstellar material 
(hydrogen and helium). This process produces also positrons, though different 
inclusive cross-sections come into play. Since spallation involves positively 
charged particles, charge conservation implies that it generates more positrons 
than electrons \citep[\eg][]{2006ApJ...647..692K}. This statement is not 
entirely accurate for neutron decay, but electrons arising from neutron decay 
have a very low energy (mostly E $<$ 10 MeV), thereby out of the energy range 
considered in this paper. In sum, the steady-state source term for secondaries
may in all cases be written as
\ben
{\cal Q}_s (E,\vec{x}) = 4\pi\sum_{i,j} 
\int dE' \,\phi_i(E',\vec{x}) \,\frac{d\sigma_{ij}(E',E)}{dE}\, 
n_j(\vec{x})\;,
\een
where $i$ flags the CR species of flux $\phi$ and $j$ the ISM gas species of 
density $n$, the latter being concentrated within the thin Galactic disk, and 
$d\sigma_{ij}(E',E)$ is the inclusive cross section for a CR-atom interaction 
to produce an electron or positron of energy $E$.

For our default computation, we selected the proton-proton cross-section 
parameterizations provided in \citet{2006ApJ...647..692K}. Any nucleus-nucleus 
cross-section (\eg~$p-He$ or $He-He$) can be derived from the latter by 
applying an empirical rescaling, usually by means of a combination of the 
involved atomic numbers. However, this rescaling is found to be 
different for the production of $\pi^{-}$ and $\pi^{+}$, or equivalently of 
$e^-$ and $e^+$. We used the prescriptions from \citet{2007NIMPB.254..187N} for 
this empirical rescaling.

Fits of the proton and $\alpha$ particle fluxes are provided in
\cite{2007APh....28..154S}, based on various measurements at the Earth. Finally,
we employed a constant density for the ISM gas, with $n_H = 0.9 \,{\rm cm^{-3}}$
and $n_{He} = 0.1  \,{\rm cm^{-3}}$, confining these species to a thin disk of
half-height $h=100$ pc. This is summarized in cylindrical coordinates by
\ben
n_j (\vec{x}) = \theta(h-|z|)\,\theta(R-r)\,n_j\;.
\label{eq:flat_disk}
\een
For this spatial distribution of the gas, the spatial integral of 
\citeeq{eq:flux_def} can be calculated analytically, following 
\citet{2009A&A...501..821D}; the solution is reported in 
\citesec{subapp:hom_disk}. We underline that this approximation is locally 
rather good over the whole energy range as long as the true gas distribution 
does not exhibit too strong spatial gradients over a distance set by the 
half-thickness $L$ --- this is discussed in more detail for primaries in 
\citesec{subsec:smooth_snrs}. In any case, this estimate is more 
reliable at high energies ($\gtrsim 100$ GeV) for which the signal is of 
local origin independent of $L$. At lower energies this approximation is 
expected to be valid for moderate $L\lesssim 4$ kpc, but much less trustworthy 
for large-halo models, as in the {\em max} propagation setup. For these extreme 
configurations, a more suitable description of the gas distribution would be 
necessary.

\begin{figure}[t]
\begin{center}
\includegraphics[width=\columnwidth,clip]{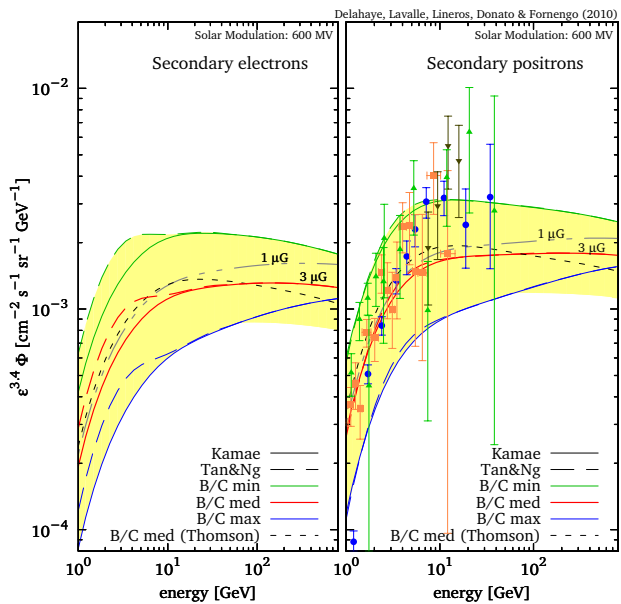}
\caption{\small Flux predictions of secondary electrons (left) and positrons 
(right) at the Earth, for the {\em min}, {\em med} and {\em max} propagation
models.}
\label{fig:secondaries}
\end{center}
\end{figure}

In Fig.~\ref{fig:secondaries}, we plot our results for the secondary 
electron (left-hand side) and positron (right-hand side) fluxes at the Earth.
For the solar modulation, we used the force-field approximation with 
a Fisk potential of $600$ MV~\citep{1971JGR....76..221F}. The solid curves 
are derived with the M1 ISRF model, a relativistic treatment of energy losses 
and nuclear cross-sections from \citet{2006ApJ...647..692K}. The yellow band 
is the flux range available for all sets of propagation parameters compatible 
with B/C constraints derived in~\citet{2001ApJ...555..585M}. We observe that 
the {\em min} ({\em max}) configuration provides the highest (lowest) and 
softest (hardest) flux, among the three beacon models. This can be clearly 
understood from~\citeeq{eq:disk_approx}, since the {\em min} configuration is 
characterized by the weakest value of diffusion coefficient normalization $K_0$ 
associated with the strongest index $\delta$, which is reversed in the {\em max}
configuration. However, we recall that these models were named so for sources 
distributed all over the diffusion halo, not only confined to the disk as is 
the case here.

The discussion in \citet{2009A&A...501..821D} on the theoretical uncertainties 
affecting secondary positrons is fully valid for secondary electrons. Aside from
energy losses, errors may originate from either uncertainties in the light 
nuclei flux, or uncertainties in nuclear cross-sections, or both. The former 
can be evaluated by using different fits of the local measurements, and by 
retro-propagating the CR nuclei flux to account for potential spatial gradients.
The latter may be estimated by considering alternative parameterizations of 
nuclear cross-sections. This is illustrated in \citefig{fig:secondaries} with 
the short-dashed curves computed with the nuclear cross sections of 
\citet{1983JPhG....9.1289T}, which are shown to differ from our default model 
only at low energy below a few GeV. All these effects were studied for positrons
in \citet{2009A&A...501..821D} and lead to an uncertainty of about 40 \%; 
this theoretical error is also valid for secondary electrons. Note, however, 
as we see later, that the electron flux is most likely to be dominated by the 
primaries, in contrast to positrons for which secondaries are a major component.
Therefore, uncertainties in the secondary contribution has more impact for 
positrons than for electrons.

Finally, we emphasize that the present results differ slightly from 
those derived in \citet{2009A&A...501..821D} because the energy losses 
are now treated in a fully relativistic formalism. Not only does this slightly
change the normalization at low energy by a factor 
$\sqrt{\tau_l/\tau_{\rm D09}}\approx 0.9$, but, more importantly, this hardens 
the spectral shape due to Klein-Nishina effects. This is more striking in 
\citefig{fig:secondaries}, where the long-dashed curves are the predictions 
calculated in the {\em med} configuration and the Thomson limit for the energy 
losses. In \citeapp{app:fit}, we provide user-friendly fitting formulae that 
closely reproduce our calculations of the secondary electron and positron 
fluxes in the \propmed~propagation setup.

\section{Primary electrons}
\label{sec:primary_el}

The GeV-TeV CR electron flux at the Earth is dominated by a primary component
originating from electrons accelerated in both SNRs and pulsar wind nebul\ae
(PWN)~\citep{1987PhR...154....1B}. CR sources are therefore connected to the 
explosion of supernov\ae (SNe), and we discuss a more general framework in 
\citesec{sec:primary_pos}. Predicting this primary contribution is a rather 
difficult exercise because it involves characterizing the energy distribution 
of these electrons at sources and their spatial distribution, in addition to 
their transport to the Earth. Moreover, since GeV-TeV electrons have a short 
range propagation scale, local sources are expected to play an important role. 
Rephrased in statistical terms, since the number of sources exhibits large 
fluctuations across short distances, the variance affecting the predictions is 
expected to increase with energy, when the effective propagation volume 
decreases. Therefore, fluctuations in the properties of local sources will have 
a strong impact. In contrast, the low energy part of the CR electron spectrum 
might be safely described in terms of the average source properties, namely a 
smooth spatial distribution associated with a mean injected energy distribution.

In the following, we estimate the primary flux of electrons and quantify the 
associated theoretical errors. More precisely, we wish to quantify the relative 
origin and impact of these uncertainties. To do so, we first discuss in 
\citesec{subsec:snr_prop} the spectral shape properties that we consider in the 
forthcoming calculations, focusing on SNRs for the moment (pulsars are discussed
in \citesec{sec:primary_pos}, together with the primary positrons); available 
spatial distributions are presented in \citesec{subsec:smooth_snrs}. Then we 
discuss the uncertainties associated with the modeling of a single source 
in~\citesec{subsec:local_snrs}. We finally discuss the primary flux and related 
uncertainties in \citesec{subsec:el_pred}, in which we make a thorough census 
of the local SNRs likely affecting the high energy part of the spectrum.

\subsection{Spectral properties of SNRs and related constraints}
\label{subsec:snr_prop}

Most SNR models \citep[\eg][]{2007ApJ...661..879E,2009A&A...499..191T} rely 
on the acceleration of CRs at non-relativistic shocks
\citep[\eg][]{2001RPPh...64..429M} and predict 
similar energy distributions for the electrons released in the ISM, which can 
be summarized as
\ben
{\cal Q}(E) = {\cal Q}_0 \epsilon^{-\gamma} 
\exp\left\{ - \frac{E}{E_c}\right\}\,.
\label{eq:spectrum}
\een
The spectral index $\gamma$ is usually found to be around 2 over a significant 
energy range \citep{2007ApJ...661..879E} --- but to exhibit large variations at 
the edges --- in agreement with radio observations. Gamma-ray observations 
suggest that the energy cut-off $E_c$ is greater than a few TeV 
\citep[\eg][]{2009ApJ...692.1500A}. These studies find rather similar indices 
for protons and electrons. Since protons of energy above a few GeV are barely 
affected by energy losses and have a long-range propagation scale, the proton 
spectrum measured at the Earth can provide information about the mean index at 
sources. Since $\phi_p \propto {\cal Q}(E)/K(E)\propto E^{-\widetilde{\gamma}}$, 
the index at source is therefore $\gamma\approx \widetilde{\gamma} - \delta$.
With $\widetilde{\gamma}\simeq 2.8$ and $\delta$ in the range 0.5-0.7, one
finds $\gamma$ in the range 2.0-2.3, in rough agreement with theoretical 
predictions. We discuss complementary constraints from radio observations at the
end of this subpart.

Aside from the spectral shape, sizing the value of the normalization 
${\cal Q}_0$ is much more problematic. To describe a distribution of sources in 
the Galaxy, one usually assumes that the high energy electron injection is 
related to the explosion rate of SNe, so that we can suppose that $Q_0$ is such 
that the total energy carried by electrons is given by
\ben
\int_{E_{\rm min}}^{\infty} dE'\, E'\, {\cal Q}(E')=f \, E_\star \, \Gamma_\star\;,
\label{eq:Q0}
\een
where $\Gamma_\star$ is the SN explosion rate, $E_\star$ is the kinetic energy 
released by the explosion, and $f$ is the fraction of this energy conferred
to electrons. Since we are only interested in the non-thermal electrons, we 
assume that $E_{\rm min}=0.1$ GeV. Note that the spectral index influences 
the normalization procedure sketched above. For a single source, the same 
expression holds but without $\Gamma_\star$ --- we discuss this case later on. 
For convenience, we further define the quantities
\ben
\label{eq:def_etilde}
\etilde &\equiv&  f \, E_\star\;, \\
\tgammae &\equiv & \Gamma_\star \, \etilde\;,\nn
\een
which also helps us to discuss the normalization issue.

Constraining $\Gamma_\star$, $E_\star$, and $f$, is a difficult exercise. 
The explosion rate of SNe is typically predicted to be a 1-5 per century and
per galaxy \citep[\eg][]{1991ARA&A..29..363V,1998MNRAS.297L..17M}, which is 
consistent with observations 
\citep[\eg][]{1998ApJ...505..134V,2006Natur.439...45D}.
Nevertheless, SNe are of different types, and may thereby lead to different 
CR acceleration processes. About $2/3$ of SNe are expected to be core-collapse 
SNe (CCSNe), the remaining $1/3$ consisting of type 1a SNe (SNe1a). This is 
at variance with the statistics derived from observations, which identify a
higher percentage of the more luminous latter type. CCSNe are often observed in 
star formation regions and spring from the collapse of massive stars 
$\gtrsim 8\,\Msol$, while SNe1a, produced by older accreting white dwarfs, are 
more modest systems, while being more widely distributed.

Explosions of CCSNe with masses $\lesssim 20\,\Msol$ can typically liberate a 
huge amount of energy, $\sim 10^{53-54}$ erg, $\sim 99$\% of which is released 
in the form of neutrinos when integrated after the cooling phase of the
proto-neutron star~\citep[\eg][]{2000Natur.403..727B,2005NatPh...1..147W,2007PhR...442...38J}. The total kinetic energy available from CCSN explosions, 
mostly due to the energy deposit of the neutrinos in the surrounding material, 
is about $\sim 10^{51}$ erg. CCSNe usually give rise quite complex systems 
characterized by SNRs beside (or inside) which one can find active neutron 
stars such as pulsars and associated wind nebul\ae~(PWN) --- we focus on 
pulsars in \citesec{sec:primary_pos}. In contrast, SNe1a are much more modest 
systems, the explosions of which blast all the material away without giving 
birth to any compact object and release an energy of about $10^{51}$ erg, which
is transferred entirely to the surrounding medium~\citep[\eg][]{1984ApJ...286..644N,2003Sci...299...77G,2007Sci...315..825M}. 
It is interesting to remark that despite the quite different natures of CCSNe 
and SNe1a, both types feed the ISM with the same typical kinetic energy. Since 
we consider only SNRs in this part, we assume that $E_\star = 10^{51}$ erg in 
the following.

The fraction of SN energy conferred to electrons was studied in 
\citet{2009A&A...499..191T}, and found to be about $f~\sim~10^{-5}-10^{-4}$. 
This result is rather independent of the exact values of the spectral index 
$\gamma$ and the cut-off energy $E_c$, but the theoretical error is still 
of about one order of magnitude.

At this stage, we emphasize that the theoretical uncertainty in $\tgammae = 
f\,E_\star\,\Gamma_\star$ already reaches about 2-3 orders of magnitude on 
average, which is quite huge and translates linearly in terms of flux.

For known single sources, it is possible to derive tighter constraints on the 
individual normalizations ${\cal Q}_0 $ from observations at wavelengths for 
which electrons are the main emitters. This is precisely the case for the 
non-thermal radio emission produced by synchrotron processes, provided the 
magnetic field is constrained independently. The synchrotron emissivity 
associated with an electron source of injection rate ${\cal Q}(E)$ is given by
\citep{1965ARA&A...3..297G,1970RvMP...42..237B,1994hea..book.....L}
\ben
J(\nu) = \frac{1}{2} \int_{0}^{\pi} d\theta\,\sin(\theta) 
\int d\nu \, P_{s}(\nu,\theta) \, {\cal Q}(E=h\nu)\;,
\een
where an average is performed over the pitch angle $\theta$, and where the 
synchrotron radiation power is defined as
\ben
P_{s}(\nu,\theta) &=& 
\frac{\sqrt{3}\, e^3\, B}{4\, \pi\, \epsilon_0 \, m\, c}\,
x\int_{x/\sin(\theta)}^{\infty} dy\,K_{5/3}(y)\nn\\
{\rm where} \;\; x &\equiv& \frac{\nu}{\nu_s}\;;\;
\nu_s\equiv \frac{3}{2}\,\gamma^2\,\nu_c = 
\frac{3\,e\,B\gamma^2}{4\,\pi\,m}\;,
\een
and $\nu_s$, which is called the {\em synchrotron peak frequency}, corresponds 
to the average frequency of the synchrotron emission arising when an electron 
of Lorentz factor $\gamma$ interacts with a magnetic field $B$, and is $\propto 
\gamma^2 \nu_c$, where $\nu_c$ is the cyclotron frequency. We note the parallel 
between the synchrotron process and the the inverse Compton process: for the 
latter, an initial photon of energy $E_k$ would indeed be boosted to 
$\sim\gamma^2 E_k$. We could calculate the radio flux from the emissivity, but 
it is also interesting to derive a more intuitive expression, which actually 
provides a fair approximation~\citep{1994hea..book.....L}
\ben
\frac{d\phi(\nu)}{d\nu}\,d\nu \simeq 
\frac{\left[b(E)\right]_{\rm sync}}{4\,\pi\,d^2\,h\nu} 
\,{\cal Q}(E)\,dE\;,
\een
where we assume that the entire radiation is emitted at the synchrotron peak 
frequency $\nu = \nu_s$, which links $\nu$ to $E$. The distance from the 
observer to the source is denoted by $d$. We remark that only the synchrotron 
part of the electron energy loss rate $b(E)$ appears --- we therefore assume 
that this is the most efficient process within the source --- and that we 
neglect the possible reabsorption of the synchrotron emission. Since $b(E)$ is 
the energy lost by an electron, it corresponds to the energy of the emitted
photon, so the factor $1/(h\nu)$ allows us to infer the number of photons.

This expression allows us to constrain ${\cal Q}_0$ by means of the source radio
brightness $B_r(\nu)$, which is usually found in catalogs
\ben
B_r(\nu) &=& \frac{1}{\delta\nu} \int_{\nu}^{\nu+\delta\nu} d\nu' \,h\nu'\,
\frac{d\phi(\nu')}{d\nu'}\\
&\overset{\delta\nu\rightarrow 0}=& 
\frac{\left[b(E)\right]_{\rm sync} }{4\,\pi\,d^2}\, 
\,{\cal Q}(E)\,\frac{dE}{d\nu}\;.\nn
\een
We readily derive
\ben
{\cal Q}_0 &= \frac{4\,\pi\,d^2}{\left[b(E)\right]_{\rm sync}}\,
\left(\frac{E}{E_0}\right)^\gamma \,\frac{d\nu}{dE}\,B_r(\nu)\;,
\een
which translates into
\begin{align}
\label{eq:q0_num}
\frac{{\cal Q}_0}{\rm GeV^{-1}}&= 1.2\times 10^{47} \times (0.79)^{\gamma}\\
&\times\,
\left[\frac{d}{\rm kpc} \right]^2\,
\left[\frac{\nu}{\rm GHz} \right]^{\frac{\gamma-1}{2}}\,
\left[\frac{B}{100\,\mu {\rm G}} \right]^{-\frac{(\gamma+1)}{2}}\,
\left[\frac{B_r(\nu)}{\rm Jy} \right]\;.\nn
\end{align}
We have just recovered the well-known relation between the radio index 
and the electron index, $\gamma_r = (\gamma-1)/2$.

\begin{figure*}[t]
\begin{center}
\includegraphics[width=\columnwidth,clip]{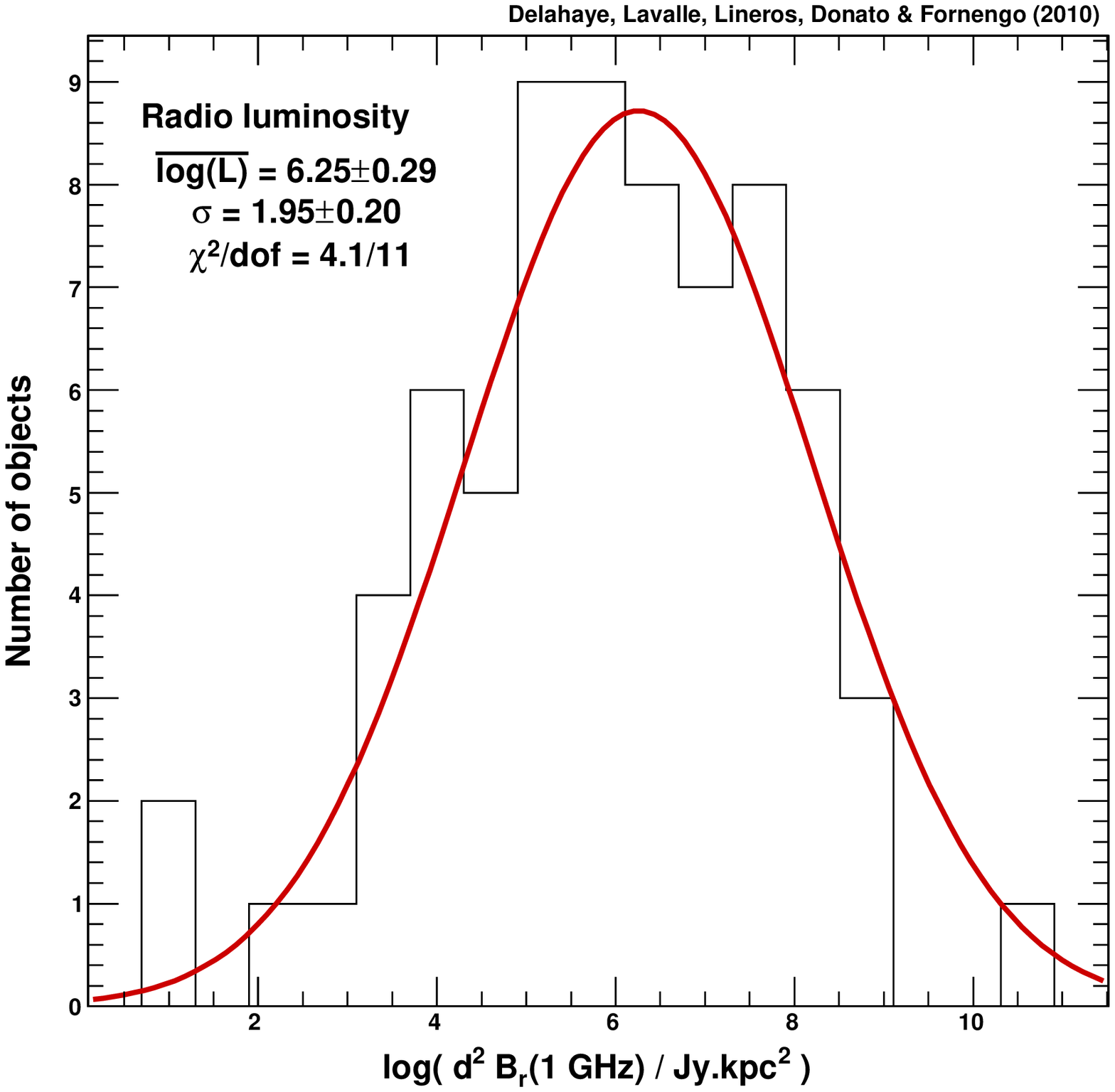}
\includegraphics[width=\columnwidth,clip]{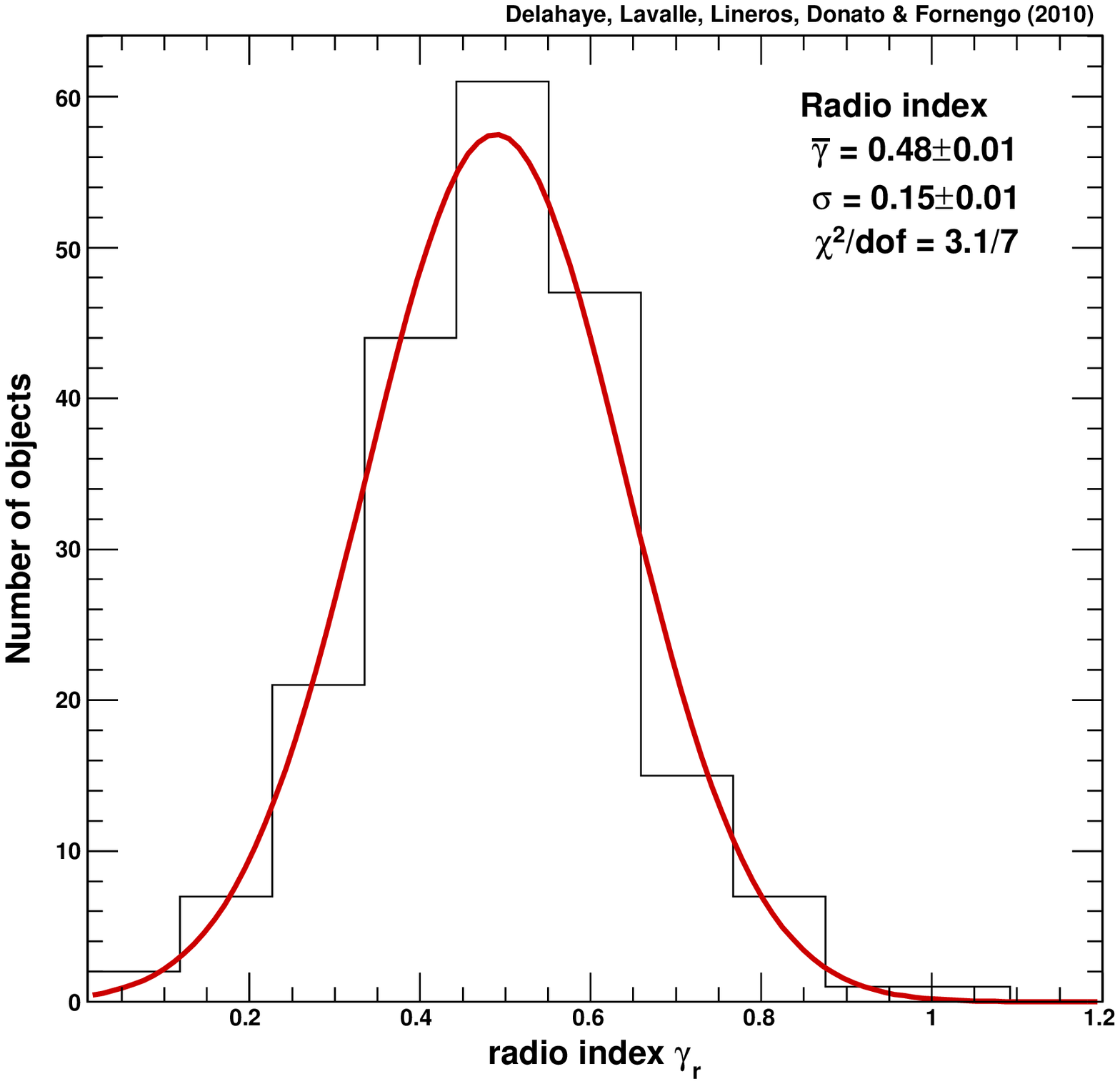}
\caption{\small Left: histogram of SNR luminosities --- 
${\cal L}/(4\,\pi) = d^2 \, B(1\,{\rm GHz})$. Right: histogram of SNR radio 
spectral indices. The SNR data are taken from the Green catalogue 
\citep{2009BASI...37...45G}.
}
\label{fig:Stats}
\end{center}
\end{figure*}

An up-to-date catalog of SNRs can be found in \citet{2009BASI...37...45G}, 
which contains $\sim$265 objects. Among these objects, only 70 have estimated 
distances to the Earth, and 207 have measured radio spectral indices. 
Observations, however, are not expected to reflect the actual statistical 
properties of the whole population of Galactic SNRs because of observational 
selection effects favoring the brightest sources and sites of fainter background
(high longitudes, towards the anticenter). Disregarding the spatial distribution
of these objects, which is probably strongly biased, this sample may still be 
fairly representative of their general spectral properties 
\citep{2005MmSAI..76..534G}.

We compiled histograms of the measured radio indices and the estimated 
intrinsic luminosities --- ${\cal L}/(4\,\pi) = d^2\,B_r(1\,{\rm GHz})$ --- in 
the right and left panels, respectively, of Fig.~\ref{fig:Stats}. The radio 
indices clearly appear to exhibit a Gaussian distribution, whereas luminosities 
follow a log-normal distribution. This points towards similar physical grounds 
for the electron properties at sources, which is obviously unsurprising. With 
these distributions, we can derive mean values and statistical ranges for the 
parameters. We find that $\langle \gamma_{r} \rangle = 0.50 \pm 0.15$ and 
$\langle d^2 \,B_r(1\,{\rm GHz}) \rangle = \exp\left\{6.26 \pm 1.95\right\}$ 
Jy.kpc$^2$. One can therefore infer that the electron index 
$\langle \gamma \rangle  =  2 \, \langle \gamma_{r} \rangle + 1 = 2.0 \pm 0.3$ 
in very good agreement with theoretical expectations. Although this relation 
between the radio index and the electron index is not entirely accurate (because
of other radio components or absorption), and although some systematic errors 
also affect the data, this provides a complementary means of sizing the 
uncertainty, which is consistent with that of theoretical results.

We use this statistical information to directly constrain the single source 
normalization ${\cal Q}_0$ from \citeeq{eq:q0_num}, but we need to estimate the 
magnetic field in SNRs. From the observational point of view, information about
the electron density and magnetic field at sources is degenerate. More insights 
may come from theoretical studies of the amplification of magnetic fields in 
sources from numerical simulations, which involve CRs themselves as seeds and 
amplifiers. The current state-of-the-art simulations 
\citep[\eg][]{2000MNRAS.314...65L} support $B\sim 100\,\mu$G, in agreement 
with observations, and that we use here. With this value, we finally find 
$\langle {\cal Q}_0 \rangle = 3.9\times 10^{49}\,{\rm GeV^{-1}} $ for an index 
$\gamma = 2$, which translates into $\langle \etilde  \rangle  \simeq  4.3
\times 10^{50}\,{\rm GeV} \simeq 6.9\times 10^{47} \,{\rm erg}$ (with a cut-off 
$E_c = 10$ TeV). This is in rough agreement with the other values derived above,
but probably biased, as expected, towards the brightest objects.

\subsection{Spatial distribution of sources}
\label{subsec:smooth_snrs}

Although GeV-TeV electrons have a short-range propagation scale, the injection 
rate of energy discussed above is insufficient to describe the Galactic CR 
electrons. We need to specify the spatial distribution of sources. For nearby 
sources, for which observational biased are less prominent, we can use 
available catalogs, which may provide a rather good description of the local 
CR injection. Nevertheless, for more distant sources, which have influence on 
the intermediate energy range $\sim$1-100 GeV, we have to rely on a distribution
model.

Since $2/3$ of SNe are expected to be CCSNe, one can use pulsars as tracers
of the SNR distribution, instead of SNRs themselves, the observed population
of which is much more modest. As an illustration, the ATNF catalog\footnotemark 
~\citep{2005AJ....129.1993M} lists more than 1800 pulsars compared to the $\sim$
265 SNRs contained in~\citet{2009BASI...37...45G}. Nevertheless, a too naive 
use of the statistics would lead to errors since it is well known that data do 
not reflect reality faithfully because of detection 
biases~\citep[\eg][]{2004IAUS..218..105L}.

\footnotetext{\url{http://www.atnf.csiro.au/research/pulsar/psrcat}}

There are few distribution models available in the literature. Since the 
energetics associated with the source injection (birth) rate has been discussed 
above, we are only interested in the {\em normalized} source distribution here. 
Consequently, the normalization coefficient in front of each model is fixed such
that it normalizes the spatial distribution to unity within the diffusion halo 
characterized by its radius $R$ and half-thickness $L$. Moreover, in the 
following, we set the position of the Sun at $\Rsol = 8$ kpc from the Galactic 
center\footnotemark.

\footnotetext{Some of the distributions listed in this paragraph are actually 
derived assuming 8.5 kpc, but we disregard this small change to make 
the discussion easier.}

Most of models have radial and vertical dependences of the form
\ben
\rho(r,z) = \rho_0 \, r^a \, \exp\left\{-\frac{r}{r_0}\right\}
\, \exp\left\{-\frac{|z|}{z_0}\right\}\;,
\label{eq:rho}
\een
where $\rho_0$ ensures the normalization to unity. For simplicity, we discuss 
only differences of the radial distributions in the following, since the 
vertical distribution is fairly similar among studies. We thereforee keep 
fixed the vertical dependence as in the above equation, with $z_0 = h = 0.1$ 
kpc, throughout the paper.

Different sets of values can be found in the literature for the pair $(a,r_0)$. 
\citet{2004IAUS..218..105L}, hereafter L04, found $(2.35,1.528\, {\rm kpc})$; 
\citet{2004A&A...422..545Y}, hereafter YK04, derived $(4,1.25\, {\rm kpc})$; 
while \citet{1990ApJ...348..485P}, hereafter P90, early determined 
$(1,4.5\, {\rm kpc})$. Finally, in contrast to the parameterization sketched 
above, we recall the distribution proposed by~\citet{1998ApJ...504..761C}, 
hereafter CB98, though it was obtained from a fit to data of poor statistics 
for 36 SNRs
\ben
\rho(r,z) = \rho_0 \, \sin\left(\pi \frac{r}{r_s} + \theta\right) \, 
\exp\left\{-\frac{r}{r_0}\right\} \, \exp\left\{-\frac{|z|}{z_0}\right\}\;,
\een
where we have added the same vertical term as in \citeeq{eq:rho}. The 
authors found  $r_0=7.7\pm 4.7$ kpc, $r_s=17.2\pm1.9$ kpc, and 
$\theta = 0.08 \pm 0.33$. This relation is only valid for 
$r< r_s(1-\theta/\pi)$, \ie~within 16.8 kpc, and null beyond. Note, however, 
that \citet{2006ApJ...639L..25B} reported the detection of 35 new remnants in 
the inner Galaxy, and suggest that former radial distribution estimations should
be revised.

\begin{figure*}[t]
\begin{center}
\includegraphics[width=0.67\columnwidth,clip]{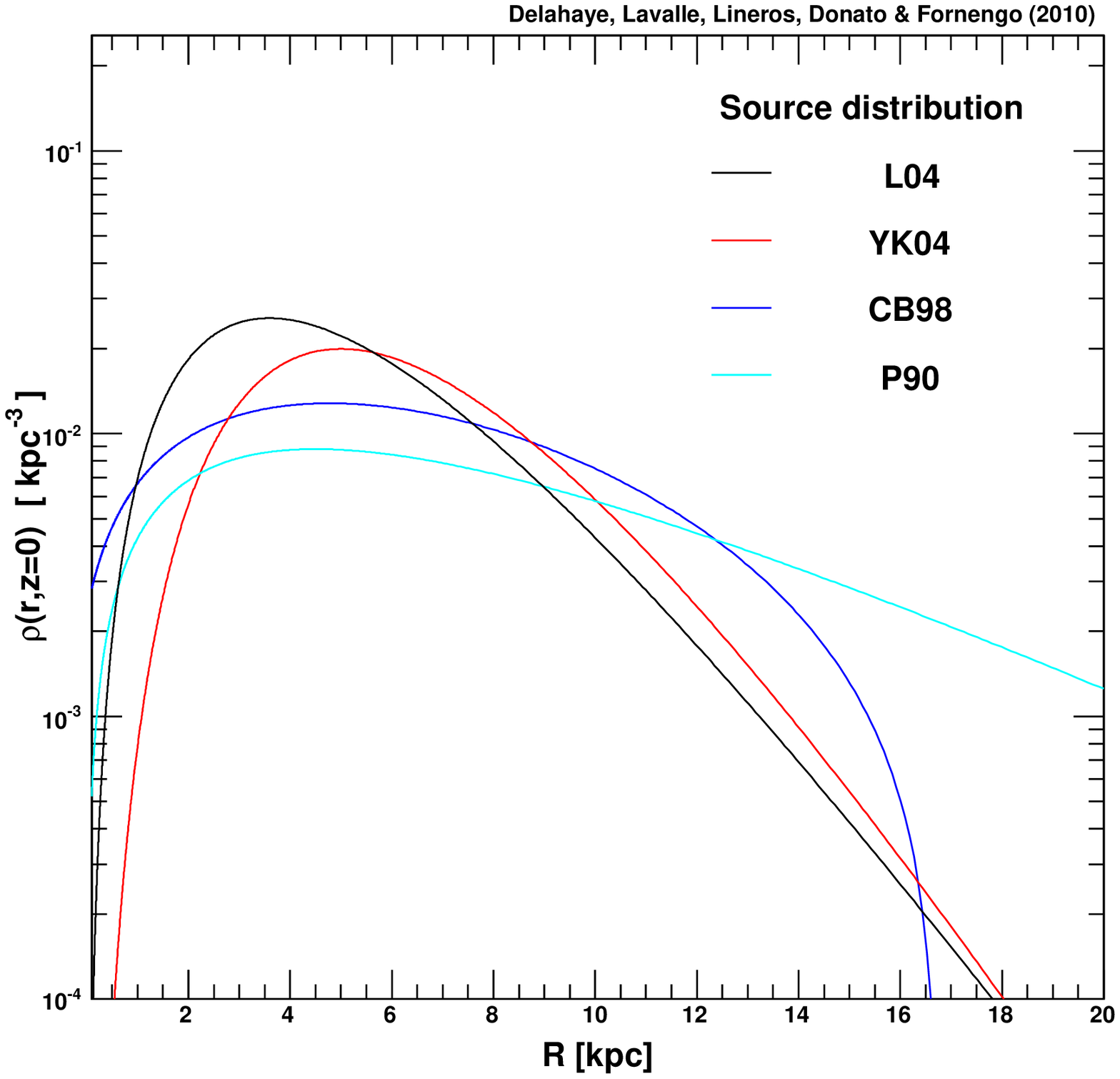}
\includegraphics[width=0.67\columnwidth,clip]{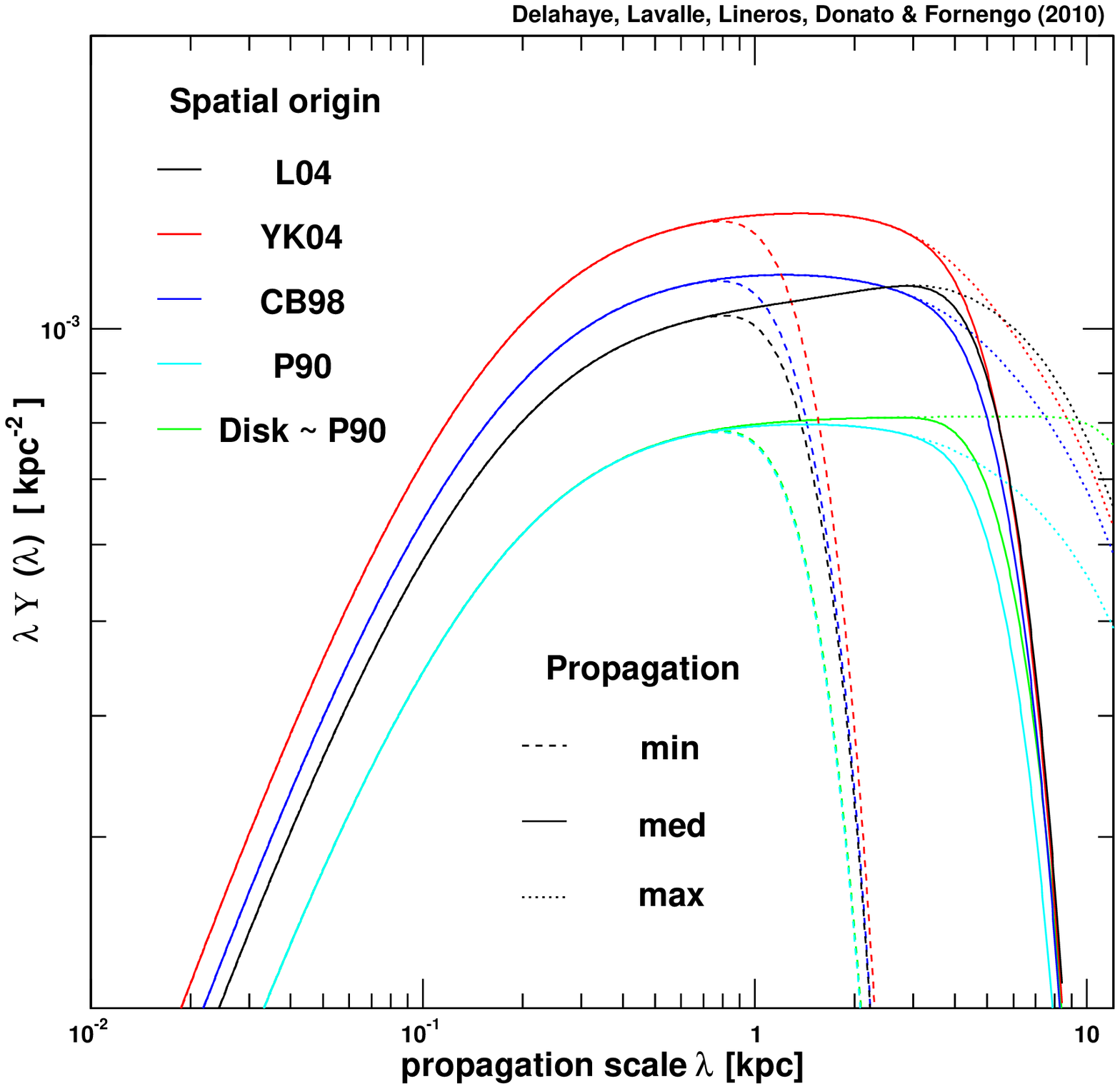}
\includegraphics[width=0.67\columnwidth,clip]{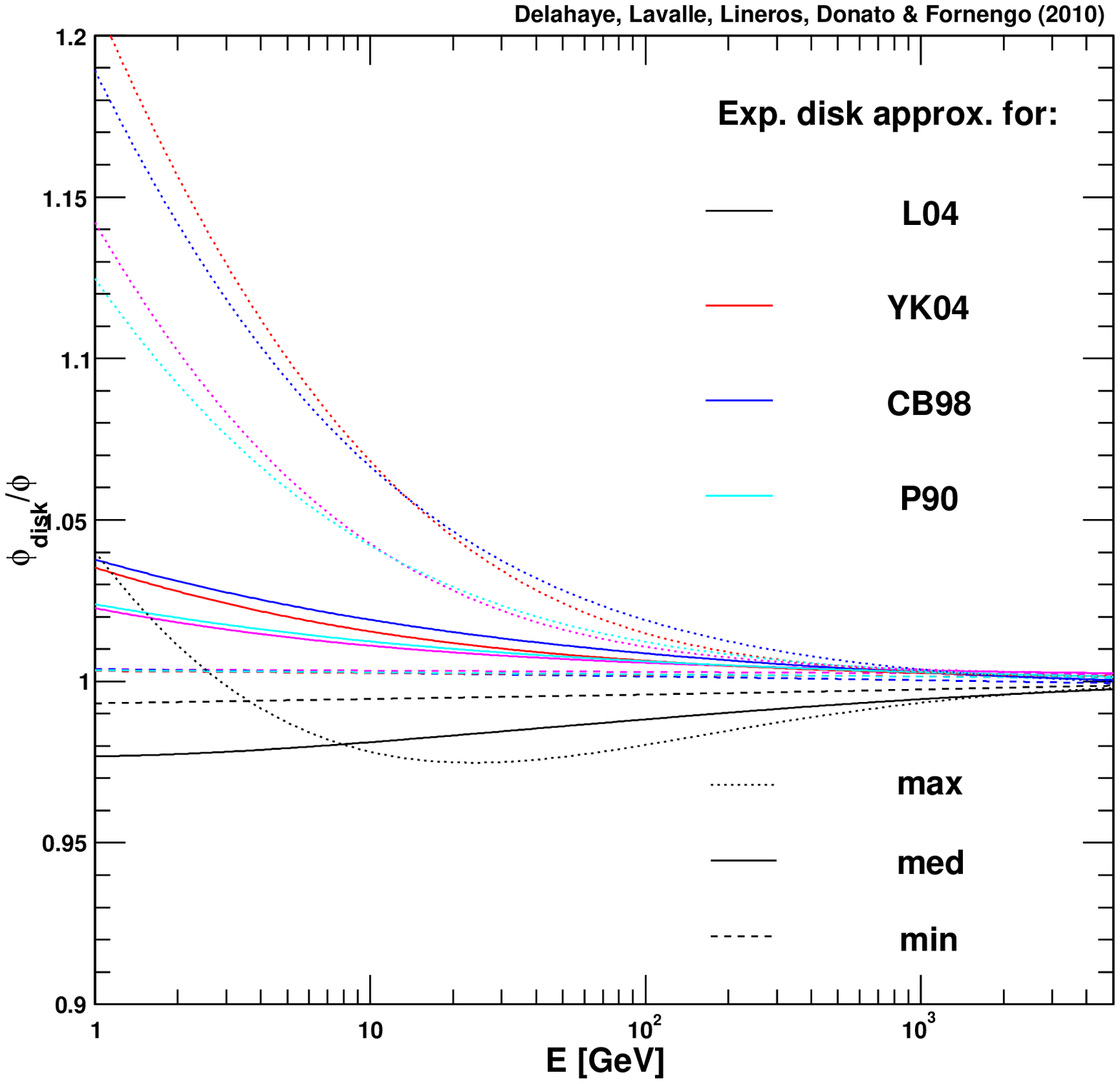}
\caption{\small Left: Spatial distribution models for SNRs and pulsars. Middle:
  Corresponding halo functions defined in \citeeq{eq:eta} and that characterize 
  the transport probability as a function of the propagation scale $\lambda$, 
  which decreases with energy. Right: Ratio of flux predictions to their 
  associated exponential disk approximations.}
\label{fig:spatial_dist}
\end{center}
\end{figure*}

To understand the deviations induced in the electron flux prediction
when using these different distributions, it is convenient to define the 
following {\em halo function}
\begin{align}
\label{eq:eta}
\Upsilon_\odot (\lambda) &= \int d^3\vec{x}_s\,\rho(r_s,z_s) \, 
\greenf_\lambda(\lambda,\vec{x}_\odot\leftarrow\vec{x}_s)\\ 
{\rm where}\;\; &\greenf_\lambda (\lambda,\vec{x}_\odot\leftarrow\vec{x}_s) 
\equiv b(E) \, \greenf(E,\vec{x}_\odot \leftarrow E_s,\vec{x}_s )\nn\;,
\end{align}
which determines the probability of an electron reaching the Earth given its 
propagation scale $\lambda$ --- see \citeeq{eq:def_lambda} --- and the 
normalized spatial distribution of source $\rho$. The electron flux is the 
the energy integral of the product of this probability and the source spectrum, 
such that the shape of this probability function provides a preliminary taste of
the final result. More importantly, it allows us to connect the spatial origin 
of the signal with energy, through the propagation scale $\lambda$.

In the left panel of \citefig{fig:spatial_dist}, we plot the spatial 
distributions listed above as functions of the galactocentric radius $r$, and in
the galactic plane ($z=0$). We see that except for the solar neighborhood, where
relative amplitudes can vary by a factor of $\sim$2 at most, the spatial 
distributions in the direction of the Galactic center and towards the anticenter
are quite different from each other. Nevertheless, these differences are 
significantly lower in terms of $\Upsilon_\odot(\lambda)$, because of the 
spatial average --- see \citeeq{eq:eta}. This is shown in the middle panel of 
Fig.~\ref{fig:spatial_dist}, where we have plotted $\lambda\times
\Upsilon_\odot(\lambda)$ as a function of $\lambda$ for the different spatial 
distributions and for the {\em min}, {\em med}, and {\em max} propagation 
setups. We see that the probability is maximal and constant --- 
$\lambda\,\Upsilon_\odot$ grows linearly with $\lambda$ --- for short 
propagation scales up to $\lambda\sim h = 0.1$ kpc. Then, the probability 
decreases linearly with $\lambda$ --- $\lambda\,\Upsilon$ exhibits a plateau ---
before shrinking exponentially when $\lambda\sim L$, $L$ being larger and larger
from the {\em min} setup to the {\em max} setup. Each spatial 
distribution model is characterized by a very similar curve that differs mostly
in terms of amplitude. This can be understood in the following manner: when 
$\lambda<h$, the source can be considered as homogeneous in 3D space, then 
$\Upsilon_\odot \propto \int dr\, r^2 \lambda^{-3}\,\exp\{-r^2/\lambda^2\} = 
{\rm cst}$; when $h<\lambda<L$, since the source distributions do not exhibit 
strong radial variations on the kpc scale, they can be considered as thin disks,
and one recovers the solution $\Upsilon\propto \lambda^{-1}$ derived in 
\citeeq{eq:disk_approx}; for $\lambda>L$, electrons escape the diffusion zone. 
This points towards the possibility of modeling, while only locally, the source 
distribution with a z-exponential infinite disk, for which full analytical 
solutions of the spatial integral exist. The green curve in the middle panel of 
Fig.~\ref{fig:spatial_dist} is the z-exponential disk approximation associated 
with P90, as an illustration, and is shown to provide a rather good 
approximation except for large diffusion thickness $L\gtrsim 4$ kpc. The 
z-exponential disk approximation is defined in cylindrical coordinates as
\ben
\rho_d(r,z) = \rho(\Rsol,0)
\,\theta(R-r)\,\exp\left\{-\frac{|z|}{z_0}\right\}\;,
\een
where $\rho(\Rsol,0)$ is the local value of the normalized density given in
\citeeq{eq:rho}. This approximation is valid for local predictions provided
the spatial distribution $\rho$ does not vary significantly over a distance 
$\sim L$, which is the case for moderate $L$. In the right-hand side panel of 
\citefig{fig:spatial_dist}, we compare the disk approximation with the full 
calculation in terms of fluxes: for different spatial distributions, we plot the
ratio {\em approximated flux} / {\em exact flux} for our three beacon 
propagation setups. We can see that the exponential disk approximation is quite 
good above a few GeV for the {\em min} and {\em med} cases, as expected, having 
an accuracy better than 5\%. Errors are obviously larger in the {\em max} case 
because of the larger spatial gradients exhibited by the spatial distributions 
within $L = 15$ kpc.

\begin{figure}[t]
\begin{center}
\includegraphics[width=0.9\columnwidth,angle=-90,clip]{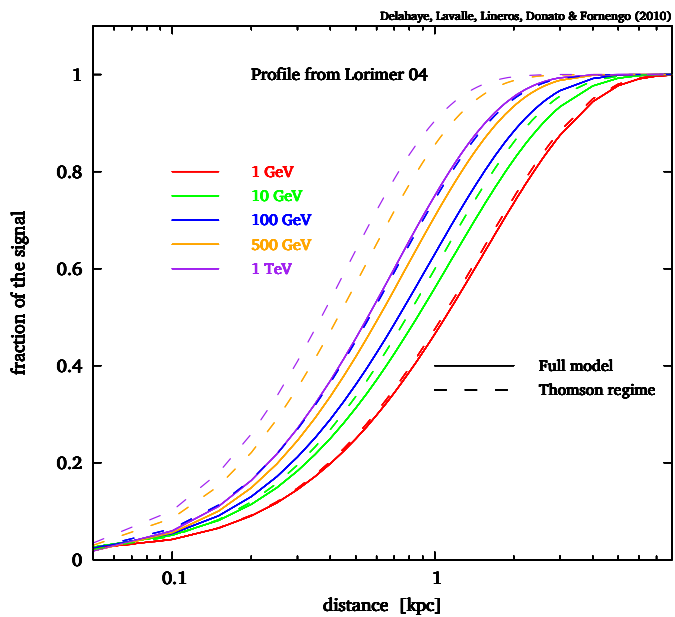}
\caption{\small Fraction of the signal reaching the Earth as a function of the
  integrated radius, for different energies and different spatial distribution
  models, using full relativistic energy losses. The Thomson approximation 
  result is reported in dashed line.}
\label{fig:signal_frac}
\end{center}
\end{figure}

A final useful exercise regarding the smooth spatial distribution modeling 
consists of checking the cumulative fraction of the IS signal received at the 
Earth as a function of the radial integration distance. In 
\citefig{fig:signal_frac}, we report this fraction for spatial model L04 at 
different energies, assuming an injection spectrum $\propto E^{-2}$, and 
for both the Thomson approximation and the relativistic energy losses. We 
see that this fraction increases more quickly at high energy than at low energy,
as expected from energy losses. This is consistent with the result obtained 
in \citet{2009A&A...501..821D} for secondary positrons. Nevertheless, above 
$\sim$10 GeV, we can observe that relativistic effects come into play and a 
difference appears between the Thomson approximation case and the relativistic 
case. Indeed, the latter induces a longer propagation scale at high energy, 
and consequently softens the rise of the cumulative fraction. This would be 
slightly less significant for a magnetic field of 3 $\mu$G instead of 1 $\mu$G, 
though still observable.

Another important piece of information that we can derive from 
\citefig{fig:signal_frac} is that the cumulative signal fraction is 
$\gtrsim 95\%$ (80\%) for $r\gtrsim 2$ kpc and $E\gtrsim 100$ (10) GeV. This 
helps us to define consistent means of including local sources in our 
predictions, as we discuss later in \citesec{subsubsec:local_el}. Indeed, we
know at present that if we replace the smooth spatial distribution within 2 kpc 
with discrete sources, these latter can affect the whole available energy range 
quite significantly: if powerful enough, local sources will dominate above a 
few tens of GeV, otherwise, flux predictions will be significantly depleted 
compared to a smooth-only description of sources, for a given normalization 
pattern.

\subsection{Sizing the uncertainties for local sources}
\label{subsec:local_snrs}

\begin{figure*}[t]
\begin{center}
\includegraphics[width=\columnwidth,clip]{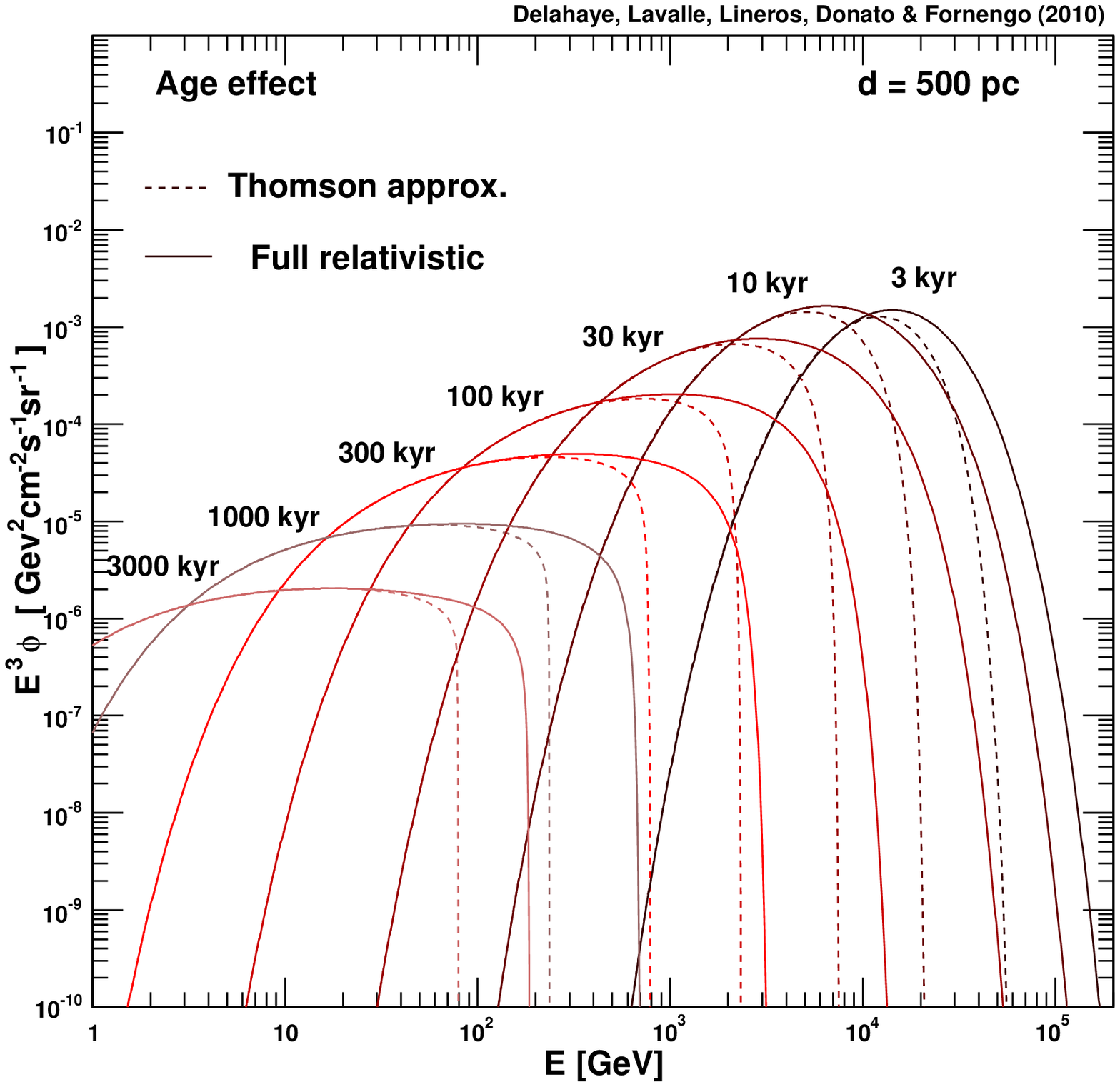}
\includegraphics[width=\columnwidth,clip]{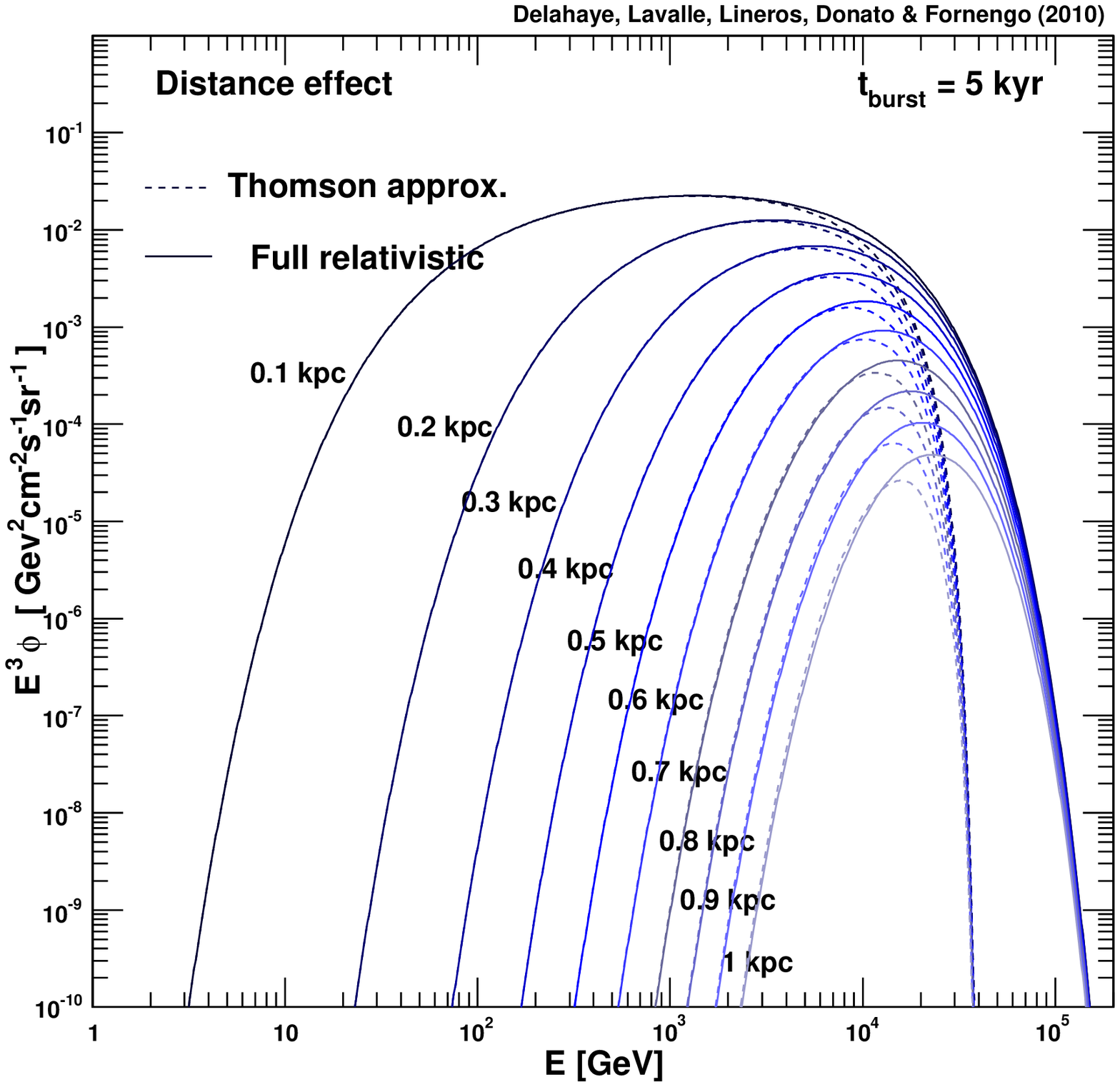}
\includegraphics[width=\columnwidth,clip]{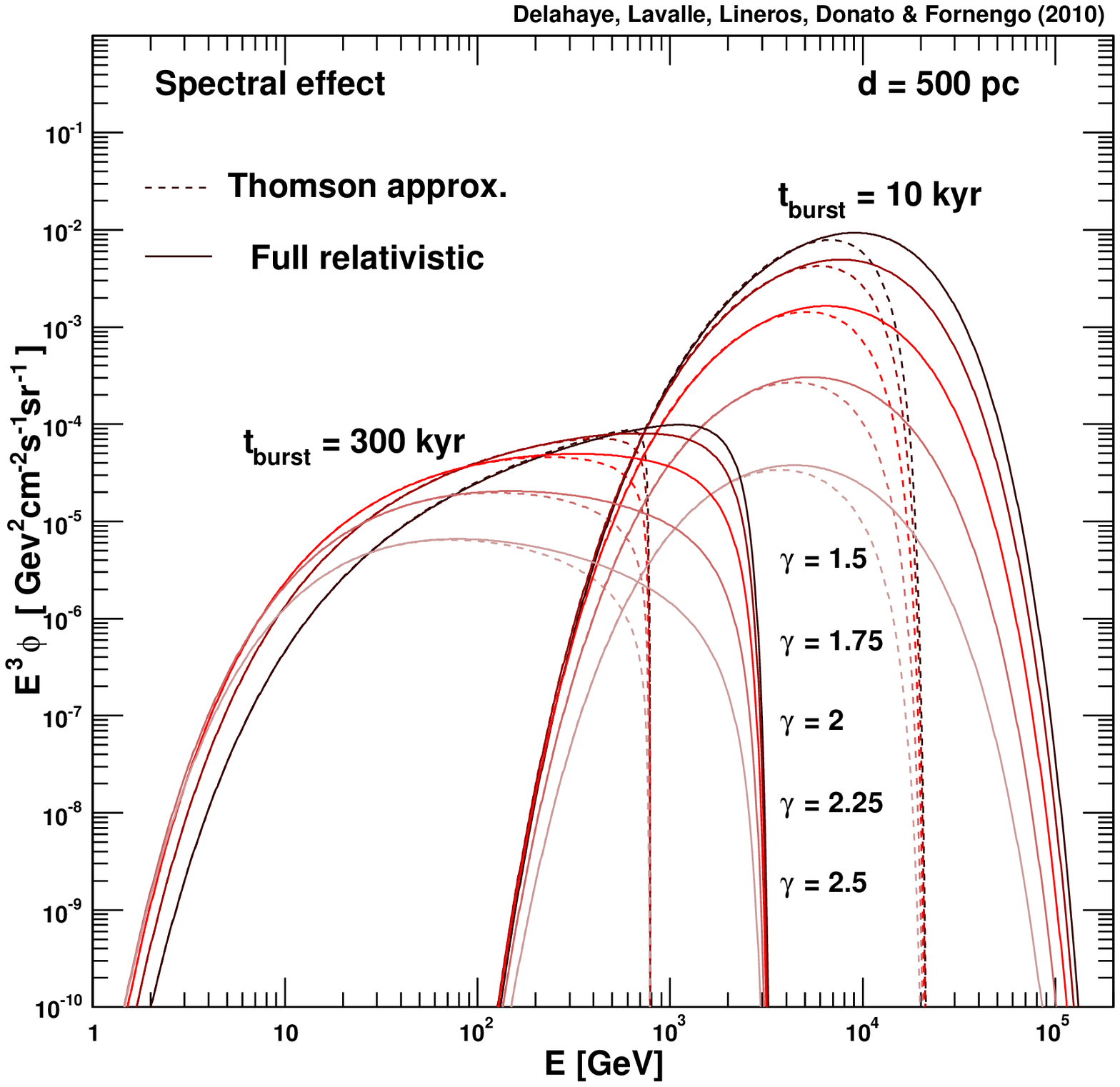}
\includegraphics[width=\columnwidth,clip]{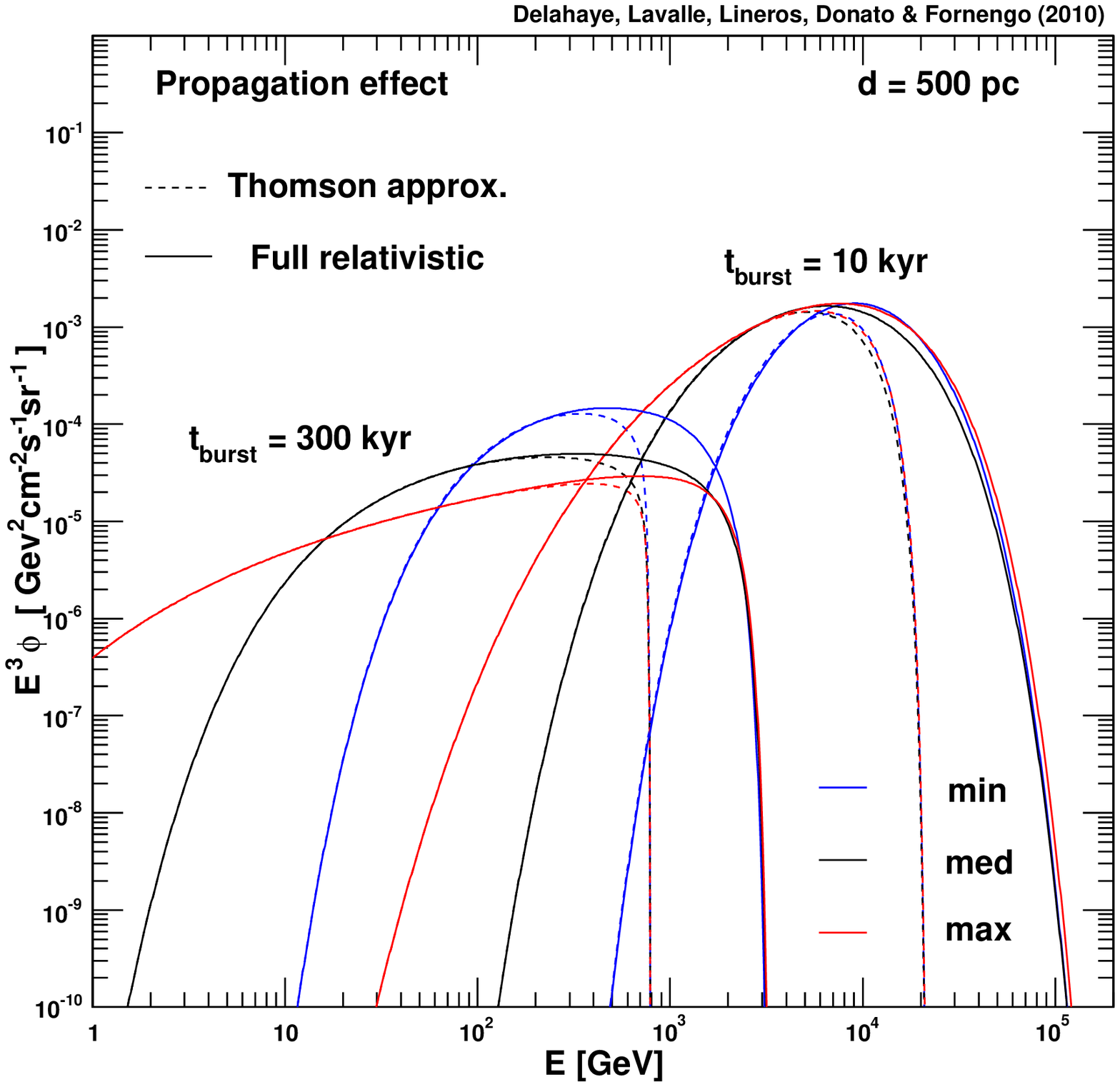}
\caption{\small Main uncertainties associated with the flux of primary 
  electrons injected from a single bursting source. Top left: varying the age 
  at a fixed 
  distance of 500 pc --- notice that taking 1 kpc would have suppressed the 3 
  kyr source for causality reasons. Top right: varying the distance for a fixed 
  age of 5 kyr. Bottom left: varying the spectral index for fixed age of 5 kyr 
  and fixed distance of 500 pc. Bottom right: varying propagation parameters, 
  with the \propmin, \propmed~and \propmax~setups from \citetab{tab:prop}.}
\label{fig:single_source}
\end{center}
\end{figure*}

Before discussing the contribution of local known SNRs to the CR electron flux 
(see \citesec{subsubsec:local_el}), it is essential to review the impact of 
uncertainties in the main parameters describing the source. They are only a few,
but their effects on the flux are shown to be important and degenerate. 

Apart from the propagation modeling and related parameters that were presented 
in \citesecs{subsec:time} and~\ref{subsec:prop_par}, theoretical errors 
may originate from uncertainties (i) in the spectral shape and normalization,
(ii) in the distance estimate, (iii) in the age estimate and (iv) in our 
understanding of the escape of cosmic rays from sources. The last point
is actually still debated and poorly known in detail 
\citep[see \eg\ ][]{2009MNRAS.396.2065C}, though it is clear that the release 
of cosmic rays in the ISM is a time- and energy-dependent process which 
takes place over $\sim 10^{3-5}$ yr, \ie\ the lifetime of the source. Since
this timescale is still almost always much lower than the diffusion timescale, 
$ t_d \simeq d^2/4K_0 \approx 0.1-10$ Myr for distances in the range 0.1-1 kpc,
we ignore the dynamical aspects of injection in this study, while we stress 
that they may lead to sizeable effects, especially in the case of very nearby 
sources. The first point was discussed in \citesec{subsec:snr_prop}, and is 
featured by two main parameters: the spectral index at source $\gamma$ and the 
energy released in the form of high energy electron $f\,E_\star$, both related 
in the normalization procedure given \citeeq{eq:Q0} that allows to derive 
${\cal Q_0}$. Points (ii) and (iii) have some impacts that can be understood
from \citesec{subsec:time} and \citesec{subsec:spectral_indices}. Although 
the consequences of varying these parameters can be understood from equations
only, we aim here to illustrate them in a more pedagogical way. To do so,
we will consider a template event-like source located in the Galactic plane 
($z=0$) at a distance $d$ to the Earth and bursting a population of electrons
a time (age) $t_\star$ ago:
\ben
{\cal Q}_\star(t_s,E_s,\vec{x}_s) = \delta(t_s-t_\star)\,\delta(z_s)\,
\delta^2(r_s - d)\,{\cal Q}(E_s)\;,
\label{eq:qstar}
\een 
where the spectrum ${\cal Q}(E_s)$ is given by \citeeq{eq:spectrum}. 
We will assume here that $f\,E_\star = 2\times 10^{47}$erg. Note that, as
emphasized in point (iv) above, a more realistic source term would not
involve a burst-like release of electrons in the ISM at time $t_\star$, but 
instead a more complex time-dependent energy spectrum. Such refinements are
beyond the scope of this paper.

In \citefig{fig:single_source}, we plot the electron flux for different
configurations of the parameters, the default configuration being defined 
by: {\em med}, $\gamma=2$, $E_c = 10$ TeV.

In the top left panel (a), we show the source age effect; in the top right panel
(b), we illustrate the distance effect; in the bottom left panel (c), we sketch 
the spectral index effect; while in the bottom right panel (d), we plot the 
propagation model effect. For all panels, we report the fluxes calculated in 
both the Thomson approximation and the full relativistic treatment of the energy
losses, as discussed in \citesecs{subsec:eloss} and~\ref{subsec:prop_par}.

As a first comment, we emphasize that the Thomson approximation can lead to 
a very strong under-estimate of the spectral break inferred from
energy losses, up to one order of magnitude in the examples shown. This is
a mere consequence of the over-estimate of the energy loss rate at high energy.
The net effect obviously depends on the magnetic field and on the actual 
cut-off considered at the source. As regards the latter, we see that using 
a value of 10 TeV already induces an underestimate by a factor of $\sim$5-10 of 
the break predicted in the non-relativistic regime. Many studies 
of the topic have employed the Thomson approximation.

The second important comment to make is that it is actually quite difficult
to relate the observed spectral index to the source spectral index, because
of the complex and degenerate effects coming from all parameters: distance,
age, source index, energy cut-off, normalization and diffusion coefficient. 
For instance, we see that a large diffusion coefficient ({\em min} model) can 
make a source of 300 kyr resemble a source of 30 kyr associated with a larger 
diffusion coefficient and a lower energy cut-off. In any case, a mere glance at 
the four panels of \citefig{fig:single_source} is striking enough.

This exhaustive analysis of the impact of the main parameters characterizing
individual sources already points towards the difficulties that we encounter in 
the interpretation of the data. Nonetheless, although this part might look 
depressing at first sight in the perspective of making predictions, it is still 
very useful to estimate the theoretical confidence level of our forthcoming 
attempts.


\subsection{Primary electron flux and theoretical uncertainties}
\label{subsec:el_pred}

In the previous parts of this section, we have discussed the main physical
quantities relevant for predictions of the primary electron flux at the Earth,
emphasizing their role as potential sources of uncertainties. Here, we implement
the full calculation and compare our results with available data on the electron
flux. We stress that pure electron data are not numerous and rather old, 
since most of recent experiments either do not distinguish electrons from 
positrons or have not yet released their charge-discriminating data. We 
therefore only use the electron data from CAPRICE \citep{2000ApJ...532..653B}, 
HEAT \citep{2001ApJ...559..296D}, and AMS-01 \citep{2000PhLB..484...10A}, to 
avoid any confused interpretation mixing positrons. The pure positron case and 
the full case are discussed in \citesecs{sec:primary_pos} and~\ref{sec:disc}, 
respectively.

We first compare the predictions arising from a smooth description of sources, 
for which we adopt the L04 spatial distribution. Indeed, we demonstrated in 
\citesec{subsec:smooth_snrs} that using different spatial distributions causes 
only small differences in the overall flux normalization locally.

We then estimate the contributions of all known local SNRs that can be 
added to a smooth and more distant component, following the method proposed in 
\citet{2004ApJ...601..340K}.

\subsubsection{Smooth description of sources}
\label{subsubsec:smooth_el}

\begin{figure*}[t]
\begin{center}
\includegraphics[width=\columnwidth,clip]{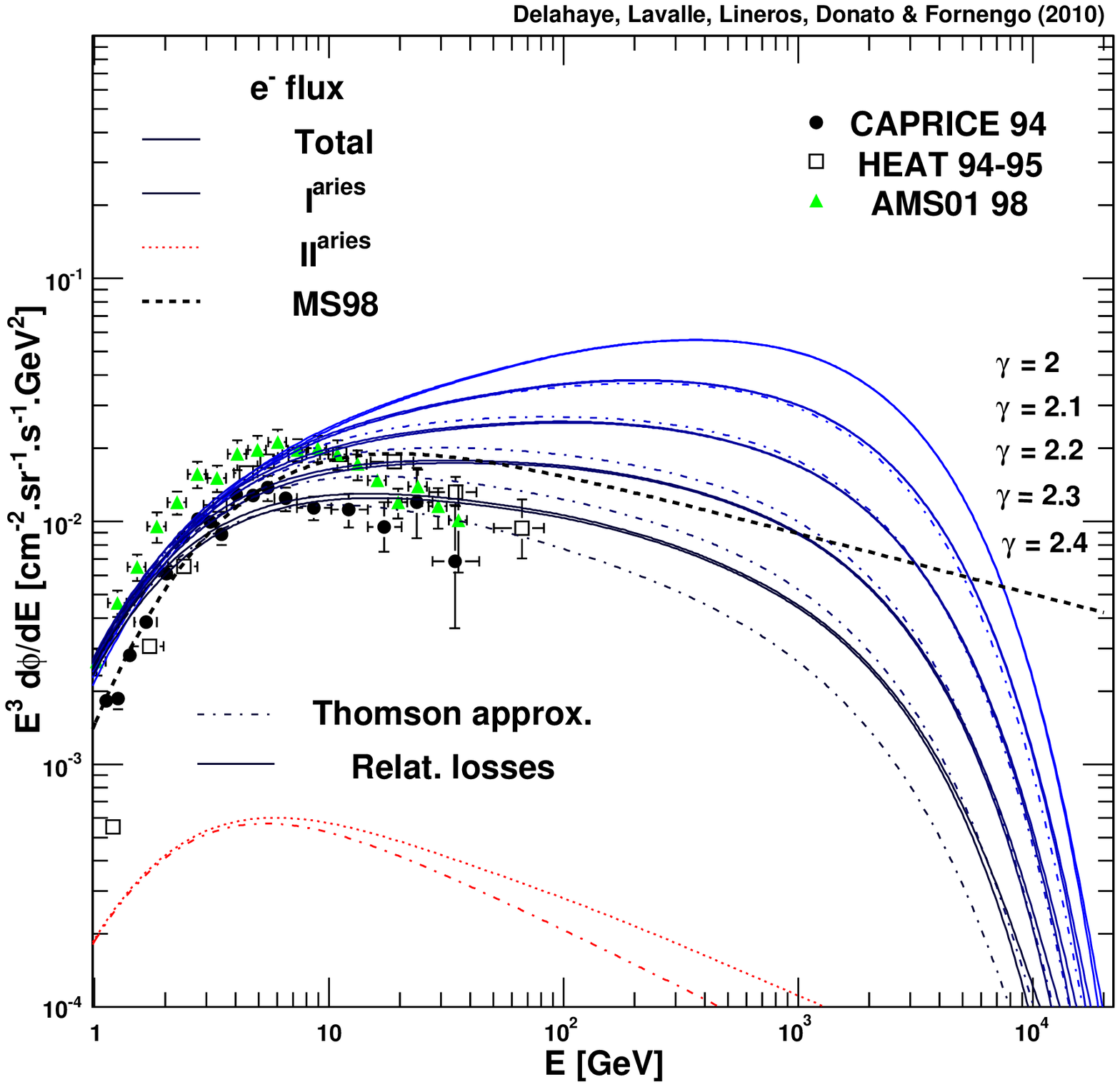}
\includegraphics[width=\columnwidth,clip]{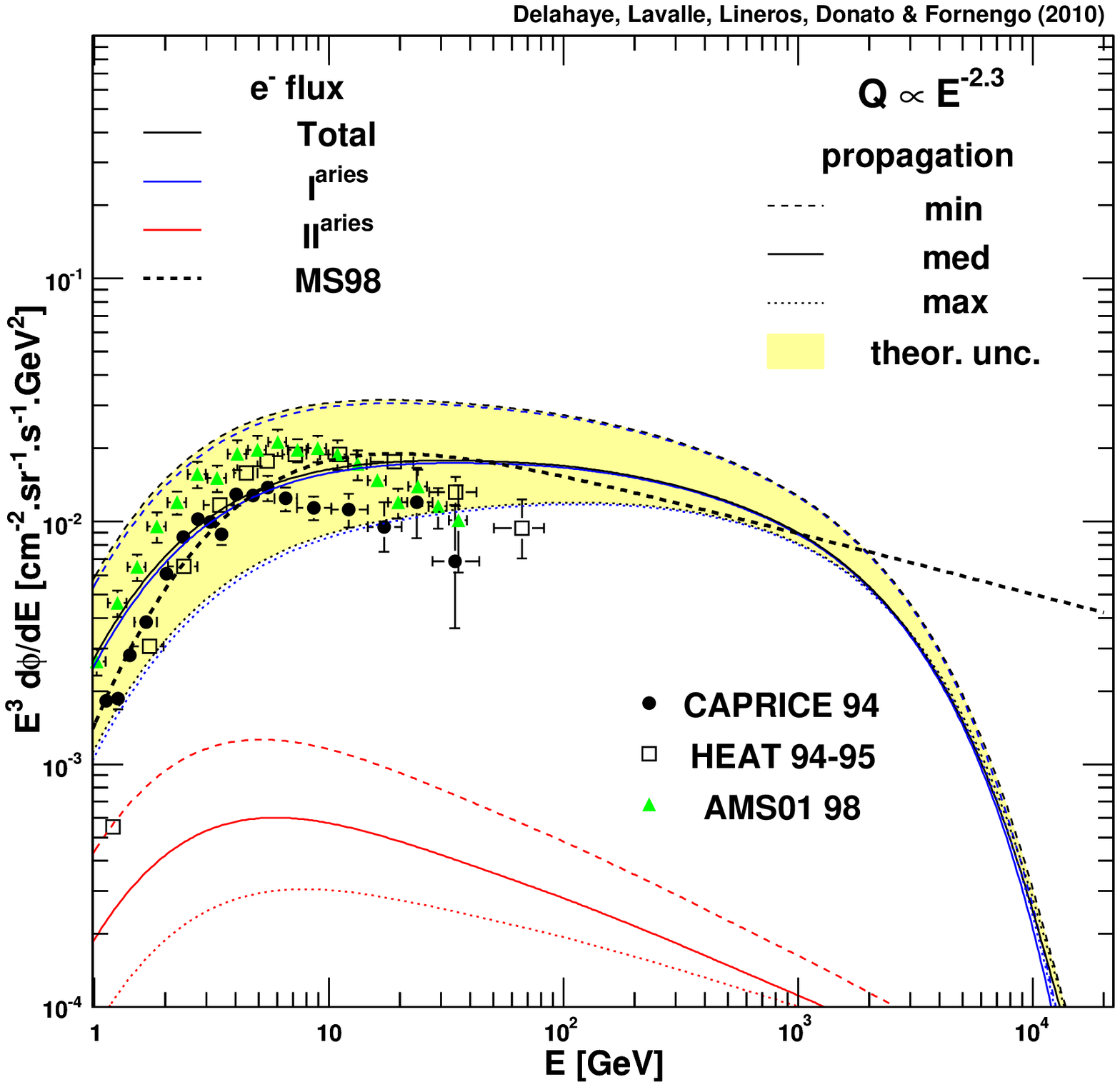}
\caption{\small primary electron flux from a smooth distribution of SNRs. 
Left: fluxes associated with injection indices $\gamma$ from 2.0 to 2.4. Right: 
propagation effect for $\gamma=2.3$. A renormalization factor of 5 has been 
applied to \tgammae~in both panels and a cut-off energy of 3 TeV has been
considered.}
\label{fig:smooth_snrs}
\end{center}
\end{figure*}

Our model for a smooth distribution of SNRs includes a propagation setup, a 
spatial distribution (here L04) and an injected spectrum, and it is interesting 
to check some of the possible configurations against the data. In particular, we
attempt to constrain the injection normalization necessary for a model to fit, 
at least roughly, the data. For the spectrum, we test different spectral 
indices, but keep the energy cut-off at 3 TeV. As a reference normalization, we 
use a SN explosion rate of 4/century, a SNR total energy of $E_\star = 10^{51}$ 
erg, of which a fraction of $f=2\times 10^{-4}$ is carried by electrons, giving 
therefore $\tgammae = 8\times 10^{47}$ erg/century.

In \citefig{fig:smooth_snrs}, we report various flux calculations, for which we
applied a solar modulation correction with a Fisk potential of 600 MV. In the 
left panel, we show the effect of varying the injected spectral index from 2 to 
2.4 for the \propmed~propagation setup, using both the Thomson approximation and
the relativistic regime for the energy losses. In this plot, we have 
renormalized \tgammae~by a factor of 5 for all indices, so that we see that 
reasonable fits to the data can be obtained within the expected normalization 
range discussed in \citesec{subsec:snr_prop}. This means that the expected 
energy budget available for electrons is in rough agreement with what is needed 
to explain the current observations. From the same plot, we could also conclude 
that the injection spectral index should be slightly softer than 2. 
Nevertheless, this also depends on the logarithmic slope $\delta$ of the 
diffusion coefficient, as seen from \citeeq{eq:gamma_obs_disk} --- 
complementary constraints on $\gamma + \delta$ could also be derived from high 
energy proton data, based on the assumption that the proton index is the same 
as the electron index after their acceleration at sources and that proton 
propagation is simply described by
diffusion (\ie~neither reacceleration nor convection). This is illustrated in 
the left panel of \citefig{fig:smooth_snrs}, where we show the effect of the 
theoretical uncertainties in the propagation parameters, using the same 
spectrum normalization and the same spectral index for all models. We see that 
the \propmin~model gives the larger amplitude because of its smaller value of 
$K_0$ and the softer observed index due to its larger diffusion slope $\delta$ 
(see \citetab{tab:prop} and \citeeq{eq:disk_approx}) --- the analysis is 
reversed in the \propmax~configuration. For a given normalization, the amplitude
uncertainty is therefore proportional to $\sqrt{K_0}$, which gives a factor of 
$\sim 7$ from the \propmin~to the \propmax~configurations. In both panels of 
\citefig{fig:smooth_snrs}, we also report the prediction obtained in 
\citet{1998ApJ...493..694M}, as fitted in~\citet{1998PhRvD..59b3511B}, where 
the authors used an injection index of 2.1 below 10 GeV, steepening to 2.4 
above. This model, very often quoted as a reference model, is shown for 
comparison.

It is noteworthy that since the data have a quite limited statistics and range 
up to $\sim 40$ GeV only, they are probably insufficient to provide strong 
constraints on the electron cosmic ray component. Moreover, we recall that this 
smooth description of the SNR contribution is not valid locally above a few 
tens of GeV, where we expect discrete effects to become important. Nevertheless,
this preliminary analysis is still useful to delineating the relevant ranges of
the spectral index and the injected energy. Likewise, it helps us to determine 
the influence of distant sources relative to local ones.

Finally, we emphasize that we only considered a single contribution from a SNR 
population. Nevertheless, the electron-positron pair injection from pulsars is 
also likely to account for a significant additional contribution to the local 
electron budget. It is not clear whether this contribution should have the same 
spectral index, and one could for instance model the smooth electron component 
with a combination of two spectral components, leading to an additional freedom 
in the normalization procedure. This electron component from pulsars is 
discussed in \citesec{sec:primary_pos}.


\subsubsection{Contributions from known local sources}
\label{subsubsec:local_el}

\begin{figure*}[t]
  \centering
  \includegraphics[angle=0,width=\columnwidth]{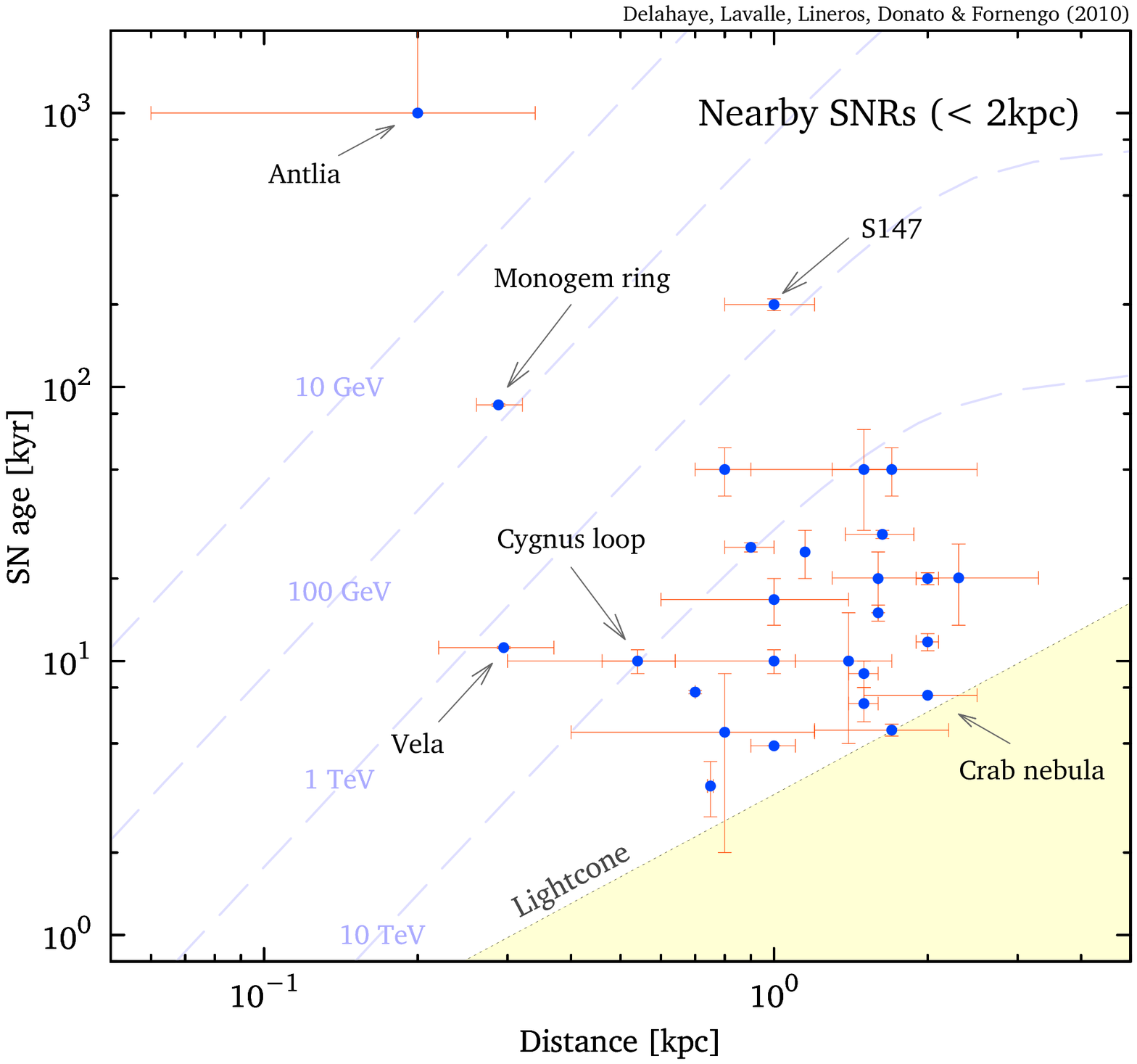}
  \includegraphics[angle=0,width=\columnwidth]{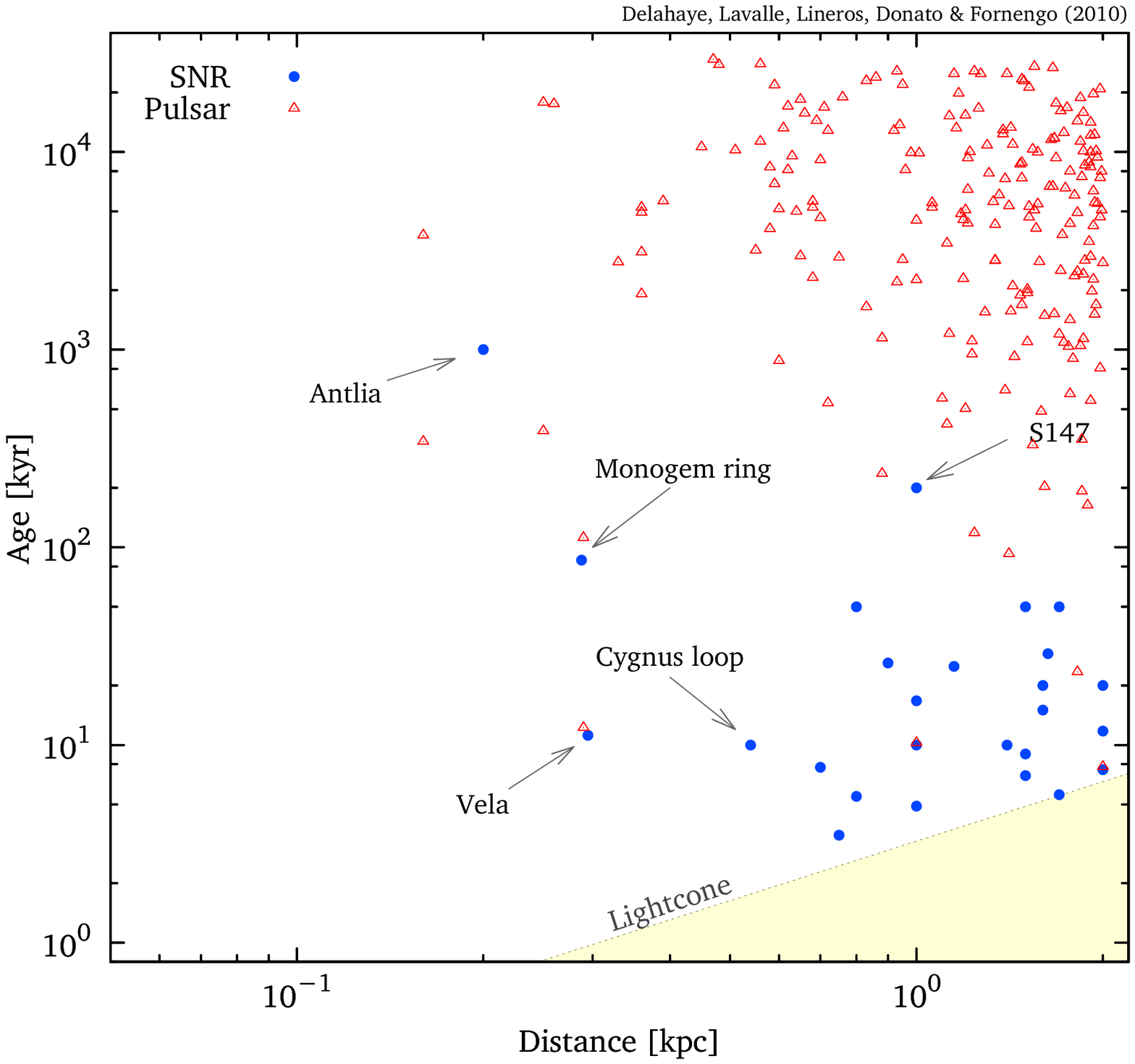}
  \caption{\small Left: Plot of the {\em observed} age versus distance to the 
    Earth for our sample of local SNRs (and associated uncertainties, see 
    \citetab{tab:SNRs}). The dashed lines correspond 
    to limits beneath which a local source cannot contribute significantly to 
    the signal at the corresponding energy (valid only in the 
    \propmed~propagation model --- see \citetab{tab:prop}). Indeed the age 
    sets an upper limit, while the distance sets a lower limit to the energy 
    range -- see \citesec{subsec:el_pred}. Right: Same plot for our complete
    sample of local SNRs and pulsars.}
  \label{fig:local_sources}
\end{figure*}

\begin{figure*}[t]
\begin{center}
\includegraphics[width=0.57\columnwidth,angle=-90,clip]{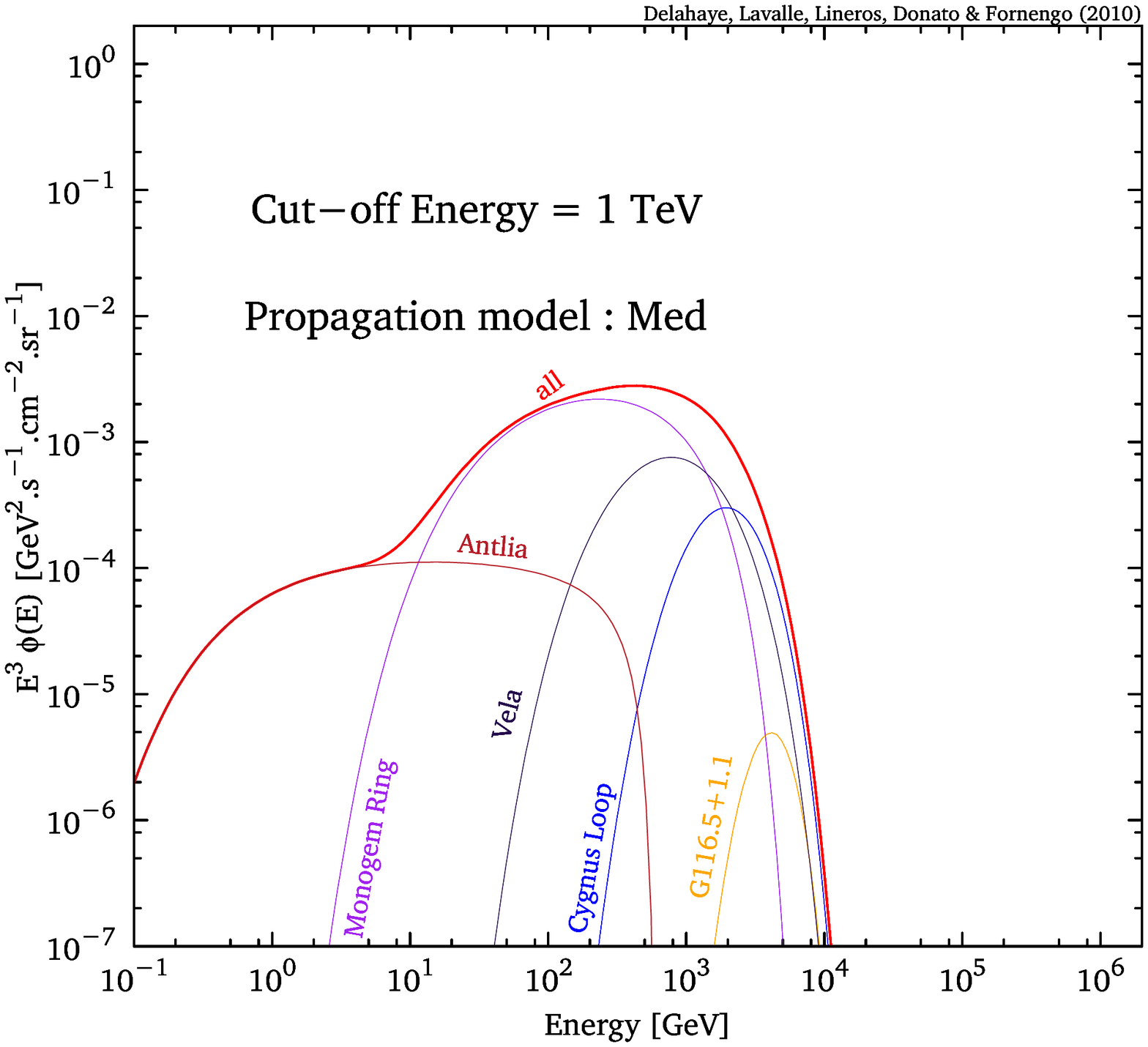}
\includegraphics[width=0.57\columnwidth,angle=-90,clip]{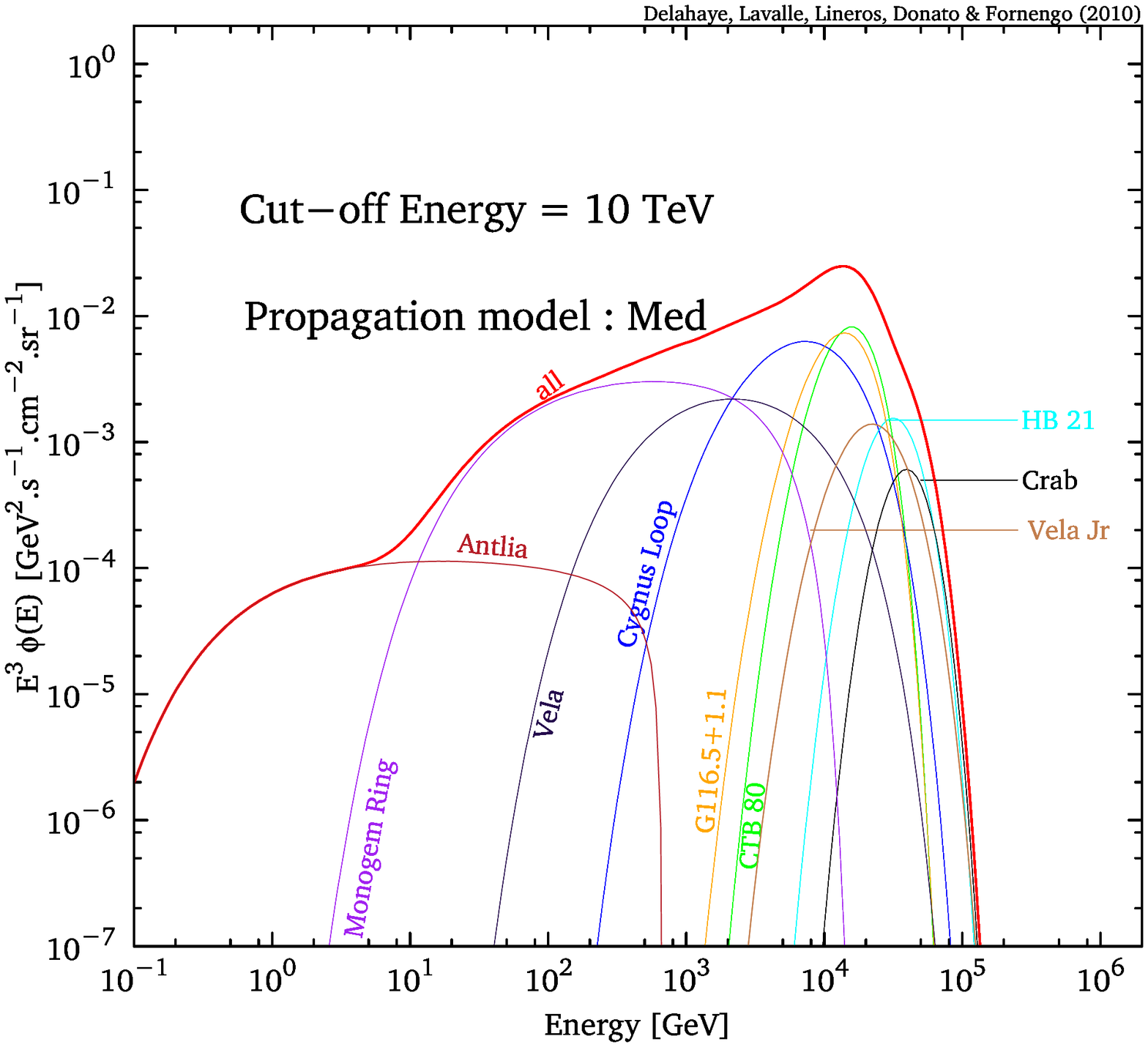}
\includegraphics[width=0.57\columnwidth,angle=-90,clip]{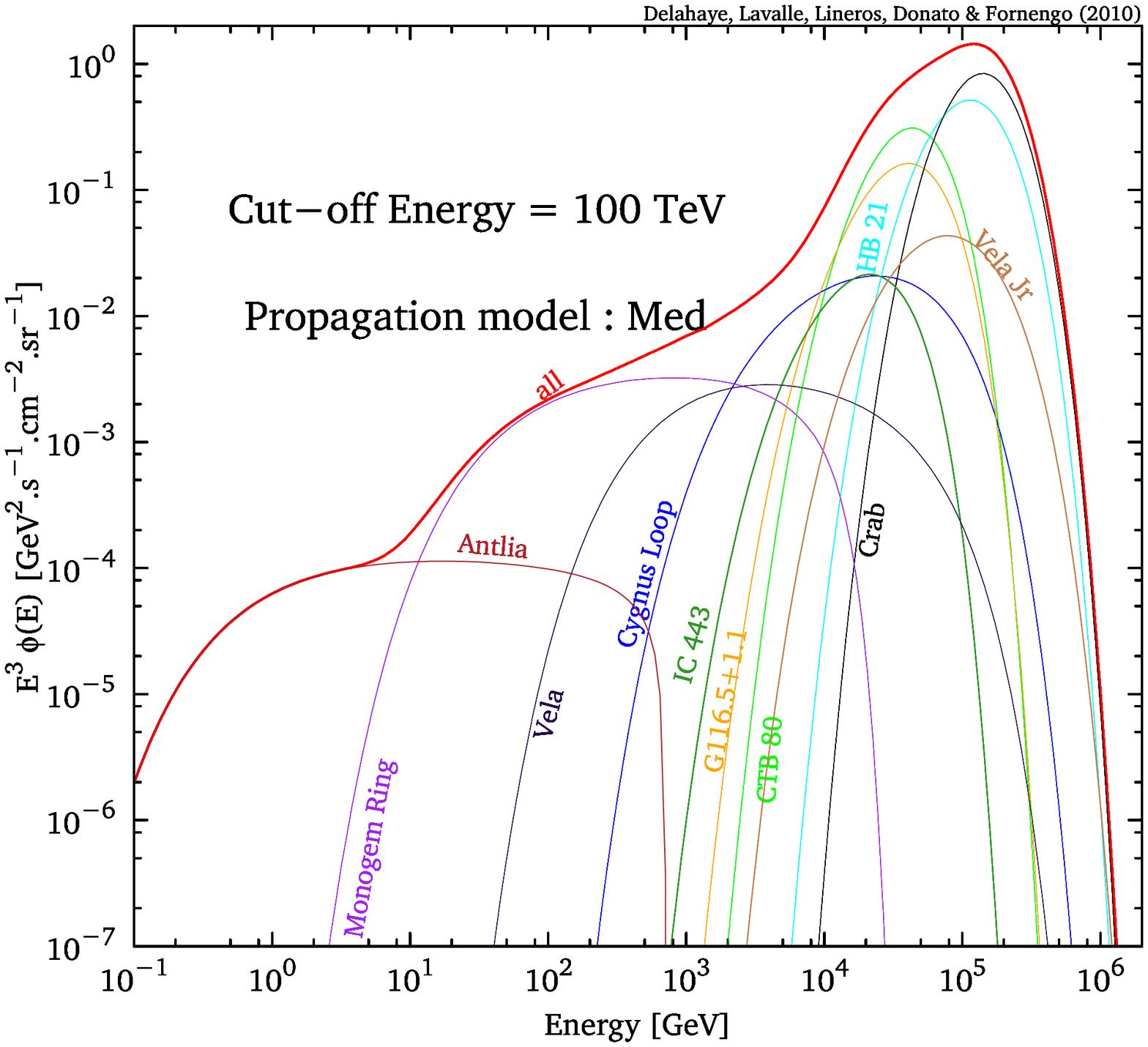}
\caption{\small Primary electron flux from local SNRs in the 
\propmed~propagation model and using radio observation constraints. Left: 
energy cut-off at $E_c = 1$ TeV. Middle: $E_c = 10$ TeV. Right: $E_c = 100$ 
TeV.}
\label{fig:local_snrs}
\end{center}
\end{figure*}

\begin{figure*}[t]
\begin{center}
\includegraphics[width=0.85\columnwidth,angle=-90,clip]{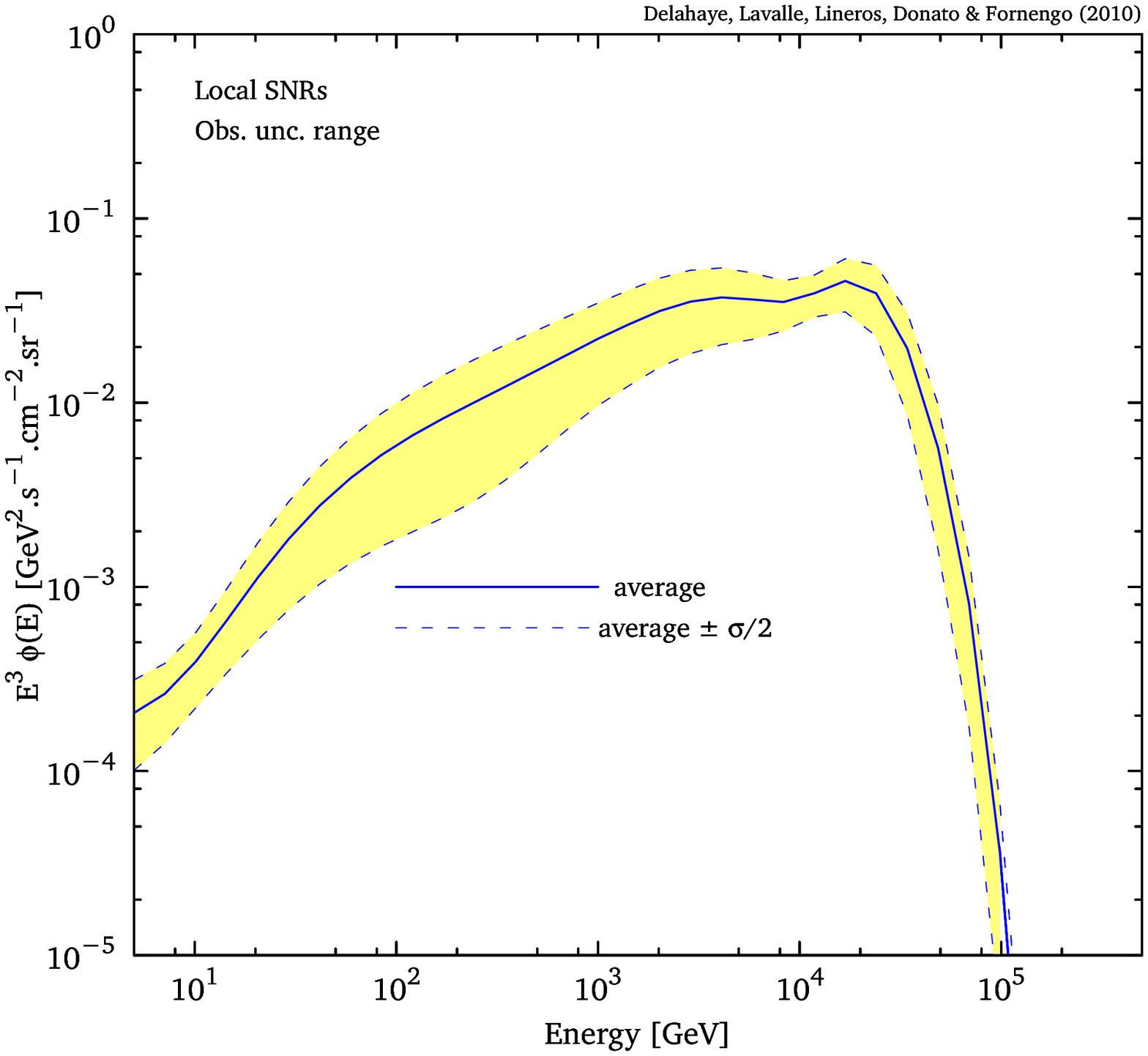}
\includegraphics[width=0.85\columnwidth,angle=-90,clip]{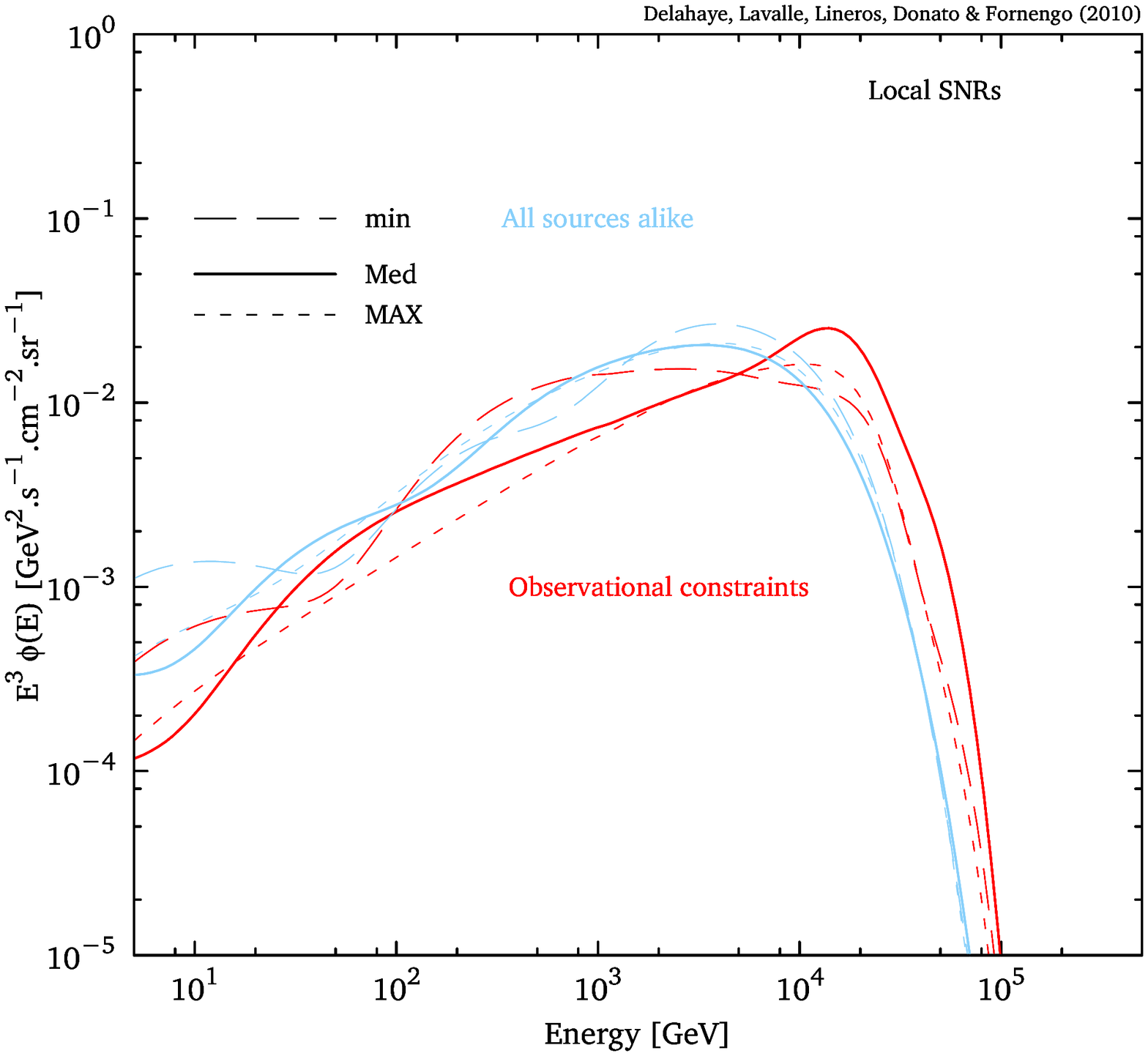}
\caption{\small Left: Electron flux from observed local SNRs, with associated 
  uncertainty band (due to observational uncertainties on ages, distances, 
  radio fluxes and spectral indices). Right: Propagation effects on the 
  electron flux 
  originating from local SNRs, using either the observational constraints or 
  the average flux and index from \citefig{fig:Stats}. In both panels, we have 
  assumed a source cut-off energy of 10 TeV.}
\label{fig:local_snrs_minmedmax}
\end{center}
\end{figure*}

\begin{figure*}[t]
\begin{center}
\includegraphics[width=0.67\columnwidth]{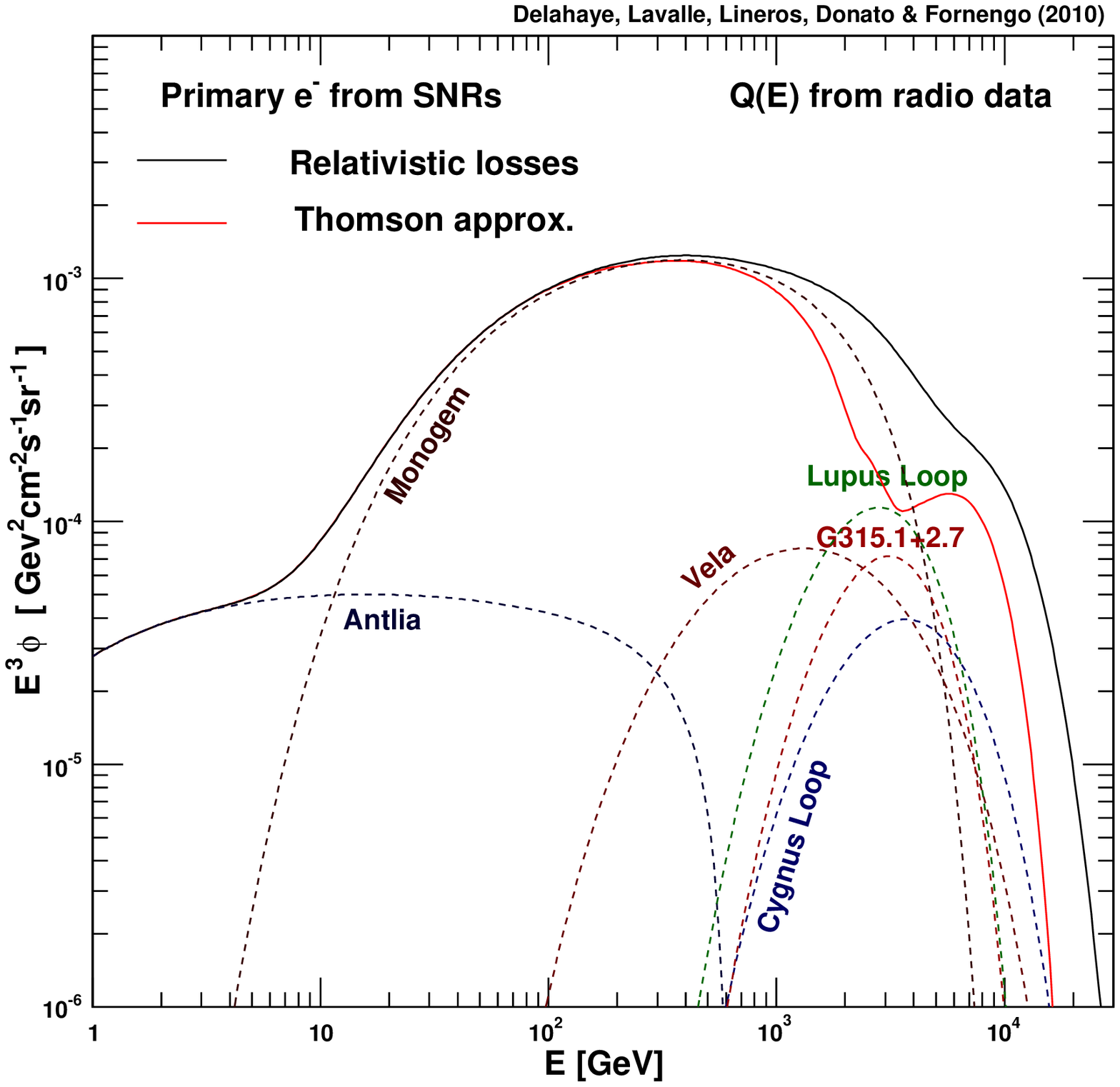}
\includegraphics[width=0.67\columnwidth]{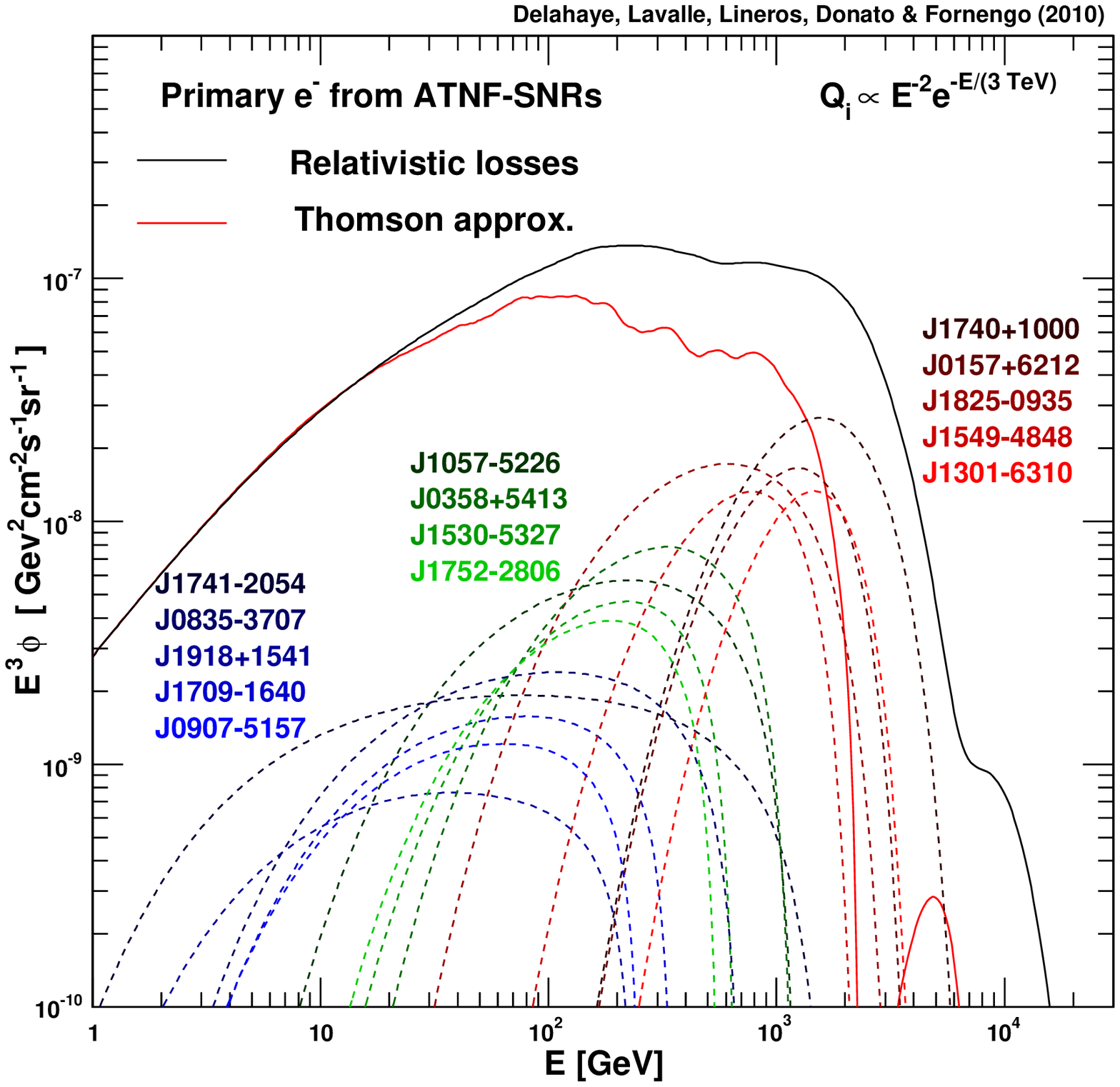}
\includegraphics[width=0.67\columnwidth]{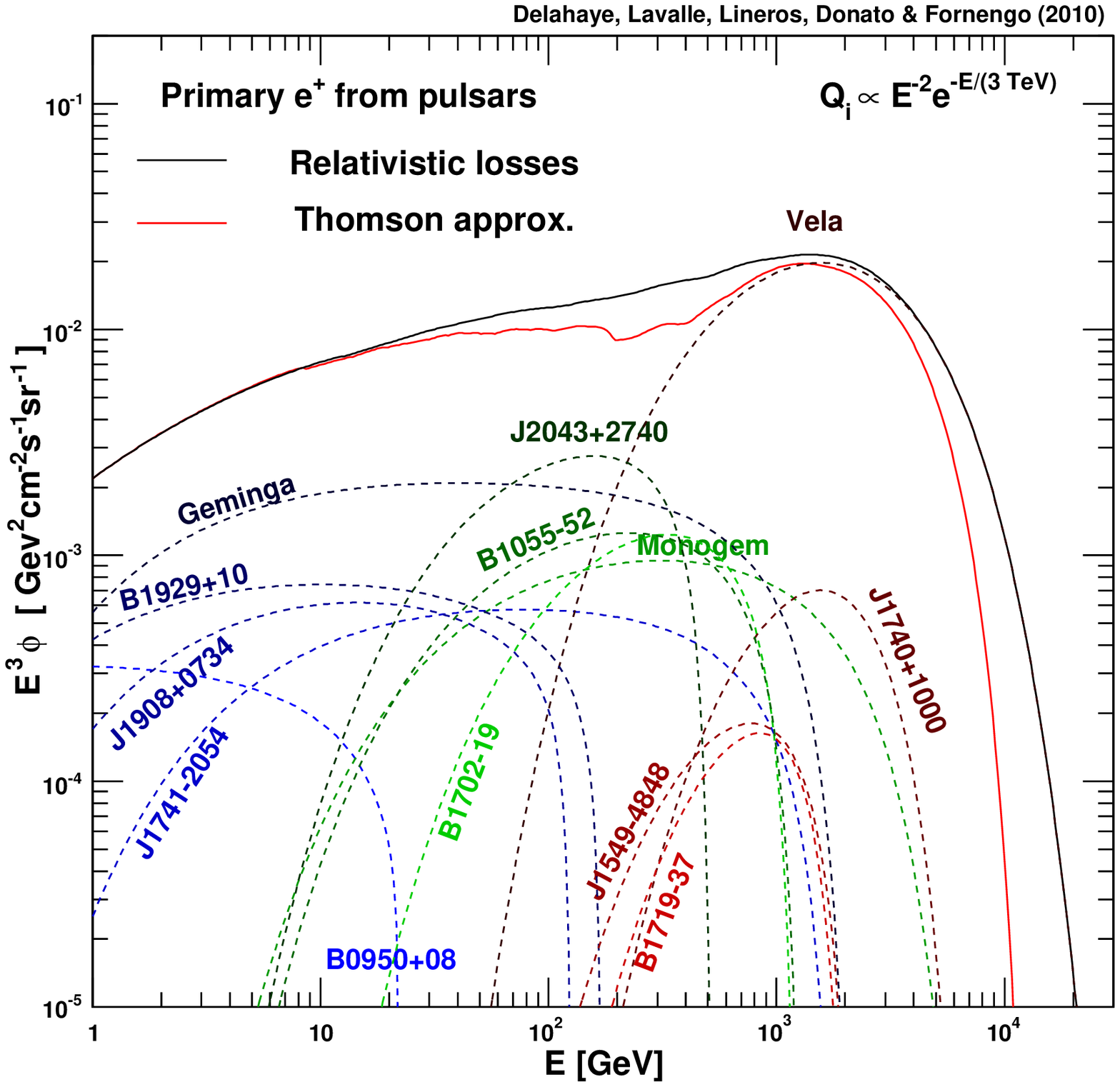} 
\caption{\small Primary electron flux from local sources located within 2 kpc 
  from the Earth. Left: from know SNRs (radio constraints). Middle: from 
  unknown SNRs that should accompany unassociated local pulsars (generic 
  spectrum for each, age and distance from the pulsar companion). Right: from 
  known pulsars (generic spectrum, observational constraints on age and 
  distance). Here, all generic spectra refer to a spectral index of 2 and an 
  energy cut-off of 3 TeV.}
\label{fig:local_snrs_1}
\end{center}
\end{figure*}

\begin{table}[h!]

\begin{center}

\begin{tabular}{c | c|p{1.9cm}|p{1.9cm}|p{1.9cm}|}

\cline{2-5}

& $E_c$ & 1 TeV& 10 TeV & 100 TeV \\

\hline

\multicolumn{1}{|
c|}{\multirow{3}{*}{\raisebox{-.75cm}{\begin{sideways} \propmin\end{sideways}}}}
& $\gamma = 2.0$ & 5, 20, 22, 24 & + 23 & +8, 18, 19, 26 \\
\cline{2-5}
\multicolumn{1}{|c|}{}& $\gamma = 2.2$ & 5, 20, 22, 24 & +8, 18 & + 17,19 \\
\cline{2-5}
\multicolumn{1}{|c|}{}& $\gamma = 2.4$ & 5, 20, 22, 24 & +8, 17, 18 & + 19 \\
\hline
\multicolumn{1}{|
c|}{\multirow{3}{*}{\raisebox{-1.cm}{\begin{sideways}\propmed \end{sideways}}}}
& $\gamma = 2.0$ & 5, 20, 22, 24, 26 & + 4, 11, 19 & +8, 18, 23 \\
\cline{2-5}
\multicolumn{1}{|c|}{}& $\gamma = 2.2$ & 5, 11, 20, 22, 24& +4, 8, 18,23 & 
+ 19 \\
\cline{2-5}
\multicolumn{1}{|c|}{}& $\gamma = 2.4$ & 5, 11, 20, 22, 24 & +4, 18, 19, 23 & 
+ 8, 17 \\
\hline
\multicolumn{1}{|
c|}{\multirow{3}{*}{\raisebox{-1.cm}{\begin{sideways}\propmax \end{sideways}}}}
& $\gamma = 2.0$ & 5, 20, 22, 24, 26 & + 8, 11, 19, 23 & +4, 18 \\
\cline{2-5}
\multicolumn{1}{|c|}{}& $\gamma = 2.2$ & 5, 20, 22, 24 & +8, 11, 18, 19
& + 4, 23, 26 \\
\cline{2-5}
\multicolumn{1}{|c|}{}& $\gamma = 2.4$ & 5, 20, 22, 24 & +8, 11, 18, 19
& +4, 23, 24 \\
\hline
\end{tabular}
\caption{Ranked {\em id} numbers (see \citetab{tab:SNRs}) of those SNRs that 
  contribute more than 10\% of the signal for various propagation models, 
  cut-off energies $\epsilon_c$, and spectral index $\gamma$. Index $\gamma$ is 
  only used for sources that are not constrained by radio data. The 
  dominant sources are the Cygnus Loop (5), Monogem Ring (20), Vela (22), and 
  Antlia (24).}
\label{tab:stronglocal}
\end{center}
\end{table}

As discussed earlier, contributions from local sources are expected to be 
significant above a few tens of GeV. Following the method proposed in 
\citet{2004ApJ...601..340K}, we take a census of all known sources of primary 
electrons located within 2 kpc from the Earth in order to compute their 
associated flux explicitly.

To proceed, we first took advantage of the information provided in the SNR 
catalog of~\citet{2009BASI...37...45G}, and we performed an extensive synthesis 
of all published properties and associated errors (mostly from radio data). We 
found 26 SNRs within 2 kpc in this catalog, to which we added an extra-object, 
Antlia SNR~\citep{2002ApJ...576L..41M,2007ApJ...670.1132S}. A full description 
of these sources including information about distance, age, spectral index, 
radio flux, associated objects, and bibliographic references is available in 
\citeapp{app:SNR}. These properties are summarized in \citetab{tab:SNRs}.

The event-like contribution of a single source is readily computed from the
results obtained in \citesec{subsec:time}. As a word of caution, however, we 
stress that the time argument used to feed the time-dependent propagator 
given in \citeeq{eq:time_dep_prop} should not be the {\em observed} age of the 
object given in catalogs, but instead the {\em actual} age, equal, in principle,
to the observed age plus $d/c$. Indeed, most of the age estimates depend on the 
dynamical properties of the objects inferred from multiwavelength observations,
which correspond to the properties the object had a time $d/c$ ago. For the 
injection spectrum, we utilize \citeeq{eq:spectrum} and set the spectral index 
$\gamma$ from the observed radio index $\gamma_r$ --- $\gamma = 2\gamma_r +1$. 
We constrain the spectrum normalization with the observed radio flux using 
\citeeq{eq:q0_num}.

Although SNRs are expected to provide an important contribution to the primary 
electron flux, we emphasize that pulsars are also expected to produce and 
accelerate electron-positron pairs. Modeling the electron injection from pulsar
is discussed in more detail in \citesec{sec:primary_pos}, to which we refer the 
reader. We found that $\sim 200$ pulsars located within a distance of 2 kpc 
from the Earth could contribute to the local electron budget, among which few 
may be dominant (see \citesec{subsec:local_pulsars}). For consistency reasons, 
we have to include the contribution of these pulsars to the electron flux.

An additional important remark should be made about the non-observed local 
sources of primary electrons that {\em should} exist. So far, we have listed 27 
SNRs and about 200 pulsars. Nevertheless, we recall, as we discuss in more 
detail in \citesec{sec:primary_pos}, that pulsars are rotating neutron stars 
originating from core-collapse supernova explosions. Therefore, each pulsar 
should be accompanied by a SNR. Such a systematic association is obviously not 
supported by observations. This is already illustrated in our object list, in 
which we find only 27 SNRs for 200 pulsars. Among these 27 local SNRs, only 10 
have a known pulsar counterpart (very often with differences in their distance 
and age estimates). However, this certainly does not mean that the theoretical 
expectation is wrong, since not all SN explosions lead to pulsars, but instead 
that the counterparts are probably not bright enough to be observed. Therefore, 
again for theoretical consistency reasons, we choose here to add a SNR 
counterpart to each non-associated pulsar, but with a brightness such that it 
could not be observed with current telescopes. We adopted $B(1\,{\rm GHz}) 
\lesssim 1 \,{\rm Jy}$ as a general criterion for non-observed SNRs. These local
statistics can be tested against predictions of the SN explosion rate 
$\Gamma_\star$. If we assume at zeroth order that sources are distributed 
homogeneously inside a flat disk of radius $R=20$ kpc, then the local explosion 
rate within a radius of $r = 2$ kpc around the observer is given by 
$\Gamma_\star(r/R)^2 = 0.01\times\Gamma_\star$, leading to $\sim$0.01 SN/century 
for usual values of $\Gamma_\star$. This can be compared with the observed local
explosion rate, which we can estimate from the number of sources in our sample 
divided by the oldest age, \ie~$\sim 200/30 \,{\rm Myr} \simeq 0.7 \times 
10^{-3}/{\rm century}$. This rough calculation leads to a difference of only a 
few, which would tend to tell us that using only the observed sources translates
into a slight underestimate of the actual local electron budget. This makes 
sense, since observations favor the brightest objects, and also since 
rapidly-rotating magnetized neutron stars with magnetic axes pointing away from 
the Earth cannot be observed as pulsars. Our samples of local SNRs and pulsars 
are shown in the age-distance plane in \citefig{fig:local_sources}.

In \citefig{fig:local_snrs}, we show the electron flux obtained in the 
\propmed~propagation setup, using the local SNR properties summarized in 
\citetab{tab:SNRs} and assuming different cut-off energies. The first important
comment to make is that the whole flux is far from being described simply by a  
smoothly injected power law, since many spectral wave-like features are evident.
Moreover, because of the interplay between the age and the maximal energy (see 
\citesec{subsec:time}), we see that varying the energy cut-off from 1 to 100 
TeV, though the latter value is probably not realistic and too high, has 
considerable effects; not only do new contributions arise at high energy when 
the cut-off value increases, but the hierarchy among other sources is also 
altered. This illustrates an additional source of theoretical uncertainty, 
beside those we discussed in \citesec{subsec:local_snrs}. Figure
\ref{fig:local_snrs_minmedmax} illustrates the theoretical and observational 
uncertainties affecting both the source modeling and the propagation modeling. 
The left panel allows us to quantify the impact of the observational 
uncertainties on the ages, distances, radio fluxes and spectral indices. This 
plot portrays the results of 1,000 Monte Carlo realizations in which we drew 
each parameter according to a flat distribution within the observational errors.
The right panel exhibits (i) the impact of varying the propagation parameters 
on the overall local SNR contribution and (ii) the differences associated with 
different injection spectrum prescriptions. For the latter point, we compared 
the flux obtained with generic spectral properties, namely a spectral index of 
2 and a fixed normalization of ${\cal Q}_0 = 3.9\times 10^{49}{\rm GeV^{-1}}$ 
(see end of \citesec{subsec:snr_prop}), with the observationally constrained
predictions. We again see that the global spectral shape is far more complex
than a mere power law, and that the overall flux can vary within a factor of 
2-5 depending on the energy.

We plot the results obtained with a template calculation for all local electron 
primaries in \citefig{fig:local_snrs_1}, where we used the \propmed~propagation 
setup with the M1 ISRF model (the large difference between the full relativistic
calculation and the Thomson approximation is evident in the plots). The three 
panels from left to right show the contributions of local known SNRs, from 
non-observed SNRs associated with observed pulsars, and from observed pulsars, 
respectively. For the non-observed SNRs, we considered the distances and ages 
of the associated pulsars, and we assumed an injection spectrum with an index 
of 2 and a non-observable radio flux of $B(1\,{\rm GHz})= 1 \,{\rm Jy}$. While 
the contribution from the non-observed SNRs is shown to be negligible with 
respect to the two others, it is interesting to note that our local pulsar 
modeling leads to a higher primary electron flux than local SNRs. For pulsars, 
we supposed than 10\% of the spin-down energy was converted into 
electron-positron pairs when defining the individual normalizations (see 
\citesec{sec:primary_pos}). Although the injection mechanism is subject to large
theoretical uncertainties, it is still rather surprising to find that even when 
accounting for observational constraints for pulsars, their local population 
can contribute as many primary electrons as known local SNRs. However, these 
calculations are subject to very large theoretical uncertainties, so no strong 
conclusions should be drawn: by no means should they be considered as 
predictions, only trends. The main conclusion at this stage is that nearby 
sources dominate the flux above energies of few tens of GeV, the spectral 
imprints of which are very difficult to predict, because of large theoretical 
uncertainties in their modeling. Finally, it is clear from the plots
that the Thomson approximation can lead to large errors in the predictions.

\subsubsection{Electron flux: sum of distant plus local sources}
\label{subsubsec:total_el}

We can now derive a full calculation including all known local sources within 
a distance of 2 kpc from the Earth and a more distant smooth component. For 
the latter, we use the L04 spatial distribution, but in contrast to what we did 
in \citesec{subsubsec:smooth_el}, we have to apply a radial cut-off to this 
smooth contribution within 2 kpc to the Earth, to avoid a possible double 
counting of the local sources. This means that the distant component has more 
impact at low energy than at high energy.

Furthermore, we can also compare our results with the available observational 
data about electrons. The top left panel of \citefig{fig:final_results} shows
an example of a calculation including all the components discussed above in
a self-consistent manner, using the \propmed~propagation setup. To proceed,
we fixed the entire parameter set from available observational constraints, and 
we tuned the other parameters to provide rough agreement between the calculated 
flux and the data. In particular, we adjusted the global normalizations and 
spectral indices of the two distant smooth components (one for SNRs, another 
for pulsars) and of the local pulsars. The parameters that we used are 
summarized in \citetab{tab:final_tab}.

\section{Primary positrons}
\label{sec:primary_pos}

The increase in the positron fraction inferred from the PAMELA data is barely
consistent with what is expected for secondary 
positrons~\citep[\eg][]{2009A&A...501..821D}, except for those secondaries 
produced and accelerated in sources~\citep{2003A&A...410..189B,2009PhRvL.103e1104B,2009PhRvL.103h1104M,2009PhRvD..80l3017A}. Additional spatial arguments might 
still be helpful in dealing with this issue~\citep[][]{2009PhRvL.103k1302S}.

Although exotic primary contributions from dark matter annihilation could 
contribute to the positron budget, the most popular dark matter particle 
candidates can hardly exceed the secondary background unless the annihilation 
rate is boosted substantially \citep[\eg][]{1998PhRvD..59b3511B,
2008PhRvD..78j3526L,2008PhRvD..77f3527D,2008PhRvD..78j3520B}, or the expansion
rate is more rapid than expected in the early universe, before the primordial 
nucleosynthesis (see \citealt{2003PhLB..571..121S} for the original idea, 
and~\eg~\citealt{2009arXiv0912.4421C} more specifically for the positron
channel). Furthermore, the potential enhancement provided by dark matter 
substructures was demonstrated to be too small~\citep[see][]{2008A&A...479..427L,2008PhRvD..78j3526L,2009arXiv0908.0195P}, and any other type of global 
enhancement was shown to be severely restricted by the companion upper limit to 
the antiproton yield~\citep[\eg][]{2009PhRvL.102g1301D}, which fully applies to 
the most popular dark matter models that have in no case exclusive couplings to 
leptons. We therefore emphasize that generic dark matter candidates are not 
expected, unfortunately, to manifest themselves in the positron spectrum in 
the GeV-TeV energy range.

For other potential sources of primary positrons, we note that since the seminal
work on electron-positron pair production in strong magnetic fields by 
\cite{1966RvMP...38..626E}, a particular class of cosmic-ray sources has long 
been predicted to provide electron-positron pairs: pulsars
\citep[][]{1970Natur.227..465S,1971ApJ...164..529S}. Interestingly enough, a 
detailed discussion about the increase in the measured positron fraction and 
the potential role of pulsars was already conducted two decades ago 
by \cite{1989ApJ...342..807B}\footnotemark. Subsequent deeper studies have 
been performed since then~\citep[\eg][]{1995A&A...294L..41A,1995PhRvD..52.3265A,1996ApJ...459L..83C}, which were recently revisited by several authors
\citep[\eg][]{2009JCAP...01..025H,2008arXiv0812.4457P,2009PhRvL.103e1101Y,2009PhRvD..80f3005M}. Since pulsars are commonly observed in the vicinity of 
the Earth, it is very likely that these astrophysical sources of positrons are 
the origin of most of the local cosmic ray positrons. Nevertheless, many 
uncertainties remain in the characterization of pulsars, and there has been 
little theoretical progress in the past decade.

\footnotetext{\cite{1989ApJ...342..807B} did also mention that dark matter 
was another, though exotic, possibility.}

Pulsars are rapidly-rotating magnetized neutron stars, the rotation axis of 
which is misaligned with the axis of the magnetic field, which produces a 
pulsed emission for an observer located on the cone scanned by the magnetic 
axis. They are potential end products of CCSN explosions, such as black holes 
for the most massive progenitors and other neutron stars unseen as pulsars 
(if the observer is always off-magnetic axis, the rotation and magnetic 
axes are aligned, or the neutron star is either non-rotating or non-magnetized).
All pulsars must in principle be associated with a companion SNR. This is 
supported by observations since pulsar wind nebulae (PWNe) are sometimes found 
inside or close to the shells of SNRs \citep[see~\eg][for recent reviews on 
PWNe]{2006ARA&A..44...17G,2008AdSpR..41..491B}, though many pulsars remain 
non-associated\footnotemark. Thus, a more general picture of pulsar modeling, at
least for cosmic-ray electrons and positrons, should involve a SNR as the main 
energy supply, and a pulsar and its PWN as a subsystem injecting additional 
high energy electrons and positrons \citep[see \eg][]{2001ApJ...563..806B}. Note
that even though a SNR should be associated with each pulsar, the reverse is not
true: explosions of SNe1a leave no compact objects at all, and some CCSNe do 
not become pulsars. Likewise, we stress that the cosmic-ray acceleration 
mechanisms are different in SNRs and in PWNe, leading to different 
observational properties. In the former case, acceleration takes place at 
non-relativistic shocks, while at relativistic shocks in the latter case, which 
might imply different spectral shapes for the accelerated electrons. Moreover, 
pulsars convert a significant part of their spin-down energy into 
electron-positron pairs that are accelerated by the PWNe; this does not occur 
inside SNRs, where the energy fraction transferred to electrons is much lower 
(see \citesec{subsec:snr_prop}).

\footnotetext{Pulsars are likely usually expelled from the SNR system, but the 
relative velocity between both objects $\lesssim 10^4$ km/s induces a distance 
$\lesssim 100$ pc between them by the SNR lifetime. This distance may be
larger for old systems, which makes clear associations more difficult.}

In this section, we complete our calculations of the electron and positron flux 
at the Earth by including the contribution of pulsars. We adopt the same 
methodology as for SNRs, and consider two populations, (i) the local one, for 
which we constrain the individual properties from observational data, and (ii) 
a more distant and smoothly described population. As in most recent studies, 
the individual properties of pulsars are derived from the ATNF catalog
\citep{2005AJ....129.1993M}. In \citesec{subsec:pulsar_model}, we sketch the 
generic model that we have adopted. In \citesec{subsec:local_pulsars}, we 
discuss the contributions of both populations to the local electron and positron
flux. We then compare some template calculations to the current measurements of 
the positron fraction in \citesec{subsec:pf}.

\subsection{Generic pulsar modeling}
\label{subsec:pulsar_model}

Independently of the specific pulsar model, such as for instance the polar gap 
\citep{1975ApJ...196...51R}, the outer gap~\citep{1976ApJ...203..209C,1986ApJ...300..500C}, or the slot gap \citep{2008ApJ...680.1378H} models, we may summarize
the physics relevant to the production of cosmic-ray electrons and positrons as 
follows. Gamma rays can be generated in the pulsar magnetosphere from inverse 
Compton interactions of electrons accelerated along the strong and rotating 
magnetic field with local synchrotron radiation, which can in turn produce 
electron-positron pairs by annihilating with photons from the local radiation 
fields. Those gamma rays can be observed as pulsed emission, such as those 
recently detected with the Fermi satellite~\citep{2009Sci...325..840A}, which 
may therefore be used to constrain the pair production. These electron-positron 
pairs are then accelerated within the surrounding and expanding shocked medium, 
namely the PWN, located inside or offset from a more extended SNR. Observations 
of young systems such as the Crab nebula tell us that this acceleration can be 
very efficient and lead to huge Lorentz factors, up to $\sim 10^8-10^9$ 
\citep{1996MNRAS.278..525A}. What is important when trying to predict the 
electron-positron yield from a pulsar is not their energy distribution and 
density close to the magnetosphere, but instead their final characteristics 
after acceleration has proceeded and when the particles are released into the 
ISM. This has already been noted and described by~\citet{2009PhRvD..80f3005M}.
It is therefore rather difficult to provide accurate predictions when 
disregarding the whole dynamics at stake there, and, in this part, we 
mostly aim to survey the roles of the main ingredients that characterize 
pulsars rather than making peremptory predictions. Indeed, we show in the
following that current uncertainties still make it difficult to derive anything 
but qualitative predictions.

Following the arguments developed in~\cite{2009PhRvD..80f3005M}, 
to which we refer the reader for further details, we define the source 
term associated with any single pulsar to be
\ben
{\cal Q}_p(E,\vec{x},t) = q_p(E,t_\star)\, \delta(t-(t_\star+\delta t_\star))\,
\delta(\vec{x}-\vec{x}_\star)\;,
\label{eq:pulsar_source}
\een
where $t_\star$ and $\vec{x}_\star$ are the pulsar age and position, 
respectively, and $\delta t_\star$ accounts for a certain delay in the release 
of cosmic ray electrons into the ISM after the supernova explosion.  Generic 
pulsars should have ceased their PWN phases after $\sim$10-100 
kyr~\citep{2006ARA&A..44...17G}; for simplicity, however, since characterizing 
the PWN evolution is far beyond our purpose here, we assume that 
$\delta t_\star = 0$ in the following. The pulsar age is usually estimated from 
the {\em spin-down age}~\citep{1969ApJ...157.1395O}, which is only relevant to 
the spin-down magnetic radiation approximation, that involves the rotation 
period $P$ and its first time derivative $\dot{P}$, given by
\ben
t_{\rm pulsar} = - \frac{P}{2\,\dot{P}}\;.
\label{eq:spindown_age}
\een
As in the SNR case, we again emphasize that this age estimate relies on 
current observations, so that the {\em actual} age used for cosmic-ray transport
calculations should have an additional $d/c$ term, where $d$ is the distance of
the pulsar to the observer. We stress, however, that using the spin-down
age for the pulsar age estimate turns out, in many cases, to be erroneous 
\citep[see \eg][]{2000Natur.406..158G}. Nevertheless, for simplicity, we adopt 
this method to deal with local pulsars in the following.

For the energy spectrum, we adopt the same general shape as used previously for 
SNRs (see Eq.~\ref{eq:spectrum}), \ie~a power-law of index $\gamma$ with an 
exponential cut-off at energy $E_c$ as
\ben
q_p(E) = {\cal Q}_0 \,
\left(\frac{E}{E_0}\right)^{-\gamma}\exp\left(-\frac{E}{E_c}\right)\;.
\label{eq:pulsar_spectrum}
\een
In contrast to the SNR case for which the spectral index can be
constrained from radio observations, the spectral index associated with 
high energy electrons from pulsar can hardly be constrained by the radio
observations of the pulsed emission. Indeed, this pulsed emission originates
in regions close to the pulsar magnetosphere, where the acceleration 
processes are not yet achieved. An alternative is to use the spectral indices 
derived from PWN observations, when available. To simplify the discussion,
we use $\gamma=2$ in the following, unless other values are specified.

The normalization ${\cal Q}_0$ is intimately linked to the total rotational 
energy $W_0$ of the pulsar, a fraction $f$ of which is released in the form of 
electron-positron pairs, such that
\ben
\int_{E_{\rm min}}^{\infty} dE\,E\,q_p(E) &=& f\, W_0 \;, 
\label{eq:pulsar_norm}
\een
where $W_0$ can be constrained from measurements assuming that the whole energy 
lost is carried by the magnetic dipole radiation, such that
\ben
W_0 = \dot{E}\, \tau_{\rm dec} 
\left(1+\frac{t_\star}{\tau_{\rm dec}}\right)^\upsilon\;,
\label{eq:pulsar_W0}
\een
where $\dot{E}$ is the spin-down luminosity and 
$\tau_{\rm dec}\equiv E_0/\dot{E}_0$ is the typical pulsar decay time. Note
that the index $\upsilon$ indicating the age dependence is in principle 
related to the braking index $k$ that defines the rotation deceleration 
$\dot{\Omega}\sim -\Omega^k$, where $\Omega$ is the angular velocity, in terms 
of $\upsilon = (k+1)/(k-1)$. In the spin-down approximation, $k=3$, and 
therefore $\upsilon = 2$. Nevertheless, $k$ can also be computed if the second 
time derivative of $\Omega$ is known, 
$k = -\Omega\ddot{\Omega}/\dot{\Omega}^2$. In that case, it is usually found 
to slightly differ from 3. 

This illustrates again the large degree of theoretical uncertainties arising 
when trying to model pulsars, even in simple approaches. We still use the 
spin-down approximation, and therefore assume that $\upsilon = 2$ in the 
following.

Since accounting for the details in the pulsar modeling is beyond the scope of 
this paper, we adopt the source model defined by Eqs.~(\ref{eq:pulsar_source}-
\ref{eq:pulsar_W0}), and assume a universal decay time of $\tau_{\rm dec}=1$ 
kyr. Using this latter input and the ATNF data for the rotation period and its 
derivative, \citet{2009PhRvD..80f3005M} found typical values of 
$W_0\sim 10^{49}$ erg, \ie~one or two orders of magnitude below the 
characteristic supernova energy release, in agreement with the picture of a 
pulsar as a subdominant energy supply beside its companion SNR. Finally, it is 
usually supposed that a rather significant fraction of the spin-down energy is 
converted into electron-positron pairs that are accelerated in the PWNe. In the 
following, we assume by default that this fraction is $f=0.1$, unless otherwise 
specified. Though the typical spin-down energy of pulsars is a hundred times 
lower than the typical kinetic energy of SNRs, the rather 
efficient pair conversion occurring in pulsars implies that the overall energy 
carried by charged leptons is similar in both cases --- we recall that the 
fraction of kinetic energy converted into leptons is around $10^{-5}-10^{-4}$ in
SNRs (see \citesec{subsec:snr_prop}).

For the nearby known pulsars, we extract the distances and the ages from the 
ATNF catalog. For more distant objects, say above a few kpc, we can safely use 
the continuous limit as we did for SNRs, and write the corresponding source term
as
\ben
\widetilde{\cal Q}_p(E,\vec{x}) = \widetilde{q}_p(E)\, f(\vec{x})\;.
\label{eq:pulsar_csource}
\een
The energy spectrum is assumed to be the same as in \citeeq{eq:pulsar_spectrum},
and the normalization $\widetilde{\cal Q}_0$ is defined by an explosion rate 
that is similar to the case of SNRs.


\subsection{Local versus distant pulsars}
\label{subsec:local_pulsars}

\begin{figure}[t]
\begin{center}
\includegraphics[width=\columnwidth]{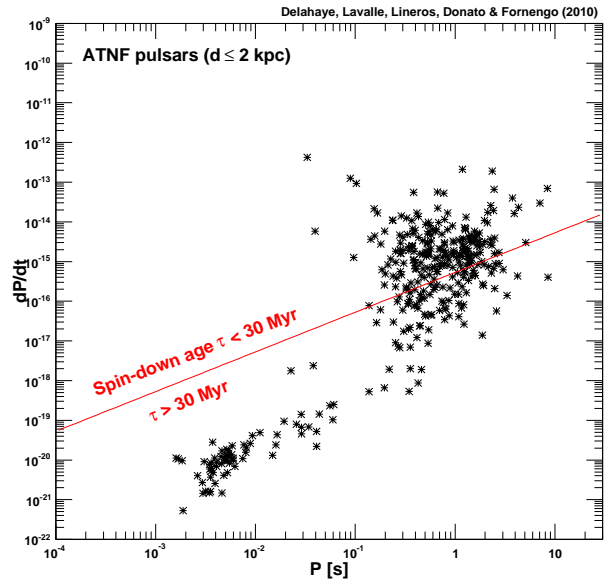}
\caption{\small Selection of local ATNF pulsars (within 2 kpc) in the 
$\dot{P}-P$ plane. Only those pulsars with a spin-down age lower than 30 Myr
were selected in our sample.}
\label{fig:pdotp}
\end{center}
\end{figure}

\begin{center}
\begin{table*}[ht] 
{\small
\hfill{}
\begin{tabular}{|c|c|c|c|c|c|c|} 
\hline
J2000  & other name & distance  & spin-down age  & spin-down energy & Rank & 
known SNR\\
  &  & [kpc] & [kyr] & [$10^{49}$ erg] & @ 5/100/1000 GeV & counterpart \\
\hline
J0633+1746 & Geminga & 0.16  & 342 &  1.25 & 1/2/4 &  \\
\hline
J1932+1059 & B1929+10 & 0.36 & 3,100 & 11.9 & 2/-/- & \\
\hline
J1908+0734 & & 0.58 & 4,080 & 17.9 & 3/-/- & \\
\hline
J1741-2054 & & 0.25 & 387 & 0.47 & 4/5/- & \\
\hline
J0953+0755 & B0950+08 & 0.26 & 17,500 & 54.2 & 5/-/- & \\
\hline
J2043+2740 &  & 1.13 & 1,200 & 25.9  & -/1/- &  \\
\hline
J1057-5226 & B1055-52 & 0.72 & 535 & 2.8 & -/3/- & \\
\hline
J0659+1414 & B0656+14 & 0.29 & 111 & 0.18 & -/4/2 & Monogem\\
\hline
J0835-4510 & B0833-45 & 0.29 & 11.3 & 0.99 & -/-/1 & Vela \\
\hline
J1740+1000 & & 1.24  & 114 & 1.1 & -/-/3 & \\
\hline
J0742-2822 & B0740-28 & 1.89 & 157 & 1.23 & -/-/5 & \\
\hline
J1549-4848 & & 1.54 & 324 & 0.8 &  -/-/6 & \\
\hline 
\end{tabular}}
\hfill{}
\caption{\small Main positron sources among the ATNF nearby pulsars. We rank
  the pulsars from the largest contribution to the flux in different energy 
  bins, assuming a spectral index of $\gamma = 2$ and a cut-off energy of 1 
  TeV. All other parameters are derived from the ATNF catalog.}
\label{tab:pulsars}
\end{table*}
\end{center}

We selected local pulsars from the ATNF catalog, imposing a few constraints. 
First, we applied a radial cut-off of 2 kpc, as for the SNR treatment, farther 
pulsars being accounted for with a smooth spatial distribution modeling. For the
latter, we adopted the L04 model, as for distant SNRs (see 
\citesec{subsubsec:smooth_el} for more details). Considering energies above the 
GeV scale imposes an upper limit of the pulsar age $\lesssim 30$ Myr, which 
decreases to 1 Myr above 100 GeV (see \citefig{fig:single_source}). We therefore
restrict our sample by requiring that the ages $\lesssim 30$ Myr. This selection
procedure is depicted in \citefig{fig:pdotp}, where the local ATNF pulsars are 
reported in the $P$-$\dot{P}$ plane. Our final sample contains a bit more than 
200 objects located at fewer than 2 kpc from the Earth, which is quite large 
compared to the number of observed SNRs, \ie~27, discussed in 
\citesec{subsubsec:local_el}.

The positron flux derived from this pulsar selection is obviously identical to 
the pulsar contribution to the electron flux discussed in 
\citesec{subsubsec:local_el}, because of the pair production mechanism, and is 
shown in the right panel of \citefig{fig:local_snrs_1}. We remind the reader 
that a SNR counterpart was systematically added to each local pulsar, except 
for pulsars that already had an observed and identified SNR counterpart 
featuring in our SNR list. In case of a non-observed SNR counterpart, we assumed
a radio flux $B(1\,{\rm GHz}) = 1 \,{\rm Jy}$ to set its spectrum normalization 
(see \citesec{fig:local_snrs_minmedmax}). These non-observed SNR counterparts 
contribute only to the electron flux, which is reported in the middle panel of 
\citefig{fig:local_snrs_1}.

As briefly mentioned in \citesec{subsubsec:local_el}, using the spin-down 
approximation to constrain the energy released in the form of electron-positron 
pairs leads pulsars to be important sources of local high energy electrons and 
positrons, as intense as the observed local SNRs. We employed a pair conversion 
efficiency of $f=0.1$ in this calculation, which might be optimistic,
but still, decreasing this efficiency to a few percent would lead to a quite 
significant contribution to the local electron and positron budget. It is 
noteworthy that the local positron flux is dominated by a few objects among our 
$\sim$200 selected objects. The main sources and their properties are listed 
in \citetab{tab:pulsars}. Although pulsar modeling is subject to many and large 
theoretical uncertainties, and despite the simplistic model we have employed to 
set the individual normalizations, our results suggest that SNRs might not be 
the only prominent sources of electron and positron cosmic rays. Nevertheless, 
this complete approach involving pulsars and SNRs in a self-consistent framework
should be studied in greater detail, which may be promising for gaining deeper 
insights into the understanding of cosmic ray leptons.

A full template calculation of the positron flux including secondaries and 
primaries from distant and local pulsars is compared with existing data in the 
bottom left panel of \citefig{fig:final_results}. We used the data from HEAT 
\citep{2001ApJ...559..296D}, CAPRICE \citep{2000ApJ...532..653B}, and AMS-01 
\citep{2000PhLB..484...10A}, which provide constraints at energies lower than 
$\sim$20 GeV only. From this plot, we see that secondaries can already account 
for a large fraction of the low energy positron flux, which constrains the 
properties of the additional primaries. We have in fact some freedom to tune 
the normalization and the spectral indices of the distant and local component, 
since they are poorly constrained. For the normalization, we have to adjust the 
pair conversion efficiency $f$ to values lower than 0.1, which reinforces 
the idea that pulsars can contribute to the local electron and positron flux 
quite naturally. This is discussed further when tackling the positron fraction 
case in \citesec{subsec:pf}. Likewise, there is no particular need to invoke 
hard spectral indices $\lesssim 2$. The parameters that we have used are 
summarized in \citetab{tab:final_tab}.

Finally, as for local contributions to the electron flux, we again point out 
that using the Thomson approximation to deal with the energy losses can lead 
to large errors and fake predictions of very peaky features in the overall 
positron spectrum.

\section{Full electron and positron results --- discussion}
\label{sec:disc}

In this section, we now perform a template calculation with reasonable 
parameters for all previously discussed ingredients to demonstrate that pure 
astrophysical processes easily account for current measurements. It is quite 
simple to find configurations in good agreement with the data of the sum of 
electrons plus positrons or/and on the positron fraction, but it is far more 
difficult to find additional agreement with the separate electron and positron 
data. Many other studies have focused only on a few of these data sets, more 
rarely on all together.

As an important and preliminary remark, we emphasize that playing with the 
parameters associated with very few local sources within reasonable ranges 
allows us to fit all the data quite easily, though roughly, simply by adjusting 
the low/high energy source hierarchy. Nevertheless, for the sake of 
illustration, we attempt to use the full set of observational constraints that 
we dispose of without turning around. This is to illustrate how difficult and 
complex this self-consistent exercise can be. However, to simplify the 
discussion, we stick to the \propmed~set of propagation parameters, keeping in 
mind that uncertainties affecting propagation lead to $\sim 1$ order of 
magnitude uncertainty in the overall predictions. For the solar modulation, we 
use a Fisk potential of 600 MV.

In addition to our template calculation, the data themselves, when 
assumed trustworthy, can provide the {\em actual} separate positron and 
electron fluxes at the Earth. If we define $f_-$ and $f_+$ as the electron flux 
and the positron flux, respectively, and assume that some fitting functions 
$f_{\rm tot}$ and $f_{\rm frac}$ for the measured total flux 
(\eg~from Fermi data) and for the positron fraction (\eg~from PAMELA data),
respectively, exist and are known, we then readily find
\ben
f_+(E) &=& f_{\rm frac}(E) \times f_{\rm tot}(E)\\
f_-(E) &=& f_{\rm tot}(E)(1-f_{\rm frac}(E) )\;.
\een
Although we do not perform the exercise in the following, we point out that 
obtaining individual predictions in good agreement with $f_+$ and $f_-$
in the relevant energy range automatically ensures close agreement with the 
measured positron fraction and total electron plus positron flux. This might 
sound tautological, but is still a helpful method if one wishes to optimize 
her/his preferred CR model.

\subsection{Electron and positron spectra: separate fluxes and sum}

\begin{table*}[t]
\begin{center}
\begin{tabular}{|c|ccc|cc|}
\hline
 & L04  & local SNRs & local SNRs & L04 & local pulsars \\
 & SNRs & (Green) & (ATNF) & pulsars & (ATNF) \\
\hline
Spectral index & 2.4 & $\dagger$ & 2.4 & 2.0 & 2.0 \\
\hline
\tgammae~ [$10^{48}$ erg/100 yr] & 21 & $6\times \dagger$ & 
from $B(1\,{\rm GHz})= 1\, {\rm Jy}$  & $3.6\times 10^{-2}$ & $\dagger$ \\
\hline
Converted fraction [\%] & - & - & - & - & 0.6 \\
\hline
$E_c$ [TeV] & 2.0 & 2.0 & 2.0 & 1.5 & 1.5 \\
\hline
\end{tabular}
\vskip 0.25cm
\caption{\small Injected energy, converted fraction, spectral indices, and 
  cut-off energies used for the overall template electron and positron flux 
  calculation. The symbol $\dagger$ indicates that we used observational 
  constraints. For local SNRs, we used a global extra-factor of 6, which 
  corresponds to assuming a magnetic field of $\sim 30$ instead of 100 $\mu$G
  in \citeeq{eq:q0_num}.}
\label{tab:final_tab}
\end{center}
\end{table*}

\begin{figure*}[t]
\begin{center}
\includegraphics[width=\columnwidth]{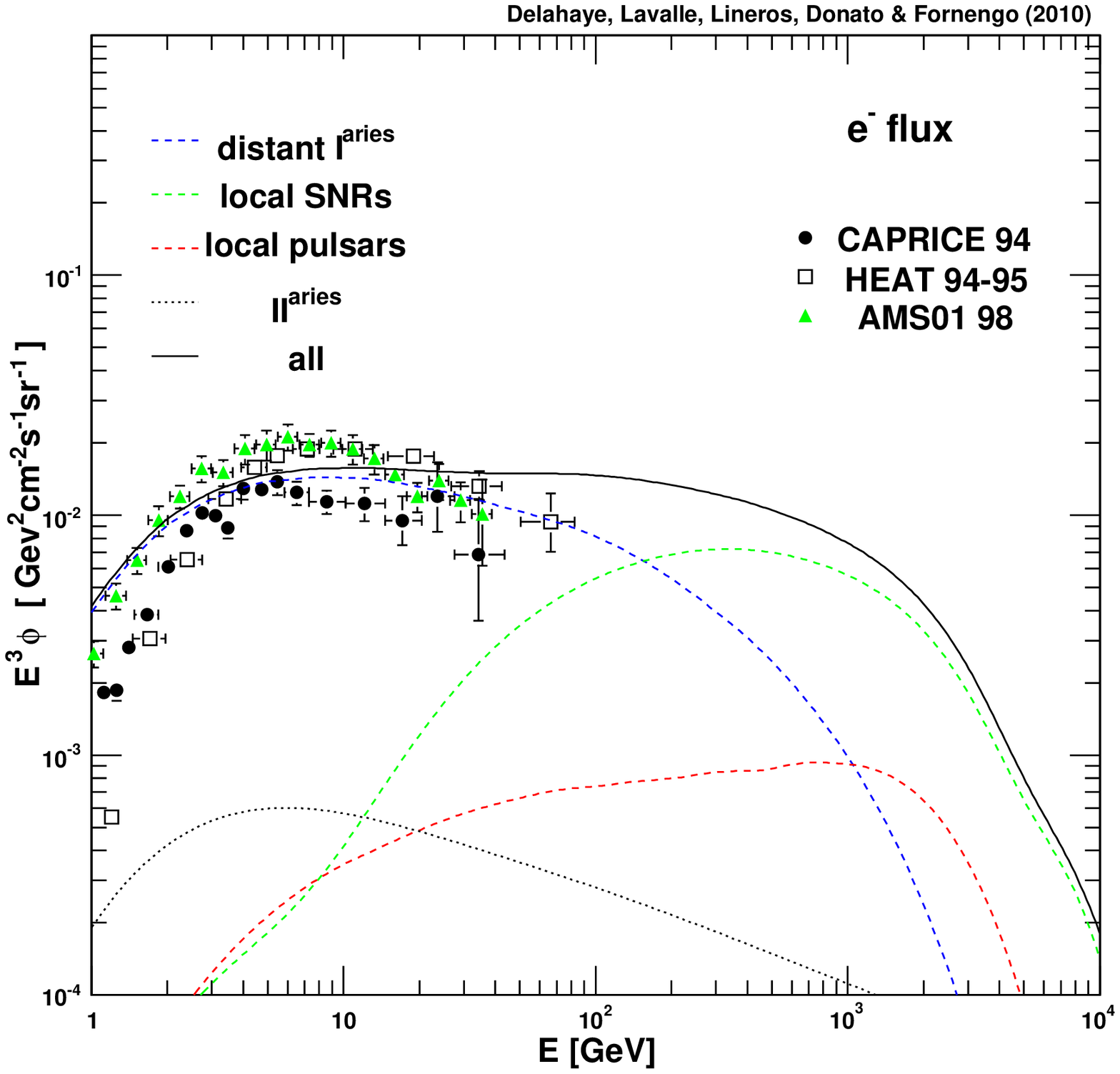}
\includegraphics[width=\columnwidth]{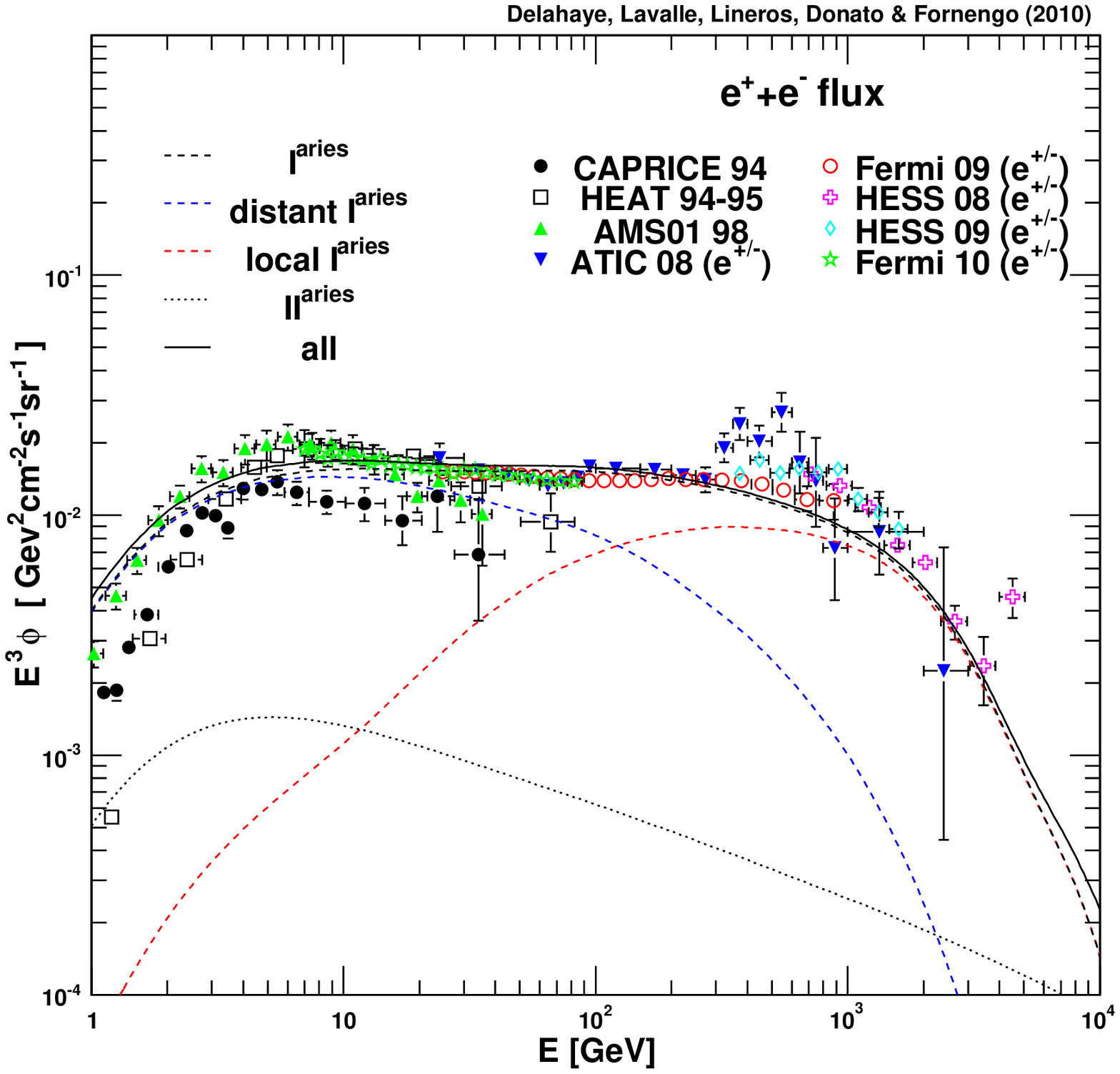} 
\includegraphics[width=\columnwidth]{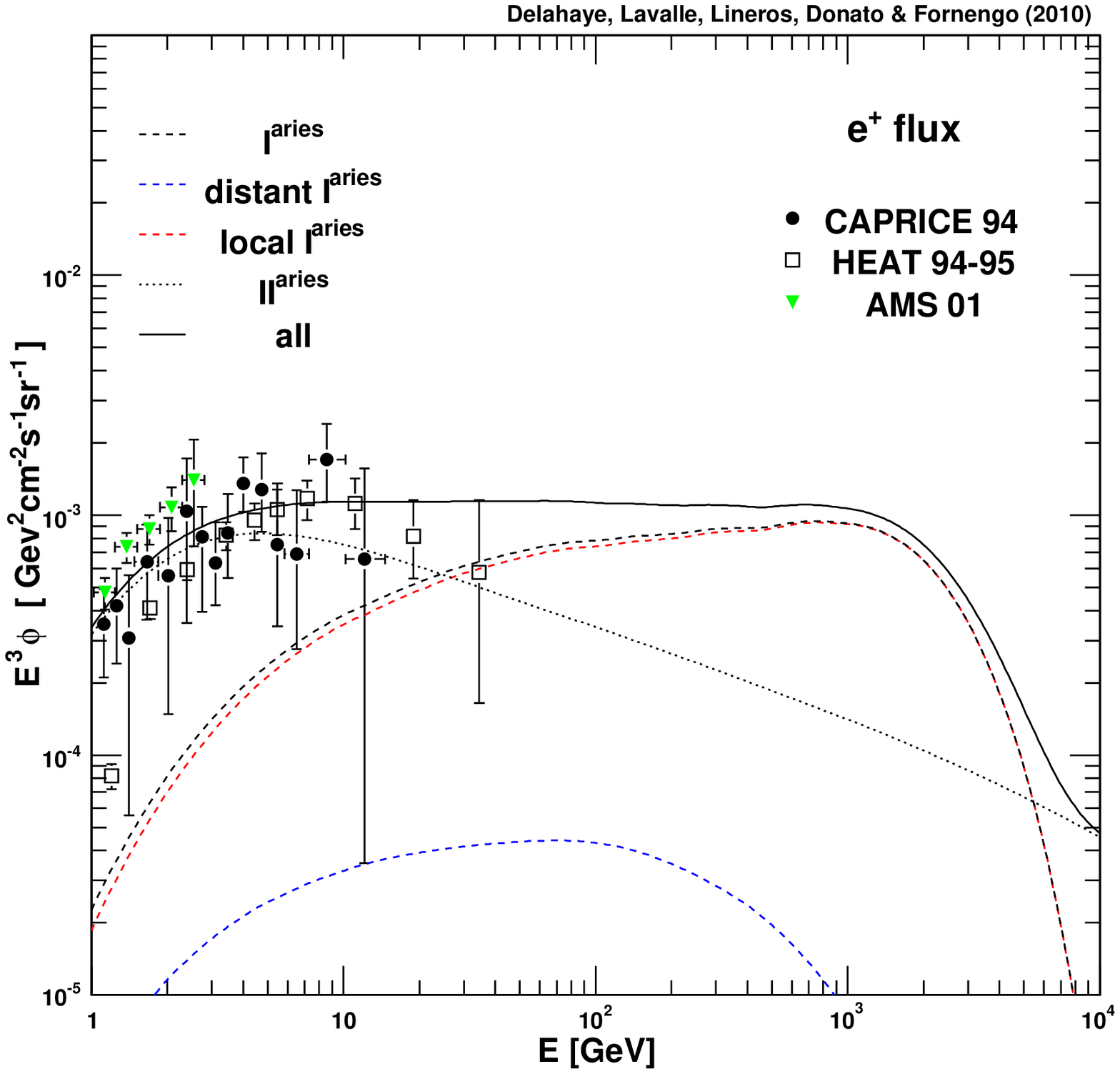}
\includegraphics[width=\columnwidth]{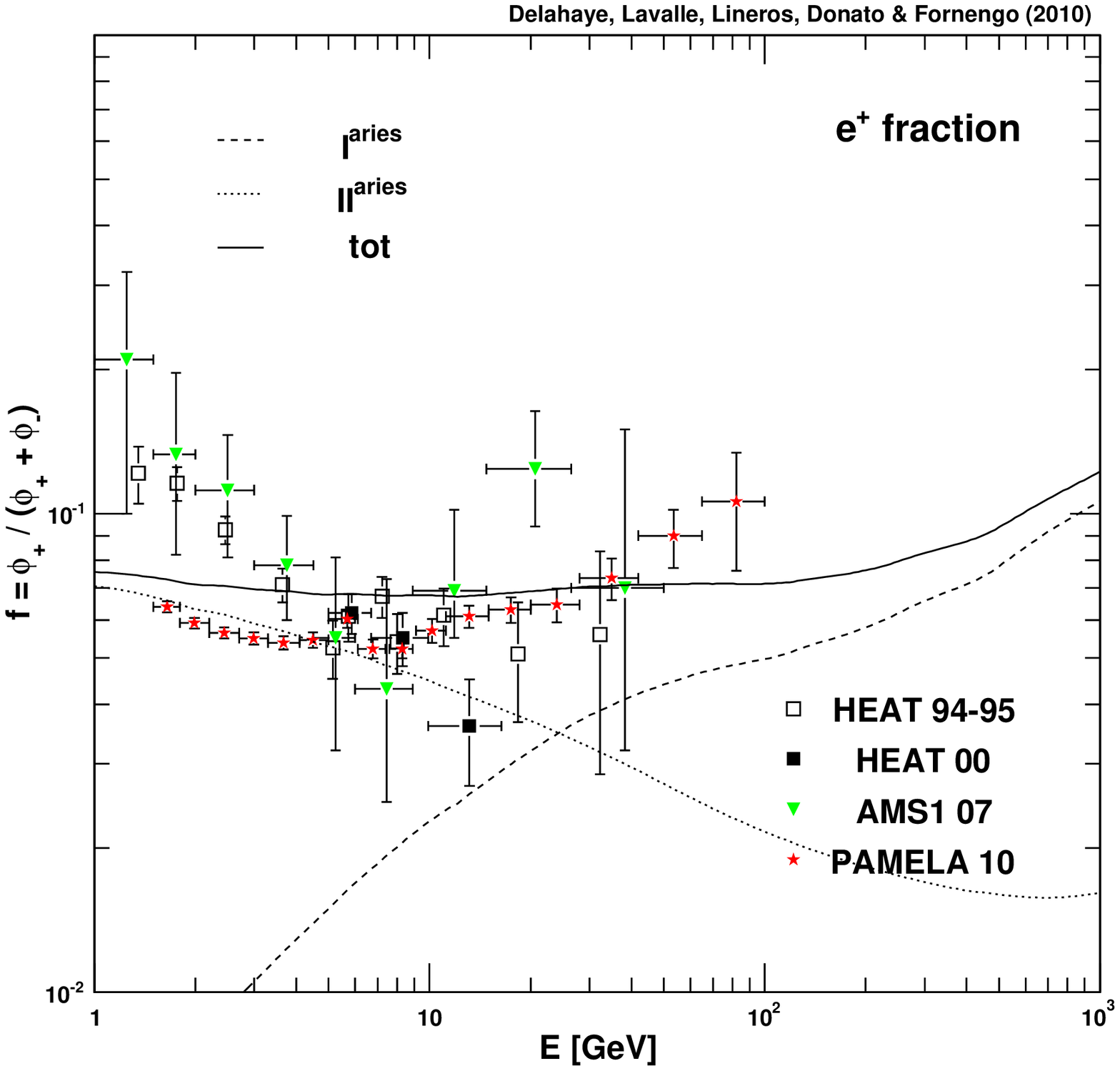} 
\caption{\small Template calculation including all primary (discrete local 
  and smooth distant) and secondary electrons and positrons in a self-consistent
  modeling, using the \propmed~propagation setup. Top left: 
  electron flux. Top right: electron and positron flux. Bottom left: positron
  flux. Bottom right: positron fraction.}
\label{fig:final_results}
\end{center}
\end{figure*}

We compare a template calculation of the electron and positron fluxes
with the available data. For electrons and positrons, as measured separately, 
we use the data from CAPRICE \citep{2000ApJ...532..653B}, HEAT 
\citep{2001ApJ...559..296D}, and AMS-01 \citep{2000PhLB..484...10A}. Note
that these data are quite poorly constraining at high energy, since they refer 
to energies below a very few tens of GeV, and are affected by large statistical 
errors for the highest available energy bins. Future data from PAMELA and 
AMS-02 may certainly provide much tighter separate constraints on both 
components. For the sum of electrons and positrons, we employ the data from 
Fermi \citep{2009PhRvL.102r1101A,2009arXiv0912.3611P}, ATIC~\citep{2008Natur.456..362C} and HESS \citep{2008PhRvL.101z1104A,2009A&A...508..561A}. It is 
meaningless to perform a fit, given the large theoretical uncertainties and the
large number of parameters. We therefore empirically select the parameter sets 
that provide the smallest statistical errors whenever possible.

The top and bottom left panels of \citefig{fig:final_results} show template 
results for the electron flux and the positron flux at the Earth, respectively.
The parameters that we used to calculate all contributions are reported in 
\citetab{tab:final_tab}.

For electrons, we see that the data of CAPRICE and AMS-01 are barely consistent,
but we can identify a median configuration. Our template model is partly 
typified by a smooth SNR-plus-pulsar component (following the L04 spatial 
distribution) that fits the low energy part --- the smooth pulsar contribution
is actually completely negligible. We used a rather soft spectral index of 2.4 
to shape this low energy smooth contribution, and a rather large value of 
\tgammae (still consistent with theoretical or observational constraints --- 
see \citesec{subsec:snr_prop}). Nevertheless, we point out that using another 
spatial distribution, for instance with a stronger/softer spatial gradient, may 
have given the same result with other appropriate spectral indices and 
normalizations (see right panel of \citefig{fig:spatial_dist}). This smooth 
component rapidly declines for energies higher than few tens GeV because of the 
2 kpc radial cut-off around the Earth that we imposed before adding the local 
component. Above 10 GeV, we see that the local SNR contribution comes into play,
clearly dominating over local pulsars. We underline that this hierarchy can
hardly be considered as a generic prediction, since both types of sources are 
expected to provide similar amounts of electrons (see left and right panels of 
\citefig{fig:local_snrs_1}). Moreover, we adjusted the local SNR flux by a 
global factor of 6 with respect to our constraint recipe given in 
\citeeq{eq:q0_num}, which corresponds to using a mean magnetic field of 
$\sim 30\,\mu$G instead of 100. We remark that only a few sources dominate the 
overall flux (see \citefig{fig:local_snrs_1} and \citetab{tab:stronglocal}),
so modifying the parameters of these latter only may have a very strong impact 
on the predictions. We also note that another pulsar to SNR contribution ratio 
could easily lead to a similar overall shape for the electron flux. Likewise, 
slightly relaxing our chosen constraints either directly on the individual 
intensities or on the ages and distances, even across quite limited ranges, 
would allow far more freedom to shape each contribution differently. 
Nevertheless, the source hierarchy is here imposed by the observational 
constraints, which are relevant to \citeeq{eq:q0_num} in the SNR case and 
\citeeq{eq:pulsar_W0} in the pulsar case; other methods based on different 
assumptions and different data would obviously provide different results. 
Finally, we stress that the local source flux is not characterized by individual
prominent spiky peaks, as would be the case in the Thomson approximation for 
the energy losses (see \citefig{fig:local_snrs_1}). Additional constraints on 
sources of electrons come from the sum of electrons and positrons and from 
the positron fraction, as we later discuss.

For the positron flux, the important point is that secondaries are most likely 
predominant below 5 GeV. The corresponding calculation is less affected by 
theoretical uncertainties, so we could assume it to be a {\em prediction}. 
Indeed, the progenitor proton flux is not expected to vary significantly within 
a few kpc, and the main uncertainty apart from propagation is the average ISM 
gas density \citep{2009A&A...501..821D}, such that the low energy positron
flux may vary by at most a few tens of \%. Beyond 5-10 GeV, there are poor 
constraints on the positron flux, so that the contribution of primaries may have
any shape. The positron fraction provides additional information and 
constraints, as we later demonstrate. Nevertheless, we see that our choice of 
parameters for pulsars (distant and local) makes their contribution significant 
above $\sim 5$ GeV. From \citetab{tab:final_tab}, it appears that assuming
less than 1\% of magnetic energy converted into electron-positron pairs is 
enough to make the pulsar contribution arise around 10 GeV; this is rather 
independent of the injected spectrum (we here assumed a spectral index of 2). 
We also remark that the smooth pulsar contribution is found to be negligible 
with respect to the local one. Finally, we emphasize that the overall spectral 
shape obtained for the positron flux weakly depends, in fact, on the assumed 
spectral index for pulsars. Indeed, as already shown in the right panel of 
\citefig{fig:local_snrs_1}, decreasing the intensity of a few old sources (such 
as Geminga or B1929+10) could easily result in a much harder spectrum at high 
energy and even a lower contribution below 10 GeV. Therefore, more tightly 
constraining the amplitude of the observed sources is a priority for future 
works.

The top right panel of \citefig{fig:final_results} shows the sum of electrons
and positrons, consistently with the separate results that we have just
discussed above. The most distinctive information one can extract from this 
plot is probably the energy cut-off that explicitly appears around a few TeV,
which obviously encourages us to set a cut-off energy around in this range. In 
our template calculation, the high energy part of the spectrum, above 
$\sim 50$ GeV, is actually dominated by our local SNR sample. This cannot
be considered as a robust prediction, since we could have obtained a dominating 
pulsar contribution by slightly modifying the injection parameters. For 
instance, we could have reduced the local SNR yield, both depleted the supply 
from Geminga and increased that from Vela, and finally allowed a larger 
fraction of $\sim 2$\% of pair conversion.

We point out that the current constraints on sources are far too weak 
to ascertain the predictive power of our template calculation. Nevertheless,
it is clear that the source modeling is the key point in the understanding of 
the high energy CR electrons, at least for identifying more clearly those few 
sources which may set most of the local flux. We also underline
that it is important to challenge any CR electron prediction with the separate
electron and positron data, since reproducing the sum does not necessarily 
ensure the relevance of the model. It will be important in the future to have 
independent sets of far higher quality data; PAMELA and AMS-02 carry many hopes.

\subsection{Positron fraction}
\label{subsec:pf}

Measurements of the positron fraction contribute additional constraints that 
should be fulfilled consistently with the previous data.
In the bottom right panel of \citefig{fig:final_results}, we have used the 
data from HEAT \citep{1997ApJ...482L.191B,2004PhRvL..93x1102B}, AMS-01 
\citep{2007PhLB..646..145A}, and PAMELA \citep{2010APh....34....1A}. 
From this plot, our template calculation turns out to be consistent with 
the data, especially above 10 GeV. At lower energies, there are large 
discrepancies among the available measurements, which renders the 
interpretation complicated, though a more realistic treatment of the solar 
modulation --- which is beyond the scope of this paper --- might improve their 
level of consistency. Note that we can accommodate a slightly increasing 
positron fraction with very reasonable parameters for the local pulsar 
modeling, \ie\ quite modest values of the converted energy fraction in $e^+e^-$
pairs and spectral indices, without the need to tune the parameters for the 
sources individually. Indeed, for our local samples of SNRs and pulsars, we 
emphasize again that we constrained the whole contributions from observational 
constraints and fixed assumptions. It would have been quite easy to make this 
fraction increasing much rapidly with energy by considering additional 
assumptions for the pulsar power. We instead considered the simplest approach of
the spin-down approximation for which the injection rate scales like the 
squared age (see Eq.~\ref{eq:pulsar_W0}). Had we adopted alternative modeling 
with a shallower dependence on the age \citep[see \eg~][]{1996ApJ...459L..83C,2001A&A...368.1063Z,2008arXiv0812.4457P},
we would have placed more weight on younger objects, and thereby to the high 
energy part of the spectrum. As for the other measurements discussed above, we 
conclude that the positron fraction can be explained by pure astrophysical 
processes, even if the predictive power of our calculation remains weak.

\subsubsection{Final comments}

We wish here to highlight a few important points. First, we can fairly state 
that standard astrophysics is capable of reproducing the existing data, despite 
the large theoretical uncertainties associated with the flux calculation. 
Second, we attempted to convince the reader that an overall check of the 
separate electron and positron flux calculations against the associated 
separate data is mandatory to determine the relevance of a model: this step
required the use of the four sets of data used in \citefig{fig:final_results} 
separately. This was not covered in the literature so far, where 
most authors focused on either the sum of CR electrons and positrons or on the 
positron fraction. Third, we have shown that the full relativistic treatment of 
the energy losses could lead to a global spectral shape devoid of peaky 
structures. Fourth, we have discussed a template example where we fixed the 
local source hierarchy based on a few assumptions and observational constraints.
It is clear, however, that a more refined analysis with a far more accurate 
source modeling will help us to reduce the theoretical uncertainties and 
clarify the local source hierarchy; it will also probably lead to different 
results. Nevertheless, we have proposed an exhaustive set of analytical tools 
together with a robust method to tackle such a detailed analysis. Finally, we 
stress that we have derived our template calculation using our \propmed~set of 
propagation parameters to facilitate the reading of the plots, and have 
therefore not discussed the uncertainties originating in the propagation side, 
which are actually large, as discussed in 
\citesec{subsec:local_snrs} for individual sources --- see bottom right panel 
of \citefig{fig:single_source}, and \citefig{fig:local_snrs_minmedmax}.

\section{Summary and conclusions}
\label{sec:concl}

We attempted to perform an exhaustive study of the main ingredients relevant to 
the calculation of the high energy CR electron and positron fluxes in the 
GeV-TeV energy range. We have underlined the complexity of the source 
description, by emphasizing the potentially strong effects of nearby sources, 
and shown that though we can hardly predict the local electron and positron 
fluxes with accuracy, reasonable parameterizations of the source and the 
propagation modeling can fairly well and simultaneously account for the current 
measurements of (i) electrons, (ii) positrons, (iii) the sum of them, and (iv) 
the positron fraction. We summarize hereafter the way in which we have 
proceeded.

In \citesec{sec:propag}, we presented our propagation modeling in detail.
We reviewed the analytical solutions to the transport equation in 
\citesecs{subsec:propag} and~\ref{subsec:time} and derived explicit links 
between the source spectrum and the propagated spectrum in 
\citesec{subsec:spectral_indices}. More importantly, we developed in 
\citesec{subsec:eloss} a method to account for the relativistic energy losses, 
providing useful fit formulae that can be used for any inverse Compton 
processes involving a black-body photon target. We have indeed shown that the 
Thomson approximation was by far insufficient and could lead to fake 
predictions of peaky signatures in the electron spectrum as potential imprints 
of local sources. In \citesec{subsec:prop_par}, we recalled our propagation 
parameters with emphasis on the energy losses. In particular, we developed a 
quite tightly constrained local ISRF modeling and explained in detail how its 
different components result in decreasing steps in the overall 
energy loss function, as a consequence of Klein-Nishina effects.

In \citesec{sec:secondaries}, we revisited the predictions for the 
secondary electron and positron fluxes, finding slight differences from 
the previous analysis we performed in \citet{2009A&A...501..821D} --- for 
positrons only, and using the Thomson approximation for the energy losses. 
These differences are caused by our novel relativistic treatment of the energy 
losses. We found a slightly harder secondary spectrum, with an overall flux 
higher by $\sim 10$\% at 100 GeV up to $\sim 50$\% at 1 TeV. We emphasized, 
however, that this can still not explain an increasing positron fraction. We
provided fitting formulae of our results in \citeapp{app:fit}, valid in the 
\propmed~propagation setup.

In \citesec{sec:primary_el}, we described in detail the calculation of the 
primary electron flux. We explored thoroughly the possible ways of normalizing 
the injection rate of CR electrons in \citesec{subsec:snr_prop}, focusing on a 
smooth distribution of sources as well as on discrete sources, and showed that 
the theoretical uncertainties are impressively large, reaching 2 or 3 orders of 
magnitude. In particular, the fraction of energy transmitted to CR electrons is 
quite an issue. Nevertheless, we sketched a method to normalize an event-like 
source injection density from its measured radio flux, which allows us to use 
observational constraints whenever available. We demonstrated in 
\citesec{subsec:smooth_snrs} that any smooth spatial distribution of sources 
could be approximated with a mere $z$-exponential disk up to very high accuracy 
for electrons of energy above a few tens of GeV, showing that the specific 
spatial distribution had negligible impact on the predictions, for which only 
the local density of sources really accounts. Since local effects are more
important above a few tens of GeV, we exhaustively studied in 
\citesec{subsec:local_snrs} the impacts of the parameters relevant to the 
local event-like source description, \ie~the age, the distance and the spectral 
index, as well as of the propagation parameters (see 
\citefig{fig:single_source}). We notably illustrated how irrelevant the 
individual spectral indices were to inferring the overall spectral shape 
produced by a population of non-identical objects. We have shown that the 
overall spectrum was mostly set by a hardly predictable hierarchy in the local 
sources. The energy range characterizing a single contribution is mostly bounded
from above by the age and cut-off energy, and from below by the distance. For 
the former bound, we demonstrated that the Thomson approximation led to 
erroneous predictions, with significant underestimates of the maximal energy 
and fake peaky features. We also demonstrated that changing the diffusion 
coefficient had a strong impact on the results. Finally, in 
\citesec{subsec:el_pred}, we discussed a template calculation including the 
constrained contributions of all known local sources lying within a distance of 
2 kpc from the Earth, showing that pulsars might contribute as many primary 
electrons as SNRs. We again emphasized the quite poor predictive power that can 
be achieved even when accounting for observational constraints. This is mostly 
due to the theoretical uncertainties affecting the source modeling. 
Nevertheless, we indicated a few objects that probably dominate the overall 
high energy flux.

In \citesec{sec:primary_pos}, we studied the contribution of pulsars to 
the primary positron flux, using the spin-down approximation to 
constrain the injection rate of local pulsars. As for primary electrons, we 
demonstrated that the very large theoretical uncertainties makes it difficult 
to claim robust predictions.

For the local sources, we proposed a novel approach including not only known 
SNRs or pulsars, but also adding a SNR counterpart to each non-associated 
pulsar. Indeed, since a pulsar is the relic neutron star of a core-collapse 
supernova explosion, it should be accompanied by a SNR for consistency reasons. 
We accounted for these non-observed SNRs needing to have a radio flux below the 
current experimental sensitivities when normalizing their injection rates of 
electrons. In contrast, all SNRs are not expected to have a pulsar 
companion, in particular those SNRs coming from SN1a explosions. A more refined 
and realistic modeling of sources including such composite contributions and 
further accounting for the dynamics of CR injection remains to be considered in 
terms of electron flux. This is, however, far beyond the scope of this paper, 
though we attempted to pave the road for such an ambitious study.

Finally, in \citesec{sec:disc}, we presented a template calculation 
including all sources of primaries, in which we used observational constraints 
whenever available (see \citefig{fig:final_results}). We considered the true 
complexity of the source description within a self-consistent framework, at 
variance with many other studies that merely added local contributions to other 
so-called {\em standard} predictions. We demonstrated that for reasonable 
assumptions about the parameters of both the local and distant components, all 
of the discussed astrophysical processes can account rather well for all 
measurements of cosmic ray electrons and positrons independently. We pointed out
that comparing any prediction with only the positron fraction or the sum of 
electrons and positrons is incomplete, and does not ensure its consistency. We 
emphasize that our template calculation cannot stand for a robust prediction 
because of the very large theoretical uncertainties underlined along this paper.
Our results should instead be considered as a proof that current measurements 
are {\em compatible} with pure astrophysical processes, so that one can hardly 
claim any prominent anomaly at the moment. More importantly, we hope to have 
shown that a {\em standard model} of CR electrons, {\em standard} in the sense 
that departures from observational data would sign something unexpected or 
exotic, is far from being reached, since we even do not control the modeling 
of the local environment (mostly the sources), which is of paramount importance 
here. This may have drastic consequences; for instance, it indicates how well 
one can control or predict the CR electron contribution to the high energy part 
of the Galactic diffuse gamma-ray emission, or to the diffuse radio emission. 
In turn, however, this diffuse emission can provide additional information 
and constraints on the astrophysical processes at stake. More accurate 
predictions may only emerge for observables that involve features averaged over 
large spatial and/or timescales, to guarantee protection against potentially 
large and hardly controllable fluctuations in the calculation results. This is 
for instance the case for high energy stable nuclei.

Such conclusions may sound disappointing but they imply many interesting 
by-products and open anyway a very broad landscape for future theoretical 
improvements. For instance, we have shown that a few local sources may dominate 
the electron flux. If so, far more refined models of these sources may allow us 
to use electrons as independent tracers of the local diffusion coefficient. 
Positive measurements of anisotropies in the CR electron flux  might also 
provide insights in the power hierarchy among local sources, provided this 
hierarchy is strong; indeed, each flux energy-bin is most likely filled by 
contributions from several sources located at different positions in the sky. 
In any case, the connection between the local CR electrons and the physics of 
sources is probably one of the most important issue to investigate further in 
future studies; low energy protons might also provide additional constraints on 
the local sources.

Finally, for the searches for exotic signatures, \eg~from dark matter 
annihilation or decay, the large theoretical uncertainties associated with the 
astrophysical processes prevent us from deriving much stronger constraints at 
the moment, unfortunately. Nevertheless, we underline that invoking any exotic 
contribution is absolutely unnecessary to understand the current data, and is 
thereby quite easily arguable. It seems therefore likely that models 
which are over-tuned to fit the positron data, aside from independent 
motivations coming from particle physics, are of very weak relevance. We 
recall that almost all well-studied dark matter particle candidates, 
\eg~in the framework of supersymmetric or extra-dimensional theories, are  
expected to manifest themselves in neither the local positron nor antiproton 
spectra \citep[\eg][]{2008PhRvD..78j3526L,2008PhRvD..77f3527D,2004PhRvD..69f3501D}.
Were dark matter annihilation really to enhance the positron budget, far more 
work would be necessary to prove it, by means of multi-messenger and 
multiwavelength approaches. This would require great efforts to face the 
complexity of the astrophysical backgrounds and, more importantly, this will 
have to rely on self-consistent calculations of both the signals and 
backgrounds.

\begin{acknowledgements}
We are deeply indebted to P. Salati for very inspiring comments and previous 
collaborations on related topics. We would like to thank A. Fiasson and 
Y. Gallant for very helpful discussions about SNRs and pulsars. We are also 
grateful to the participants of the two Workshops on Diffuse $\gamma$-ray 
emission, organized in 2008 and 2009 in the frame of the French GDR PCHE, for 
fruitful debates related to the topic\footnotemark. TD thanks the 
International Doctorate on Astro-Particle Physics (IDAPP) and the Rh\^one-Alpes 
region (Explora'Doc program) for financial support. JL is grateful to LAPTH 
for hospitality during parts of this study. NF acknowledges support of the 
Spanish MICINN’s Consolider--Ingenio 2010 Programme under grant MULTIDARK 
CSD2009-00064. This work was partly supported by research grants funded jointly 
by Ministero dell'Istruzione, dell'Universit\`a e della Ricerca (MIUR), by the 
University of Torino (UniTO), by the Istituto Nazionale di Fisica Nucleare 
(INFN) within the {\sl Astroparticle Physics Project}, by the Italian Space 
Agency (ASI) under contract N$^{\circ}$ I/088/06/0 and by the French Programme 
National de Cosmologie (PNC).
\end{acknowledgements}

\footnotetext{A short summary of the workshops and the website addresses are
  available in \citet{2009sf2a.conf..165L}.}

\bibliographystyle{aa}
\bibliography{lavalle_bib}

\begin{thebibliography}{187}
\expandafter\ifx\csname natexlab\endcsname\relax\def\natexlab#1{#1}\fi

\bibitem[{{Abdo} {et~al.}(2009{\natexlab{a}}){Abdo}, {Ackermann}, {Ajello},
  {Anderson}, {Atwood}, {Axelsson}, {Baldini}, {Ballet}, {Barbiellini},
  {Baring}, {Bastieri}, {Baughman}, {Bechtol}, {Bellazzini}, {Berenji},
  {Bignami}, {Blandford}, {Bloom}, {Bonamente}, {Borgland}, {Bregeon}, {Brez},
  {Brigida}, {Bruel}, {Burnett}, {Caliandro}, {Cameron}, {Caraveo},
  {Casandjian}, {Cecchi}, {{\c C}elik}, {Chekhtman}, {Cheung}, {Chiang},
  {Ciprini}, {Claus}, {Cohen-Tanugi}, {Conrad}, {Cutini}, {Dermer}, {de
  Angelis}, {de Luca}, {de Palma}, {Digel}, {Dormody}, {do Couto e Silva},
  {Drell}, {Dubois}, {Dumora}, {Farnier}, {Favuzzi}, {Fegan}, {Fukazawa},
  {Funk}, {Fusco}, {Gargano}, {Gasparrini}, {Gehrels}, {Germani}, {Giebels},
  {Giglietto}, {Giommi}, {Giordano}, {Glanzman}, {Godfrey}, {Grenier},
  {Grondin}, {Grove}, {Guillemot}, {Guiriec}, {Gwon}, {Hanabata}, {Harding},
  {Hayashida}, {Hays}, {Hughes}, {J{\'o}hannesson}, {Johnson}, {Johnson},
  {Johnson}, {Kamae}, {Katagiri}, {Kataoka}, {Kawai}, {Kerr}, {Kn{\"o}dlseder},
  {Kocian}, {Kuss}, {Lande}, {Latronico}, {Lemoine-Goumard}, {Longo},
  {Loparco}, {Lott}, {Lovellette}, {Lubrano}, {Madejski}, {Makeev}, {Marelli},
  {Mazziotta}, {McConville}, {McEnery}, {Meurer}, {Michelson}, {Mitthumsiri},
  {Mizuno}, {Monte}, {Monzani}, {Morselli}, {Moskalenko}, {Murgia}, {Nolan},
  {Norris}, {Nuss}, {Ohsugi}, {Omodei}, {Orlando}, {Ormes}, {Paneque},
  {Parent}, {Pelassa}, {Pepe}, {Pesce-Rollins}, {Pierbattista}, {Piron},
  {Porter}, {Primack}, {Rain{\`o}}, {Rando}, {Ray}, {Razzano}, {Rea}, {Reimer},
  {Reimer}, {Reposeur}, {Ritz}, {Rochester}, {Rodriguez}, {Romani}, {Ryde},
  {Sadrozinski}, {Sanchez}, {Sander}, {Parkinson}, {Scargle}, {Sgr{\`o}},
  {Siskind}, {Smith}, {Smith}, {Spandre}, {Spinelli}, {Starck}, {Strickman},
  {Suson}, {Tajima}, {Takahashi}, {Takahashi}, {Tanaka}, {Thayer}, {Thompson},
  {Tibaldo}, {Tibolla}, {Torres}, {Tosti}, {Tramacere}, {Uchiyama}, {Usher},
  {Van Etten}, {Vasileiou}, {Vilchez}, {Vitale}, {Waite}, {Wang}, {Watters},
  {Winer}, {Wolff}, {Wood}, {Ylinen}, \& {Ziegler}}]{2009Sci...325..840A}
{Abdo}, A.~A., {Ackermann}, M., {Ajello}, M., {et~al.} 2009{\natexlab{a}},
  Science, 325, 840

\bibitem[{{Abdo} {et~al.}(2009{\natexlab{b}}){Abdo}, {Ackermann}, {Ajello},
  {Atwood}, {Axelsson}, {Baldini}, {Ballet}, {Barbiellini}, {Bastieri},
  {Battelino}, {Baughman}, {Bechtol}, {Bellazzini}, {Berenji}, {Blandford},
  {Bloom}, {Bogaert}, {Bonamente}, {Borgland}, {Bregeon}, {Brez}, {Brigida},
  {Bruel}, {Burnett}, {Caliandro}, {Cameron}, {Caraveo}, {Carlson},
  {Casandjian}, {Cecchi}, {Charles}, {Chekhtman}, {Cheung}, {Chiang},
  {Ciprini}, {Claus}, {Cohen-Tanugi}, {Cominsky}, {Conrad}, {Cutini}, {Dermer},
  {de Angelis}, {de Palma}, {Digel}, {di Bernardo}, {Do Couto E Silva},
  {Drell}, {Dubois}, {Dumora}, {Edmonds}, {Farnier}, {Favuzzi}, {Focke},
  {Frailis}, {Fukazawa}, {Funk}, {Fusco}, {Gaggero}, {Gargano}, {Gasparrini},
  {Gehrels}, {Germani}, {Giebels}, {Giglietto}, {Giordano}, {Glanzman},
  {Godfrey}, {Grasso}, {Grenier}, {Grondin}, {Grove}, {Guillemot}, {Guiriec},
  {Hanabata}, {Harding}, {Hartman}, {Hayashida}, {Hays}, {Hughes},
  {J{\'o}hannesson}, {Johnson}, {Johnson}, {Johnson}, {Kamae}, {Katagiri},
  {Kataoka}, {Kawai}, {Kerr}, {Kn{\"o}dlseder}, {Kocevski}, {Kuehn}, {Kuss},
  {Lande}, {Latronico}, {Lemoine-Goumard}, {Longo}, {Loparco}, {Lott},
  {Lovellette}, {Lubrano}, {Madejski}, {Makeev}, {Massai}, {Mazziotta},
  {McConville}, {McEnery}, {Meurer}, {Michelson}, {Mitthumsiri}, {Mizuno},
  {Moiseev}, {Monte}, {Monzani}, {Moretti}, {Morselli}, {Moskalenko}, {Murgia},
  {Nolan}, {Norris}, {Nuss}, {Ohsugi}, {Omodei}, {Orlando}, {Ormes}, {Ozaki},
  {Paneque}, {Panetta}, {Parent}, {Pelassa}, {Pepe}, {Pesce-Rollins}, {Piron},
  {Pohl}, {Porter}, {Profumo}, {Rain{\`o}}, {Rando}, {Razzano}, {Reimer},
  {Reimer}, {Reposeur}, {Ritz}, {Rochester}, {Rodriguez}, {Romani}, {Roth},
  {Ryde}, {Sadrozinski}, {Sanchez}, {Sander}, {Saz Parkinson}, {Scargle},
  {Schalk}, {Sellerholm}, {Sgr{\`o}}, {Smith}, {Smith}, {Spandre}, {Spinelli},
  {Starck}, {Stephens}, {Strickman}, {Strong}, {Suson}, {Tajima}, {Takahashi},
  {Takahashi}, {Tanaka}, {Thayer}, {Thayer}, {Thompson}, {Tibaldo}, {Tibolla},
  {Torres}, {Tosti}, {Tramacere}, {Uchiyama}, {Usher}, {van Etten},
  {Vasileiou}, {Vilchez}, {Vitale}, {Waite}, {Wallace}, {Wang}, {Winer},
  {Wood}, {Ylinen}, \& {Ziegler}}]{2009PhRvL.102r1101A}
{Abdo}, A.~A., {Ackermann}, M., {Ajello}, M., {et~al.} 2009{\natexlab{b}},
  Physical Review Letters, 102, 181101

\bibitem[{{Abdo} {et~al.}(2008){Abdo}, {Ackermann}, {Atwood}, {Baldini},
  {Ballet}, {Barbiellini}, {Baring}, {Bastieri}, {Baughman}, {Bechtol},
  {Bellazzini}, {Berenji}, {Blandford}, {Bloom}, {Bogaert}, {Bonamente},
  {Borgland}, {Bregeon}, {Brez}, {Brigida}, {Bruel}, {Burnett}, {Caliandro},
  {Cameron}, {Caraveo}, {Carlson}, {Casandjian}, {Cecchi}, {Charles},
  {Chekhtman}, {Cheung}, {Chiang}, {Ciprini}, {Claus}, {Cohen-Tanugi},
  {Cominsky}, {Conrad}, {Cutini}, {Davis}, {Dermer}, {de Angelis}, {de Palma},
  {Digel}, {Dormody}, {do Couto e Silva}, {Drell}, {Dubois}, {Dumora},
  {Edmonds}, {Farnier}, {Focke}, {Fukazawa}, {Funk}, {Fusco}, {Gargano},
  {Gasparrini}, {Gehrels}, {Germani}, {Giebels}, {Giglietto}, {Giordano},
  {Glanzman}, {Godfrey}, {Grenier}, {Grondin}, {Grove}, {Guillemot}, {Guiriec},
  {Harding}, {Hartman}, {Hays}, {Hughes}, {J{\'o}hannesson}, {Johnson},
  {Johnson}, {Johnson}, {Johnson}, {Kamae}, {Kanai}, {Kanbach}, {Katagiri},
  {Kawai}, {Kerr}, {Kishishita}, {Kiziltan}, {Kn{\"o}dlseder}, {Kocian},
  {Komin}, {Kuehn}, {Kuss}, {Latronico}, {Lemoine-Goumard}, {Longo}, {Lonjou},
  {Loparco}, {Lott}, {Lovellette}, {Lubrano}, {Makeev}, {Marelli}, {Mazziotta},
  {McEnery}, {McGlynn}, {Meurer}, {Michelson}, {Mineo}, {Mitthumsiri},
  {Mizuno}, {Moiseev}, {Monte}, {Monzani}, {Morselli}, {Moskalenko}, {Murgia},
  {Nakamori}, {Nolan}, {Nuss}, {Ohno}, {Ohsugi}, {Okumura}, {Omodei},
  {Orlando}, {Ormes}, {Ozaki}, {Paneque}, {Panetta}, {Parent}, {Pelassa},
  {Pepe}, {Pesce-Rollins}, {Piano}, {Pieri}, {Piron}, {Porter}, {Rain{\`o}},
  {Rando}, {Ray}, {Razzano}, {Reimer}, {Reimer}, {Reposeur}, {Ritz},
  {Rochester}, {Rodriguez}, {Romani}, {Roth}, {Ryde}, {Sadrozinski}, {Sanchez},
  {Sander}, {Parkinson}, {Schalk}, {Sellerholm}, {Sgr{\`o}}, {Siskind},
  {Smith}, {Smith}, {Spandre}, {Spinelli}, {Starck}, {Strickman}, {Suson},
  {Tajima}, {Takahashi}, {Takahashi}, {Tanaka}, {Thayer}, {Thayer}, {Thompson},
  {Thorsett}, {Tibaldo}, {Torres}, {Tosti}, {Tramacere}, {Usher}, {Van Etten},
  {Vilchez}, {Vitale}, {Wang}, {Watters}, {Winer}, {Wood}, {Yasuda}, {Ylinen},
  \& {Ziegler}}]{2008Sci...322.1218A}
{Abdo}, A.~A., {Ackermann}, M., {Atwood}, W.~B., {et~al.} 2008, Science, 322,
  1218

\bibitem[{{Aharonian} {et~al.}(2009{\natexlab{a}}){Aharonian}, {Akhperjanian},
  {Anton}, {Barres de Almeida}, {Bazer-Bachi}, {Becherini}, {Behera},
  {Bernl{\"o}hr}, {Bochow}, {Boisson}, {Bolmont}, {Borrel}, {Brucker}, {Brun},
  {Brun}, {B{\"u}hler}, {Bulik}, {B{\"u}sching}, {Boutelier}, {Chadwick},
  {Charbonnier}, {Chaves}, {Cheesebrough}, {Chounet}, {Clapson}, {Coignet},
  {Dalton}, {Daniel}, {Davids}, {Degrange}, {Deil}, {Dickinson},
  {Djannati-Ata{\"i}}, {Domainko}, {O'C.~Drury}, {Dubois}, {Dubus}, {Dyks},
  {Dyrda}, {Egberts}, {Emmanoulopoulos}, {Espigat}, {Farnier}, {Feinstein},
  {Fiasson}, {F{\"o}rster}, {Fontaine}, {F{\"u}{\ss}ling}, {Gabici}, {Gallant},
  {G{\'e}rard}, {Gerbig}, {Giebels}, {Glicenstein}, {Gl{\"u}ck}, {Goret},
  {G{\"o}ring}, {Hauser}, {Hauser}, {Heinz}, {Heinzelmann}, {Henri}, {Hermann},
  {Hinton}, {Hoffmann}, {Hofmann}, {Holleran}, {Hoppe}, {Horns},
  {Jacholkowska}, {de Jager}, {Jahn}, {Jung}, {Katarzy{\'n}ski}, {Katz},
  {Kaufmann}, {Kendziorra}, {Kerschhaggl}, {Khangulyan}, {Kh{\'e}lifi},
  {Keogh}, {Klu{\'z}niak}, {Kneiske}, {Komin}, {Kosack}, {Kossakowski},
  {Lamanna}, {Lenain}, {Lohse}, {Marandon}, {Martin}, {Martineau-Huynh},
  {Marcowith}, {Masbou}, {Maurin}, {McComb}, {Medina}, {Moderski}, {Moulin},
  {Naumann-Godo}, {de Naurois}, {Nedbal}, {Nekrassov}, {Nicholas}, {Niemiec},
  {Nolan}, {Ohm}, {Olive}, {de O{\~n}a Wilhelmi}, {Orford}, {Ostrowski},
  {Panter}, {Paz Arribas}, {Pedaletti}, {Pelletier}, {Petrucci}, {Pita},
  {P{\"u}hlhofer}, {Punch}, {Quirrenbach}, {Raubenheimer}, {Raue}, {Rayner},
  {Reimer}, {Renaud}, {Rieger}, {Ripken}, {Rob}, {Rosier-Lees}, {Rowell},
  {Rudak}, {Rulten}, {Ruppel}, {Sahakian}, {Santangelo}, {Schlickeiser},
  {Sch{\"o}ck}, {Schr{\"o}der}, {Schwanke}, {Schwarzburg}, {Schwemmer},
  {Shalchi}, {Sikora}, {Skilton}, {Sol}, {Spangler}, {Stawarz}, {Steenkamp},
  {Stegmann}, {Stinzing}, {Superina}, {Szostek}, {Tam}, {Tavernet}, {Terrier},
  {Tibolla}, {Tluczykont}, {van Eldik}, {Vasileiadis}, {Venter}, {Venter},
  {Vialle}, {Vincent}, {Vivier}, {V{\"o}lk}, {Volpe}, {Wagner}, {Ward},
  {Zdziarski}, \& {Zech}}]{2009A&A...508..561A}
{Aharonian}, F., {Akhperjanian}, A.~G., {Anton}, G., {et~al.}
  2009{\natexlab{a}}, \aap, 508, 561

\bibitem[{{Aharonian} {et~al.}(2008){Aharonian}, {Akhperjanian}, {Barres de
  Almeida}, {Bazer-Bachi}, {Becherini}, {Behera}, {Benbow}, {Bernl{\"o}hr},
  {Boisson}, {Bochow}, {Borrel}, {Braun}, {Brion}, {Brucker}, {Brun},
  {B{\"u}hler}, {Bulik}, {B{\"u}sching}, {Boutelier}, {Carrigan}, {Chadwick},
  {Charbonnier}, {Chaves}, {Cheesebrough}, {Chounet}, {Clapson}, {Coignet},
  {Costamante}, {Dalton}, {Degrange}, {Deil}, {Dickinson}, {Djannati-Ata{\"i}},
  {Domainko}, {Drury}, {Dubois}, {Dubus}, {Dyks}, {Dyrda}, {Egberts},
  {Emmanoulopoulos}, {Espigat}, {Farnier}, {Feinstein}, {Fiasson},
  {F{\"o}rster}, {Fontaine}, {F{\"u}{\ss}ling}, {Gabici}, {Gallant},
  {G{\'e}rard}, {Giebels}, {Glicenstein}, {Gl{\"u}ck}, {Goret},
  {Hadjichristidis}, {Hauser}, {Hauser}, {Heinz}, {Heinzelmann}, {Henri},
  {Hermann}, {Hinton}, {Hoffmann}, {Hofmann}, {Holleran}, {Hoppe}, {Horns},
  {Jacholkowska}, {de Jager}, {Jung}, {Katarzy{\'n}ski}, {Kaufmann},
  {Kendziorra}, {Kerschhaggl}, {Khangulyan}, {Kh{\'e}lifi}, {Keogh}, {Komin},
  {Kosack}, {Lamanna}, {Lenain}, {Lohse}, {Marandon}, {Martin},
  {Martineau-Huynh}, {Marcowith}, {Maurin}, {McComb}, {Medina}, {Moderski},
  {Moulin}, {Naumann-Godo}, {de Naurois}, {Nedbal}, {Nekrassov}, {Niemiec},
  {Nolan}, {Ohm}, {Olive}, {de O{\~n}a Wilhelmi}, {Orford}, {Osborne},
  {Ostrowski}, {Panter}, {Pedaletti}, {Pelletier}, {Petrucci}, {Pita},
  {P{\"u}hlhofer}, {Punch}, {Quirrenbach}, {Raubenheimer}, {Raue}, {Rayner},
  {Renaud}, {Rieger}, {Ripken}, {Rob}, {Rosier-Lees}, {Rowell}, {Rudak},
  {Rulten}, {Ruppel}, {Sahakian}, {Santangelo}, {Schlickeiser}, {Sch{\"o}ck},
  {Schr{\"o}der}, {Schwanke}, {Schwarzburg}, {Schwemmer}, {Shalchi}, {Skilton},
  {Sol}, {Spangler}, {Stawarz}, {Steenkamp}, {Stegmann}, {Superina}, {Tam},
  {Tavernet}, {Terrier}, {Tibolla}, {van Eldik}, {Vasileiadis}, {Venter},
  {Vialle}, {Vincent}, {Vivier}, {V{\"o}lk}, {Volpe}, {Wagner}, {Ward},
  {Zdziarski}, \& {Zech}}]{2008PhRvL.101z1104A}
{Aharonian}, F., {Akhperjanian}, A.~G., {Barres de Almeida}, U., {et~al.} 2008,
  Physical Review Letters, 101, 261104

\bibitem[{{Aharonian} {et~al.}(2009{\natexlab{b}}){Aharonian}, {Akhperjanian},
  {de Almeida}, {Bazer-Bachi}, {Behera}, {Beilicke}, {Benbow}, {Bernl{\"o}hr},
  {Boisson}, {Bochow}, {Borrel}, {Braun}, {Brion}, {Brucker}, {B{\"u}hler},
  {Bulik}, {B{\"u}sching}, {Boutelier}, {Carrigan}, {Chadwick}, {Charbonnier},
  {Chaves}, {Chounet}, {Clapson}, {Coignet}, {Costamante}, {Dalton},
  {Degrange}, {Dickinson}, {Djannati-Ata{\"i}}, {Domainko}, {Drury}, {Dubois},
  {Dubus}, {Dyks}, {Egberts}, {Emmanoulopoulos}, {Espigat}, {Farnier},
  {Feinstein}, {Fiasson}, {F{\"o}rster}, {Fontaine}, {F{\"u}{\ss}ling},
  {Gabici}, {Gallant}, {G{\'e}rard}, {Giebels}, {Glicenstein}, {Gl{\"u}ck},
  {Goret}, {Hadjichristidis}, {Hauser}, {Hauser}, {Heinzelmann}, {Henri},
  {Hermann}, {Hinton}, {Hoffmann}, {Hofmann}, {Holleran}, {Hoppe}, {Horns},
  {Jacholkowska}, {de Jager}, {Jung}, {Katarzy{\'n}ski}, {Kaufmann},
  {Kendziorra}, {Kerschhaggl}, {Khangulyan}, {Kh{\'e}lifi}, {Keogh}, {Komin},
  {Kosack}, {Lamanna}, {Latham}, {Lemoine-Goumard}, {Lenain}, {Lohse},
  {Marandon}, {Martin}, {Martineau-Huynh}, {Marcowith}, {Masterson}, {Maurin},
  {McComb}, {Medina}, {Moderski}, {Moulin}, {Naumann-Godo}, {de Naurois},
  {Nedbal}, {Nekrassov}, {Niemiec}, {Nolan}, {Ohm}, {Olive}, {de O{\~n}a
  Wilhelmi}, {Orford}, {Osborne}, {Ostrowski}, {Panter}, {Pedaletti},
  {Pelletier}, {Petrucci}, {Pita}, {P{\"u}hlhofer}, {Punch}, {Quirrenbach},
  {Raubenheimer}, {Raue}, {Rayner}, {Renaud}, {Rieger}, {Ripken}, {Rob},
  {Rosier-Lees}, {Rowell}, {Rudak}, {Ruppel}, {Sahakian}, {Santangelo},
  {Schlickeiser}, {Sch{\"o}ck}, {Schr{\"o}der}, {Schwanke}, {Schwarzburg},
  {Schwemmer}, {Shalchi}, {Skilton}, {Sol}, {Spangler}, {Stawarz}, {Steenkamp},
  {Stegmann}, {Superina}, {Tam}, {Tavernet}, {Terrier}, {Tibolla}, {van Eldik},
  {Vasileiadis}, {Venter}, {Vialle}, {Vincent}, {Vink}, {Vivier}, {V{\"o}lk},
  {Volpe}, {Wagner}, {Ward}, {Zdziarski}, \& {Zech}}]{2009ApJ...692.1500A}
{Aharonian}, F., {Akhperjanian}, A.~G., {de Almeida}, U.~B., {et~al.}
  2009{\natexlab{b}}, \apj, 692, 1500

\bibitem[{{Aharonian} {et~al.}(1995){Aharonian}, {Atoyan}, \&
  {Voelk}}]{1995A&A...294L..41A}
{Aharonian}, F.~A., {Atoyan}, A.~M., \& {Voelk}, H.~J. 1995, \aap, 294, L41

\bibitem[{{Ahlers} {et~al.}(2009){Ahlers}, {Mertsch}, \&
  {Sarkar}}]{2009PhRvD..80l3017A}
{Ahlers}, M., {Mertsch}, P., \& {Sarkar}, S. 2009, \prd, 80, 123017

\bibitem[{{Alcaraz} {et~al.}(2000){Alcaraz}, {Alpat}, {Ambrosi}, {Anderhub},
  {Ao}, {Arefiev}, {Azzarello}, {Babucci}, {Baldini}, {Basile}, {Barancourt},
  {Barao}, {Barbier}, {Barreira}, {Battiston}, {Becker}, {Becker},
  {Bellagamba}, {B{\'e}n{\'e}}, {Berdugo}, {Berges}, {Bertucci}, {Biland},
  {Bizzaglia}, {Blasko}, {Boella}, {Boschini}, {Bourquin}, {Brocco}, {Bruni},
  {Buenerd}, {Burger}, {Burger}, {Cai}, {Camps}, {Cannarsa}, {Capell},
  {Casadei}, {Casaus}, {Castellini}, {Cecchi}, {Chang}, {Chen}, {Chen}, {Chen},
  {Chernoplekov}, {Chiueh}, {Chuang}, {Cindolo}, {Commichau}, {Contin},
  {Crespo}, {Cristinziani}, {da Cunha}, {Dai}, {Deus}, {Dinu}, {Djambazov},
  {D'Antone}, {Dong}, {Emonet}, {Engelberg}, {Eppling}, {Eronen}, {Esposito},
  {Extermann}, {Favier}, {Fiandrini}, {Fisher}, {Fluegge}, {Fouque},
  {Galaktionov}, {Gervasi}, {Giusti}, {Grandi}, {Grimm}, {Gu}, {Hangarter},
  {Hasan}, {Hermel}, {Hofer}, {Huang}, {Hungerford}, {Ionica}, {Ionica},
  {Jongmanns}, {Karlamaa}, {Karpinski}, {Kenney}, {Kenny}, {Kim}, {Klimentov},
  {Kossakowski}, {Koutsenko}, {Kraeber}, {Laborie}, {Laitinen}, {Lamanna},
  {Laurenti}, {Lebedev}, {Lee}, {Levi}, {Levtchenko}, {Liu}, {Liu}, {Lopes},
  {Lu}, {Lu}, {L{\"u}belsmeyer}, {Luckey}, {Lustermann}, {Ma{\~n}a},
  {Margotti}, {Mayet}, {McNeil}, {Meillon}, {Menichelli}, {Mihul}, {Mourao},
  {Mujunen}, {Palmonari}, {Papi}, {Park}, {Pauluzzi}, {Pauss}, {Perrin},
  {Pesci}, {Pevsner}, {Pimenta}, {Plyaskin}, {Pojidaev}, {Postolache},
  {Produit}, {Rancoita}, {Rapin}, {Raupach}, {Ren}, {Ren}, {Ribordy},
  {Richeux}, {Riihonen}, {Ritakari}, {Roeser}, {Roissin}, {Sagdeev},
  {Sartorelli}, {Schultz von Dratzig}, {Schwering}, {Scolieri}, {Seo},
  {Shoutko}, {Shoumilov}, {Siedling}, {Son}, {Song}, {Steuer}, {Sun}, {Suter},
  {Tang}, {Ting}, {Ting}, {Tornikoski}, {Torsti}, {Tr{\"u}mper}, {Ulbricht},
  {Urpo}, {Usoskin}, {Valtonen}, {Vandenhirtz}, {Velcea}, {Velikhov},
  {Verlaat}, {Vetlitsky}, {Vezzu}, {Vialle}, {Viertel}, {Vit{\'e}}, {Von
  Gunten}, {Waldmeier Wicki}, {Wallraff}, {Wang}, {Wang}, {Wang}, {Wiik},
  {Williams}, {Wu}, {Xia}, {Yan}, {Yan}, {Yang}, {Yang}, {Ye}, {Yeh}, {Xu},
  {Zhang}, {Zhang}, {Zhao}, {Zhu}, {Zhu}, {Zhuang}, {Zichichi}, \&
  {Zimmermann}}]{2000PhLB..484...10A}
{Alcaraz}, J., {Alpat}, B., {Ambrosi}, G., {et~al.} 2000, Physics Letters B,
  484, 10

\bibitem[{{Alvarez} {et~al.}(2001){Alvarez}, {Aparici}, {May}, \&
  {Reich}}]{2001A&A...372..636A}
{Alvarez}, H., {Aparici}, J., {May}, J., \& {Reich}, P. 2001, \aap, 372, 636

\bibitem[{{AMS-01 Collaboration} {et~al.}(2007){AMS-01 Collaboration},
  {Aguilar}, {Alcaraz}, {Allaby}, {Alpat}, {Ambrosi}, {Anderhub}, {Ao},
  {Arefiev}, {Azzarello}, {Baldini}, {Basile}, {Barancourt}, {Barao},
  {Barbier}, {Barreira}, {Battiston}, {Becker}, {Becker}, {Bellagamba},
  {B{\'e}n{\'e}}, {Berdugo}, {Berges}, {Bertucci}, {Biland}, {Blasko},
  {Boella}, {Boschini}, {Bourquin}, {Brocco}, {Bruni}, {Bu{\'e}nerd}, {Burger},
  {Burger}, {Cai}, {Camps}, {Cannarsa}, {Capell}, {Cardano}, {Casadei},
  {Casaus}, {Castellini}, {Chang}, {Chen}, {Chen}, {Chen}, {Chernoplekov},
  {Chiueh}, {Cho}, {Choi}, {Choi}, {Cindolo}, {Commichau}, {Contin},
  {Cortina-Gil}, {Cristinziani}, {Dai}, {Delgado}, {Difalco}, {Djambazov},
  {D'Antone}, {Dong}, {Emonet}, {Engelberg}, {Eppling}, {Eronen}, {Esposito},
  {Extermann}, {Favier}, {Fiandrini}, {Fisher}, {Fl{\"u}gge}, {Fouque},
  {Galaktionov}, {Gast}, {Gervasi}, {Giusti}, {Grandi}, {Grimm}, {Gu},
  {Hangarter}, {Hasan}, {Hermel}, {Hofer}, {Hungerford}, {Jongmanns},
  {Karlamaa}, {Karpinski}, {Kenney}, {Kim}, {Kim}, {Kim}, {Kim}, {Klimentov},
  {Kossakowski}, {Kounine}, {Koutsenko}, {Kraeber}, {Laborie}, {Laitinen},
  {Lamanna}, {Lanciotti}, {Laurenti}, {Lebedev}, {Lechanoine-Leluc}, {Lee},
  {Lee}, {Levi}, {Liu}, {Liu}, {Lu}, {Lu}, {L{\"u}belsmeyer}, {Luckey},
  {Lustermann}, {Ma{\~n}a}, {Margotti}, {Mayet}, {McNeil}, {Meillon},
  {Menichelli}, {Mihul}, {Mujunen}, {Oliva}, {Olzem}, {Palmonari}, {Park},
  {Park}, {Pauluzzi}, {Pauss}, {Perrin}, {Pesci}, {Pevsner}, {Pilo}, {Pimenta},
  {Plyaskin}, {Pojidaev}, {Pohl}, {Produit}, {Rancoita}, {Rapin}, {Raupach},
  {Ren}, {Ren}, {Ribordy}, {Richeux}, {Riihonen}, {Ritakari}, {Ro}, {Roeser},
  {Rossin}, {Sagdeev}, {Santos}, {Sartorelli}, {Sbarra}, {Schael}, {Schultz von
  Dratzig}, {Schwering}, {Seo}, {Shin}, {Shoumilov}, {Shoutko}, {Siedenburg},
  {Siedling}, {Son}, {Song}, {Spinella}, {Steuer}, {Sun}, {Suter}, {Tang},
  {Ting}, {Ting}, {Tornikoski}, {Torsti}, {Tr{\"u}mper}, {Ulbricht}, {Urpo},
  {Valtonen}, {Vandenhirtz}, {Velikhov}, {Verlaat}, {Vetlitsky}, {Vezzu},
  {Vialle}, {Viertel}, {Vit{\'e}}, {von Gunten}, {Waldmeier Wicki}, {Wallraff},
  {Wang}, {Wang}, {Wiik}, {Williams}, {Wu}, {Xia}, {Xu}, {Yan}, {Yan}, {Yang},
  {Yang}, {Yang}, {Ye}, {Xu}, {Zhang}, {Zhang}, {Zhao}, {Zhou}, {Zhu}, {Zhu},
  {Zhuang}, {Zichichi}, {Zimmermann}, \& {Zuccon}}]{2007PhLB..646..145A}
{AMS-01 Collaboration}, {Aguilar}, M., {Alcaraz}, J., {et~al.} 2007, Physics
  Letters B, 646, 145

\bibitem[{{Anderson} {et~al.}(1996){Anderson}, {Cadwell}, {Jacoby},
  {Wolszczan}, {Foster}, \& {Kramer}}]{1996ApJ...468L..55A}
{Anderson}, S.~B., {Cadwell}, B.~J., {Jacoby}, B.~A., {et~al.} 1996, \apjl,
  468, L55+

\bibitem[{{Asano} {et~al.}(2007){Asano}, {Matsumoto}, {Okada}, \&
  {Okada}}]{2007PhRvD..75f3506A}
{Asano}, M., {Matsumoto}, S., {Okada}, N., \& {Okada}, Y. 2007, \prd, 75,
  063506

\bibitem[{{Atoyan} \& {Aharonian}(1996)}]{1996MNRAS.278..525A}
{Atoyan}, A.~M. \& {Aharonian}, F.~A. 1996, \mnras, 278, 525

\bibitem[{{Atoyan} {et~al.}(1995){Atoyan}, {Aharonian}, \&
  {V{\"o}lk}}]{1995PhRvD..52.3265A}
{Atoyan}, A.~M., {Aharonian}, F.~A., \& {V{\"o}lk}, H.~J. 1995, \prd, 52, 3265

\bibitem[{{Baltz} \& {Edsj{\"o}}(1998)}]{1998PhRvD..59b3511B}
{Baltz}, E.~A. \& {Edsj{\"o}}, J. 1998, \prd, 59, 023511

\bibitem[{{Baltz} \& {Wai}(2004)}]{2004PhRvD..70b3512B}
{Baltz}, E.~A. \& {Wai}, L. 2004, \prd, 70, 023512

\bibitem[{{Barwick} {et~al.}(1997){Barwick}, {Beatty}, {Bhattacharyya},
  {Bower}, {Chaput}, {Coutu}, {de Nolfo}, {Knapp}, {Lowder}, {McKee},
  {Mueller}, {Musser}, {Nutter}, {Schneider}, {Swordy}, {Tarle}, {Tomasch},
  {Torbet}, \& {The HEAT Collaboration}}]{1997ApJ...482L.191B}
{Barwick}, S.~W., {Beatty}, J.~J., {Bhattacharyya}, A., {et~al.} 1997, \apjl,
  482, L191+

\bibitem[{{Battiston}(2007)}]{battiston_07}
{Battiston}, R. 2007, Nuclear Physics B Proceedings Supplements, 166, 19

\bibitem[{{Beatty} {et~al.}(2004){Beatty}, {Bhattacharyya}, {Bower}, {Coutu},
  {Duvernois}, {McKee}, {Minnick}, {M{\"u}ller}, {Musser}, {Nutter},
  {Labrador}, {Schubnell}, {Swordy}, {Tarl{\'e}}, \&
  {Tomasch}}]{2004PhRvL..93x1102B}
{Beatty}, J.~J., {Bhattacharyya}, A., {Bower}, C., {et~al.} 2004, Physical
  Review Letters, 93, 241102

\bibitem[{{Berezhko} {et~al.}(2003){Berezhko}, {Ksenofontov}, {Ptuskin},
  {Zirakashvili}, \& {V{\"o}lk}}]{2003A&A...410..189B}
{Berezhko}, E.~G., {Ksenofontov}, L.~T., {Ptuskin}, V.~S., {Zirakashvili},
  V.~N., \& {V{\"o}lk}, H.~J. 2003, \aap, 410, 189

\bibitem[{{Berezinskii} {et~al.}(1990){Berezinskii}, {Bulanov}, {Dogiel}, \&
  {Ptuskin}}]{berezinsky_book_90}
{Berezinskii}, V.~S., {Bulanov}, S.~V., {Dogiel}, V.~A., \& {Ptuskin}, V.~S.
  1990, {Astrophysics of cosmic rays} (Amsterdam: North-Holland, 1990, edited
  by Ginzburg, V.L.)

\bibitem[{{Bergstr{\"o}m} {et~al.}(2008){Bergstr{\"o}m}, {Bringmann}, \&
  {Edsj{\"o}}}]{2008PhRvD..78j3520B}
{Bergstr{\"o}m}, L., {Bringmann}, T., \& {Edsj{\"o}}, J. 2008, \prd, 78, 103520

\bibitem[{{Blair} {et~al.}(2005){Blair}, {Sankrit}, \&
  {Raymond}}]{2005AJ....129.2268B}
{Blair}, W.~P., {Sankrit}, R., \& {Raymond}, J.~C. 2005, \aj, 129, 2268

\bibitem[{{Blair} {et~al.}(2009){Blair}, {Sankrit}, {Torres}, {Chayer}, \&
  {Danforth}}]{2009ApJ...692..335B}
{Blair}, W.~P., {Sankrit}, R., {Torres}, S.~I., {Chayer}, P., \& {Danforth},
  C.~W. 2009, \apj, 692, 335

\bibitem[{{Blandford} \& {Eichler}(1987)}]{1987PhR...154....1B}
{Blandford}, R. \& {Eichler}, D. 1987, \physrep, 154, 1

\bibitem[{{Blasi}(2009)}]{2009PhRvL.103e1104B}
{Blasi}, P. 2009, Physical Review Letters, 103, 051104

\bibitem[{{Blasi} \& {Serpico}(2009)}]{2009PhRvL.103h1103B}
{Blasi}, P. \& {Serpico}, P.~D. 2009, Physical Review Letters, 103, 081103

\bibitem[{{Blondin} {et~al.}(2001){Blondin}, {Chevalier}, \&
  {Frierson}}]{2001ApJ...563..806B}
{Blondin}, J.~M., {Chevalier}, R.~A., \& {Frierson}, D.~M. 2001, \apj, 563, 806

\bibitem[{{Blumenthal} \& {Gould}(1970)}]{1970RvMP...42..237B}
{Blumenthal}, G.~R. \& {Gould}, R.~J. 1970, Reviews of Modern Physics, 42, 237

\bibitem[{{Boezio} {et~al.}(2000){Boezio}, {Carlson}, {Francke}, {Weber},
  {Suffert}, {Hof}, {Menn}, {Simon}, {Stephens}, {Bellotti}, {Cafagna},
  {Castellano}, {Circella}, {De Marzo}, {Finetti}, {Papini}, {Piccardi},
  {Spillantini}, {Ricci}, {Casolino}, {De Pascale}, {Morselli}, {Picozza},
  {Sparvoli}, {Barbiellini}, {Bravar}, {Schiavon}, {Vacchi}, {Zampa},
  {Grimani}, {Mitchell}, {Ormes}, {Streitmatter}, {Golden}, \&
  {Stochaj}}]{2000ApJ...532..653B}
{Boezio}, M., {Carlson}, P., {Francke}, T., {et~al.} 2000, \apj, 532, 653

\bibitem[{{Bonatto} \& {Bica}(2009)}]{2009MNRAS.tmp..278B}
{Bonatto}, C. \& {Bica}, E. 2009, \mnras, 278

\bibitem[{{Borka Jovanovi{\'c}} \& {Uro{\v
  s}evi{\'c}}(2009)}]{2009arXiv0904.2261B}
{Borka Jovanovi{\'c}}, V. \& {Uro{\v s}evi{\'c}}, D. 2009, ArXiv e-prints
  \href{http://arxiv.org/abs/0904.2261}{0904.2261}

\bibitem[{{Boulares}(1989)}]{1989ApJ...342..807B}
{Boulares}, A. 1989, \apj, 342, 807

\bibitem[{{Brisken} {et~al.}(2003){Brisken}, {Thorsett}, {Golden}, \&
  {Goss}}]{2003ApJ...593L..89B}
{Brisken}, W.~F., {Thorsett}, S.~E., {Golden}, A., \& {Goss}, W.~M. 2003,
  \apjl, 593, L89

\bibitem[{{Brogan} {et~al.}(2006){Brogan}, {Gelfand}, {Gaensler}, {Kassim}, \&
  {Lazio}}]{2006ApJ...639L..25B}
{Brogan}, C.~L., {Gelfand}, J.~D., {Gaensler}, B.~M., {Kassim}, N.~E., \&
  {Lazio}, T.~J.~W. 2006, \apjl, 639, L25

\bibitem[{{Bucciantini}(2008)}]{2008AdSpR..41..491B}
{Bucciantini}, N. 2008, Advances in Space Research, 41, 491

\bibitem[{{Bulanov} \& {Dogel}(1974)}]{1974Ap&SS..29..305B}
{Bulanov}, S.~V. \& {Dogel}, V.~A. 1974, \apss, 29, 305

\bibitem[{{Burrows}(2000)}]{2000Natur.403..727B}
{Burrows}, A. 2000, \nat, 403, 727

\bibitem[{{Bykov} {et~al.}(2004){Bykov}, {Krassilchtchikov}, {Uvarov},
  {Bloemen}, {Chevalier}, {Gustov}, {Hermsen}, {Lebrun}, {Lozinskaya}, {Rauw},
  {Smirnova}, {Sturner}, {Swings}, {Terrier}, \&
  {Toptygin}}]{2004A&A...427L..21B}
{Bykov}, A.~M., {Krassilchtchikov}, A.~M., {Uvarov}, Y.~A., {et~al.} 2004,
  \aap, 427, L21

\bibitem[{{Byun} {et~al.}(2006){Byun}, {Koo}, {Tatematsu}, \&
  {Sunada}}]{2006ApJ...637..283B}
{Byun}, D.-Y., {Koo}, B.-C., {Tatematsu}, K., \& {Sunada}, K. 2006, \apj, 637,
  283

\bibitem[{{Caprioli} {et~al.}(2009){Caprioli}, {Blasi}, \&
  {Amato}}]{2009MNRAS.396.2065C}
{Caprioli}, D., {Blasi}, P., \& {Amato}, E. 2009, \mnras, 396, 2065

\bibitem[{{Caraveo} {et~al.}(2001){Caraveo}, {De Luca}, {Mignani}, \&
  {Bignami}}]{2001ApJ...561..930C}
{Caraveo}, P.~A., {De Luca}, A., {Mignani}, R.~P., \& {Bignami}, G.~F. 2001,
  \apj, 561, 930

\bibitem[{{Case} \& {Bhattacharya}(1998)}]{1998ApJ...504..761C}
{Case}, G.~L. \& {Bhattacharya}, D. 1998, \apj, 504, 761

\bibitem[{{Castelletti} \& {Dubner}(2005)}]{2005A&A...440..171C}
{Castelletti}, G. \& {Dubner}, G. 2005, \aap, 440, 171

\bibitem[{{Castelletti} {et~al.}(2003){Castelletti}, {Dubner}, {Golap}, {Goss},
  {Vel{\'a}zquez}, {Holdaway}, \& {Rao}}]{2003AJ....126.2114C}
{Castelletti}, G., {Dubner}, G., {Golap}, K., {et~al.} 2003, \aj, 126, 2114

\bibitem[{{Catena} {et~al.}(2009){Catena}, {Fornengo}, {Pato}, {Pieri}, \&
  {Masiero}}]{2009arXiv0912.4421C}
{Catena}, R., {Fornengo}, N., {Pato}, M., {Pieri}, L., \& {Masiero}, A. 2009,
  ArXiv e-prints

\bibitem[{{Cha} {et~al.}(1999){Cha}, {Sembach}, \&
  {Danks}}]{1999ApJ...515L..25C}
{Cha}, A.~N., {Sembach}, K.~R., \& {Danks}, A.~C. 1999, \apjl, 515, L25

\bibitem[{{Chang} {et~al.}(2008){Chang}, {Adams}, {Ahn}, {Bashindzhagyan},
  {Christl}, {Ganel}, {Guzik}, {Isbert}, {Kim}, {Kuznetsov}, {Panasyuk},
  {Panov}, {Schmidt}, {Seo}, {Sokolskaya}, {Watts}, {Wefel}, {Wu}, \&
  {Zatsepin}}]{2008Natur.456..362C}
{Chang}, J., {Adams}, J.~H., {Ahn}, H.~S., {et~al.} 2008, \nat, 456, 362

\bibitem[{{Cheng} {et~al.}(1976){Cheng}, {Ruderman}, \&
  {Sutherland}}]{1976ApJ...203..209C}
{Cheng}, A., {Ruderman}, M., \& {Sutherland}, P. 1976, \apj, 203, 209

\bibitem[{{Cheng} {et~al.}(1986){Cheng}, {Ho}, \&
  {Ruderman}}]{1986ApJ...300..500C}
{Cheng}, K.~S., {Ho}, C., \& {Ruderman}, M. 1986, \apj, 300, 500

\bibitem[{{Chi} {et~al.}(1996){Chi}, {Cheng}, \& {Young}}]{1996ApJ...459L..83C}
{Chi}, X., {Cheng}, K.~S., \& {Young}, E.~C.~M. 1996, \apjl, 459, L83+

\bibitem[{{Cirelli} {et~al.}(2008){Cirelli}, {Franceschini}, \&
  {Strumia}}]{2008NuPhB.800..204C}
{Cirelli}, M., {Franceschini}, R., \& {Strumia}, A. 2008, Nuclear Physics B,
  800, 204

\bibitem[{{Cowsik} \& {Lee}(1979)}]{1979ApJ...228..297C}
{Cowsik}, R. \& {Lee}, M.~A. 1979, \apj, 228, 297

\bibitem[{{Delahaye} {et~al.}(2009){Delahaye}, {Donato}, {Fornengo}, {Lavalle},
  {Lineros}, {Salati}, \& {Taillet}}]{2009A&A...501..821D}
{Delahaye}, T., {Donato}, F., {Fornengo}, N., {et~al.} 2009, \aap, 501, 821

\bibitem[{{Delahaye} {et~al.}(2008){Delahaye}, {Lineros}, {Donato}, {Fornengo},
  \& {Salati}}]{2008PhRvD..77f3527D}
{Delahaye}, T., {Lineros}, R., {Donato}, F., {Fornengo}, N., \& {Salati}, P.
  2008, \prd, 77, 063527

\bibitem[{{Diehl} {et~al.}(2006){Diehl}, {Halloin}, {Kretschmer}, {Lichti},
  {Sch{\"o}nfelder}, {Strong}, {von Kienlin}, {Wang}, {Jean}, {Kn{\"o}dlseder},
  {Roques}, {Weidenspointner}, {Schanne}, {Hartmann}, {Winkler}, \&
  {Wunderer}}]{2006Natur.439...45D}
{Diehl}, R., {Halloin}, H., {Kretschmer}, K., {et~al.} 2006, \nat, 439, 45

\bibitem[{{Donato} {et~al.}(2004){Donato}, {Fornengo}, {Maurin}, {Salati}, \&
  {Taillet}}]{2004PhRvD..69f3501D}
{Donato}, F., {Fornengo}, N., {Maurin}, D., {Salati}, P., \& {Taillet}, R.
  2004, \prd, 69, 063501

\bibitem[{{Donato} {et~al.}(2009){Donato}, {Maurin}, {Brun}, {Delahaye}, \&
  {Salati}}]{2009PhRvL.102g1301D}
{Donato}, F., {Maurin}, D., {Brun}, P., {Delahaye}, T., \& {Salati}, P. 2009,
  Physical Review Letters, 102, 071301

\bibitem[{{Donato} {et~al.}(2002){Donato}, {Maurin}, \&
  {Taillet}}]{2002A&A...381..539D}
{Donato}, F., {Maurin}, D., \& {Taillet}, R. 2002, Astronomy and Astrophys.,
  381, 539

\bibitem[{{Duncan} {et~al.}(1997){Duncan}, {Stewart}, {Haynes}, \&
  {Jones}}]{1997MNRAS.287..722D}
{Duncan}, A.~R., {Stewart}, R.~T., {Haynes}, R.~F., \& {Jones}, K.~L. 1997,
  \mnras, 287, 722

\bibitem[{{DuVernois} {et~al.}(2001){DuVernois}, {Barwick}, {Beatty},
  {Bhattacharyya}, {Bower}, {Chaput}, {Coutu}, {de Nolfo}, {Lowder}, {McKee},
  {M{\"u}ller}, {Musser}, {Nutter}, {Schneider}, {Swordy}, {Tarl{\'e}},
  {Tomasch}, \& {Torbet}}]{2001ApJ...559..296D}
{DuVernois}, M.~A., {Barwick}, S.~W., {Beatty}, J.~J., {et~al.} 2001, \apj,
  559, 296

\bibitem[{{Ellison} \& {Cassam-Chena{\"i}}(2005)}]{2005ApJ...632..920E}
{Ellison}, D.~C. \& {Cassam-Chena{\"i}}, G. 2005, \apj, 632, 920

\bibitem[{{Ellison} {et~al.}(2007){Ellison}, {Patnaude}, {Slane}, {Blasi}, \&
  {Gabici}}]{2007ApJ...661..879E}
{Ellison}, D.~C., {Patnaude}, D.~J., {Slane}, P., {Blasi}, P., \& {Gabici}, S.
  2007, \apj, 661, 879

\bibitem[{{Erber}(1966)}]{1966RvMP...38..626E}
{Erber}, T. 1966, Reviews of Modern Physics, 38, 626

\bibitem[{{Ferri{\`e}re}(2001)}]{2001RvMP...73.1031F}
{Ferri{\`e}re}, K.~M. 2001, Reviews of Modern Physics, 73, 1031

\bibitem[{{Fiasson} {et~al.}(2008){Fiasson}, {Hinton}, {Gallant}, \& {et
  al.}}]{2008ICRC....2..719F}
{Fiasson}, A., {Hinton}, J.~A., {Gallant}, Y., \& {et al.} 2008, in
  International Cosmic Ray Conference, Vol.~2, 719--722

\bibitem[{{Fisk}(1971)}]{1971JGR....76..221F}
{Fisk}, L.~A. 1971, \jgr, 76, 221

\bibitem[{{Fixsen}(2009)}]{2009arXiv0911.1955F}
{Fixsen}, D.~J. 2009, ArXiv e-prints

\bibitem[{{Fuerst} {et~al.}(1997){Fuerst}, {Reich}, \&
  {Aschenbach}}]{1997A&A...319..655F}
{Fuerst}, E., {Reich}, W., \& {Aschenbach}, B. 1997, \aap, 319, 655

\bibitem[{{Furst} {et~al.}(1989){Furst}, {Hummel}, {Reich}, {Sofue}, {Sieber},
  {Reif}, \& {Dettmar}}]{1989A&A...209..361F}
{Furst}, E., {Hummel}, E., {Reich}, W., {et~al.} 1989, \aap, 209, 361

\bibitem[{{Gaensler} \& {Frail}(2000)}]{2000Natur.406..158G}
{Gaensler}, B.~M. \& {Frail}, D.~A. 2000, \nat, 406, 158

\bibitem[{{Gaensler} \& {Slane}(2006)}]{2006ARA&A..44...17G}
{Gaensler}, B.~M. \& {Slane}, P.~O. 2006, \araa, 44, 17

\bibitem[{{Gamezo} {et~al.}(2003){Gamezo}, {Khokhlov}, {Oran}, {Chtchelkanova},
  \& {Rosenberg}}]{2003Sci...299...77G}
{Gamezo}, V.~N., {Khokhlov}, A.~M., {Oran}, E.~S., {Chtchelkanova}, A.~Y., \&
  {Rosenberg}, R.~O. 2003, Science, 299, 77

\bibitem[{{Gerardy} \& {Fesen}(2007)}]{2007MNRAS.376..929G}
{Gerardy}, C.~L. \& {Fesen}, R.~A. 2007, \mnras, 376, 929

\bibitem[{{Ginzburg} \& {Syrovatskii}(1964)}]{1964ocr..book.....G}
{Ginzburg}, V.~L. \& {Syrovatskii}, S.~I. 1964, {The Origin of Cosmic Rays}
  (The Origin of Cosmic Rays, New York: Macmillan, 1964)

\bibitem[{{Ginzburg} \& {Syrovatskii}(1965)}]{1965ARA&A...3..297G}
{Ginzburg}, V.~L. \& {Syrovatskii}, S.~I. 1965, \araa, 3, 297

\bibitem[{{Gorham} {et~al.}(1996){Gorham}, {Ray}, {Anderson}, {Kulkarni}, \&
  {Prince}}]{1996ApJ...458..257G}
{Gorham}, P.~W., {Ray}, P.~S., {Anderson}, S.~B., {Kulkarni}, S.~R., \&
  {Prince}, T.~A. 1996, \apj, 458, 257

\bibitem[{{Graham} {et~al.}(1982){Graham}, {Haslam}, {Salter}, \&
  {Wilson}}]{1982A&A...109..145G}
{Graham}, D.~A., {Haslam}, C.~G.~T., {Salter}, C.~J., \& {Wilson}, W.~E. 1982,
  \aap, 109, 145

\bibitem[{{Green}(2005)}]{2005MmSAI..76..534G}
{Green}, D.~A. 2005, Memorie della Societa Astronomica Italiana, 76, 534

\bibitem[{{Green}(2009)}]{2009BASI...37...45G}
{Green}, D.~A. 2009, Bull. Astron. Soc. Ind., 37, 45

\bibitem[{{Harding} {et~al.}(2008){Harding}, {Stern}, {Dyks}, \&
  {Frackowiak}}]{2008ApJ...680.1378H}
{Harding}, A.~K., {Stern}, J.~V., {Dyks}, J., \& {Frackowiak}, M. 2008, \apj,
  680, 1378

\bibitem[{{Harrus} {et~al.}(2004){Harrus}, {Slane}, {Hughes}, \&
  {Plucinsky}}]{2004ApJ...603..152H}
{Harrus}, I.~M., {Slane}, P.~O., {Hughes}, J.~P., \& {Plucinsky}, P.~P. 2004,
  \apj, 603, 152

\bibitem[{{Hooper} {et~al.}(2009){Hooper}, {Blasi}, \& {Dario
  Serpico}}]{2009JCAP...01..025H}
{Hooper}, D., {Blasi}, P., \& {Dario Serpico}, P. 2009, Journal of Cosmology
  and Astro-Particle Physics, 1, 25

\bibitem[{{Hooper} \& {Kribs}(2004)}]{2004PhRvD..70k5004H}
{Hooper}, D. \& {Kribs}, G.~D. 2004, \prd, 70, 115004

\bibitem[{{Humphreys}(1978)}]{1978ApJS...38..309H}
{Humphreys}, R.~M. 1978, \apjs, 38, 309

\bibitem[{{Jaffe} {et~al.}(2010){Jaffe}, {Leahy}, {Banday}, {Leach}, {Lowe}, \&
  {Wilkinson}}]{2010MNRAS.401.1013J}
{Jaffe}, T.~R., {Leahy}, J.~P., {Banday}, A.~J., {et~al.} 2010, \mnras, 401,
  1013

\bibitem[{{Janka} {et~al.}(2007){Janka}, {Langanke}, {Marek},
  {Mart{\'{\i}}nez-Pinedo}, \& {M{\"u}ller}}]{2007PhR...442...38J}
{Janka}, H., {Langanke}, K., {Marek}, A., {Mart{\'{\i}}nez-Pinedo}, G., \&
  {M{\"u}ller}, B. 2007, \physrep, 442, 38

\bibitem[{{Jansson} {et~al.}(2009){Jansson}, {Farrar}, {Waelkens}, \&
  {En{\ss}lin}}]{2009JCAP...07..021J}
{Jansson}, R., {Farrar}, G.~R., {Waelkens}, A.~H., \& {En{\ss}lin}, T.~A. 2009,
  Journal of Cosmology and Astro-Particle Physics, 7, 21

\bibitem[{{Jones}(1965)}]{1965PhRv..137.1306J}
{Jones}, F.~C. 1965, Physical Review, 137, 1306

\bibitem[{{Kamae} {et~al.}(2006){Kamae}, {Karlsson}, {Mizuno}, {Abe}, \&
  {Koi}}]{2006ApJ...647..692K}
{Kamae}, T., {Karlsson}, N., {Mizuno}, T., {Abe}, T., \& {Koi}, T. 2006, \apj,
  647, 692

\bibitem[{{Kaplan} {et~al.}(2008){Kaplan}, {Chatterjee}, {Gaensler}, \&
  {Anderson}}]{2008ApJ...677.1201K}
{Kaplan}, D.~L., {Chatterjee}, S., {Gaensler}, B.~M., \& {Anderson}, J. 2008,
  \apj, 677, 1201

\bibitem[{{Kaplan} {et~al.}(2004){Kaplan}, {Frail}, {Gaensler}, {Gotthelf},
  {Kulkarni}, {Slane}, \& {Nechita}}]{2004ApJS..153..269K}
{Kaplan}, D.~L., {Frail}, D.~A., {Gaensler}, B.~M., {et~al.} 2004, \apjs, 153,
  269

\bibitem[{{Kaplan} {et~al.}(2006){Kaplan}, {Gaensler}, {Kulkarni}, \&
  {Slane}}]{2006ApJS..163..344K}
{Kaplan}, D.~L., {Gaensler}, B.~M., {Kulkarni}, S.~R., \& {Slane}, P.~O. 2006,
  \apjs, 163, 344

\bibitem[{{Katsuda} {et~al.}(2009){Katsuda}, {Petre}, {Hwang}, {Yamaguchi},
  {Mori}, \& {Tsunemi}}]{2009PASJ...61..155K}
{Katsuda}, S., {Petre}, R., {Hwang}, U., {et~al.} 2009, \pasj, 61, 155

\bibitem[{{Katsuda} {et~al.}(2008){Katsuda}, {Tsunemi}, \&
  {Mori}}]{2008ApJ...678L..35K}
{Katsuda}, S., {Tsunemi}, H., \& {Mori}, K. 2008, \apjl, 678, L35

\bibitem[{{Kobayashi} {et~al.}(2004){Kobayashi}, {Komori}, {Yoshida}, \&
  {Nishimura}}]{2004ApJ...601..340K}
{Kobayashi}, T., {Komori}, Y., {Yoshida}, K., \& {Nishimura}, J. 2004, \apj,
  601, 340

\bibitem[{{Kothes} {et~al.}(2006){Kothes}, {Fedotov}, {Foster}, \&
  {Uyan{\i}ker}}]{2006A&A...457.1081K}
{Kothes}, R., {Fedotov}, K., {Foster}, T.~J., \& {Uyan{\i}ker}, B. 2006, \aap,
  457, 1081

\bibitem[{{Kothes} {et~al.}(2008){Kothes}, {Landecker}, {Reich}, {Safi-Harb},
  \& {Arzoumanian}}]{2008ApJ...687..516K}
{Kothes}, R., {Landecker}, T.~L., {Reich}, W., {Safi-Harb}, S., \&
  {Arzoumanian}, Z. 2008, \apj, 687, 516

\bibitem[{{Ladouceur} \& {Pineault}(2008)}]{2008A&A...490..197L}
{Ladouceur}, Y. \& {Pineault}, S. 2008, \aap, 490, 197

\bibitem[{{Lavalle} {et~al.}(2009){Lavalle}, {Marcowith}, \&
  {Maurin}}]{2009sf2a.conf..165L}
{Lavalle}, J., {Marcowith}, A., \& {Maurin}, D. 2009, in Proc. SF2A-2009, ed.
  {M.~Heydari-Malayeri, C.~Reyl'E, \& R.~Samadi}, 165--+

\bibitem[{{Lavalle} {et~al.}(2008{\natexlab{a}}){Lavalle}, {Nezri},
  {Athanassoula}, {Ling}, \& {Teyssier}}]{2008PhRvD..78j3526L}
{Lavalle}, J., {Nezri}, E., {Athanassoula}, E., {Ling}, F.-S., \& {Teyssier},
  R. 2008{\natexlab{a}}, \prd, 78, 103526

\bibitem[{{Lavalle} {et~al.}(2007){Lavalle}, {Pochon}, {Salati}, \&
  {Taillet}}]{2007A&A...462..827L}
{Lavalle}, J., {Pochon}, J., {Salati}, P., \& {Taillet}, R. 2007, \aap, 462,
  827

\bibitem[{{Lavalle} {et~al.}(2008{\natexlab{b}}){Lavalle}, {Yuan}, {Maurin}, \&
  {Bi}}]{2008A&A...479..427L}
{Lavalle}, J., {Yuan}, Q., {Maurin}, D., \& {Bi}, X.-J. 2008{\natexlab{b}},
  \aap, 479, 427

\bibitem[{{Lazendic} \& {Slane}(2006)}]{2006ApJ...647..350L}
{Lazendic}, J.~S. \& {Slane}, P.~O. 2006, \apj, 647, 350

\bibitem[{{Leahy} \& {Tian}(2006)}]{2006A&A...451..251L}
{Leahy}, D. \& {Tian}, W. 2006, \aap, 451, 251

\bibitem[{{Leahy} {et~al.}(1986){Leahy}, {Naranan}, \&
  {Singh}}]{1986MNRAS.220..501L}
{Leahy}, D.~A., {Naranan}, S., \& {Singh}, K.~P. 1986, \mnras, 220, 501

\bibitem[{{Leahy} {et~al.}(1991){Leahy}, {Nousek}, \&
  {Hamilton}}]{1991ApJ...374..218L}
{Leahy}, D.~A., {Nousek}, J., \& {Hamilton}, A.~J.~S. 1991, \apj, 374, 218

\bibitem[{{Leahy} \& {Tian}(2007)}]{2007A&A...461.1013L}
{Leahy}, D.~A. \& {Tian}, W.~W. 2007, \aap, 461, 1013

\bibitem[{{Longair}(1994)}]{1994hea..book.....L}
{Longair}, M.~S. 1994, {High energy astrophysics. Vol.2: Stars, the galaxy and
  the interstellar medium} (Cambridge: Cambridge University Press, |c1994, 2nd
  ed.)

\bibitem[{{Lorimer}(2004)}]{2004IAUS..218..105L}
{Lorimer}, D.~R. 2004, in IAU Symposium, Vol. 218, Young Neutron Stars and
  Their Environments, ed. {F.~Camilo \& B.~M.~Gaensler}, 105--+

\bibitem[{{Lucek} \& {Bell}(2000)}]{2000MNRAS.314...65L}
{Lucek}, S.~G. \& {Bell}, A.~R. 2000, \mnras, 314, 65

\bibitem[{{Madau} {et~al.}(1998){Madau}, {della Valle}, \&
  {Panagia}}]{1998MNRAS.297L..17M}
{Madau}, P., {della Valle}, M., \& {Panagia}, N. 1998, \mnras, 297, L17+

\bibitem[{{Malkov} \& {O'C Drury}(2001)}]{2001RPPh...64..429M}
{Malkov}, M.~A. \& {O'C Drury}, L. 2001, Reports on Progress in Physics, 64,
  429

\bibitem[{{Malyshev} {et~al.}(2009){Malyshev}, {Cholis}, \&
  {Gelfand}}]{2009PhRvD..80f3005M}
{Malyshev}, D., {Cholis}, I., \& {Gelfand}, J. 2009, \prd, 80, 063005

\bibitem[{{Manchester} {et~al.}(2005){Manchester}, {Hobbs}, {Teoh}, \&
  {Hobbs}}]{2005AJ....129.1993M}
{Manchester}, R.~N., {Hobbs}, G.~B., {Teoh}, A., \& {Hobbs}, M. 2005, \aj, 129,
  1993

\bibitem[{{Mantovani} {et~al.}(1982){Mantovani}, {Nanni}, {Salter}, \&
  {Tomasi}}]{1982A&A...105..176M}
{Mantovani}, F., {Nanni}, M., {Salter}, C.~J., \& {Tomasi}, P. 1982, \aap, 105,
  176

\bibitem[{{Maurin} {et~al.}(2001){Maurin}, {Donato}, {Taillet}, \&
  {Salati}}]{2001ApJ...555..585M}
{Maurin}, D., {Donato}, F., {Taillet}, R., \& {Salati}, P. 2001, \apj, 555, 585

\bibitem[{{Mavromatakis}(2003)}]{2003A&A...408..237M}
{Mavromatakis}, F. 2003, \aap, 408, 237

\bibitem[{{Mavromatakis} {et~al.}(2004){Mavromatakis}, {Aschenbach}, {Boumis},
  \& {Papamastorakis}}]{2004A&A...415.1051M}
{Mavromatakis}, F., {Aschenbach}, B., {Boumis}, P., \& {Papamastorakis}, J.
  2004, \aap, 415, 1051

\bibitem[{{Mavromatakis} {et~al.}(2002){Mavromatakis}, {Boumis},
  {Papamastorakis}, \& {Ventura}}]{2002A&A...388..355M}
{Mavromatakis}, F., {Boumis}, P., {Papamastorakis}, J., \& {Ventura}, J. 2002,
  \aap, 388, 355

\bibitem[{{Mazzali} {et~al.}(2007){Mazzali}, {R{\"o}pke}, {Benetti}, \&
  {Hillebrandt}}]{2007Sci...315..825M}
{Mazzali}, P.~A., {R{\"o}pke}, F.~K., {Benetti}, S., \& {Hillebrandt}, W. 2007,
  Science, 315, 825

\bibitem[{{McCullough} {et~al.}(2002){McCullough}, {Fields}, \&
  {Pavlidou}}]{2002ApJ...576L..41M}
{McCullough}, P.~R., {Fields}, B.~D., \& {Pavlidou}, V. 2002, \apjl, 576, L41

\bibitem[{{Mertsch} \& {Sarkar}(2009)}]{2009PhRvL.103h1104M}
{Mertsch}, P. \& {Sarkar}, S. 2009, Physical Review Letters, 103, 081104

\bibitem[{{Miceli} {et~al.}(2008){Miceli}, {Bocchino}, \&
  {Reale}}]{2008ApJ...676.1064M}
{Miceli}, M., {Bocchino}, F., \& {Reale}, F. 2008, \apj, 676, 1064

\bibitem[{{Morlino} {et~al.}(2009){Morlino}, {Amato}, \&
  {Blasi}}]{2009MNRAS.392..240M}
{Morlino}, G., {Amato}, E., \& {Blasi}, P. 2009, \mnras, 392, 240

\bibitem[{{Moskalenko} \& {Strong}(1998)}]{1998ApJ...493..694M}
{Moskalenko}, I.~V. \& {Strong}, A.~W. 1998, \apj, 493, 694

\bibitem[{{Mufson} {et~al.}(1986){Mufson}, {McCollough}, {Dickel}, {Petre},
  {White}, \& {Chevalier}}]{1986AJ.....92.1349M}
{Mufson}, S.~L., {McCollough}, M.~L., {Dickel}, J.~R., {et~al.} 1986, \aj, 92,
  1349

\bibitem[{{Neufeld} {et~al.}(2007){Neufeld}, {Hollenbach}, {Kaufman}, {Snell},
  {Melnick}, {Bergin}, \& {Sonnentrucker}}]{2007ApJ...664..890N}
{Neufeld}, D.~A., {Hollenbach}, D.~J., {Kaufman}, M.~J., {et~al.} 2007, \apj,
  664, 890

\bibitem[{{Nomoto} {et~al.}(1984){Nomoto}, {Thielemann}, \&
  {Yokoi}}]{1984ApJ...286..644N}
{Nomoto}, K., {Thielemann}, F., \& {Yokoi}, K. 1984, \apj, 286, 644

\bibitem[{{Norbury} \& {Townsend}(2007)}]{2007NIMPB.254..187N}
{Norbury}, J.~W. \& {Townsend}, L.~W. 2007, Nuclear Instruments and Methods in
  Physics Research B, 254, 187

\bibitem[{{Odegard}(1986)}]{1986ApJ...301..813O}
{Odegard}, N. 1986, \apj, 301, 813

\bibitem[{{Ostriker} \& {Gunn}(1969)}]{1969ApJ...157.1395O}
{Ostriker}, J.~P. \& {Gunn}, J.~E. 1969, \apj, 157, 1395

\bibitem[{{Paczynski}(1990)}]{1990ApJ...348..485P}
{Paczynski}, B. 1990, \apj, 348, 485

\bibitem[{{PAMELA Collaboration} {et~al.}(2010){PAMELA Collaboration},
  {Adriani}, {Barbarino}, {Bazilevskaya}, {Bellotti}, {Boezio}, {Bogomolov},
  {Bonechi}, {Bongi}, {Bonvicini}, {Borisov}, {Bottai}, {Bruno}, {Cafagna},
  {Campana}, {Carbone}, {Carlson}, {Casolino}, {Castellini}, {Consiglio}, {de
  Pascale}, {de Santis}, {de Simone}, {di Felice}, {Galper}, {Gillard},
  {Grishantseva}, {Hofverberg}, {Jerse}, {Koldashov}, {Krutkov}, {Kvashnin},
  {Leonov}, {Malvezzi}, {Marcelli}, {Menn}, {Mikhailov}, {Mocchiutti},
  {Monaco}, {Mori}, {Nikonov}, {Osteria}, {Papini}, {Pearce}, {Picozza},
  {Ricci}, {Ricciarini}, {Rossetto}, {Simon}, {Sparvoli}, {Spillantini},
  {Stozhkov}, {Vacchi}, {Vannuccini}, {Vasilyev}, {Voronov}, {Wu}, {Yurkin},
  {Zampa}, {Zampa}, {Zverev}, \& {Marinucci}}]{2010APh....34....1A}
{PAMELA Collaboration}, {Adriani}, O., {Barbarino}, G.~C., {et~al.} 2010,
  Astroparticle Physics, 34, 1

\bibitem[{{PAMELA Collaboration} {et~al.}(2009){PAMELA Collaboration},
  {Adriani}, {Barbarino}, {Bazilevskaya}, {Bellotti}, {Boezio}, {Bogomolov},
  {Bonechi}, {Bongi}, {Bonvicini}, {Bottai}, {Bruno}, {Cafagna}, {Campana},
  {Carlson}, {Casolino}, {Castellini}, {de Pascale}, {de Rosa}, {de Simone},
  {di Felice}, {Galper}, {Grishantseva}, {Hofverberg}, {Koldashov}, {Krutkov},
  {Kvashnin}, {Leonov}, {Malvezzi}, {Marcelli}, {Menn}, {Mikhailov},
  {Mocchiutti}, {Orsi}, {Osteria}, {Papini}, {Pearce}, {Picozza}, {Ricci},
  {Ricciarini}, {Simon}, {Sparvoli}, {Spillantini}, {Stozhkov}, {Vacchi},
  {Vannuccini}, {Vasilyev}, {Voronov}, {Yurkin}, {Zampa}, {Zampa}, \&
  {Zverev}}]{2009Natur.458..607A}
{PAMELA Collaboration}, {Adriani}, O., {Barbarino}, G.~C., {et~al.} 2009, \nat,
  458, 607

\bibitem[{{Pannuti} \& {Allen}(2004)}]{2004AdSpR..33..434P}
{Pannuti}, T.~G. \& {Allen}, G.~E. 2004, Advances in Space Research, 33, 434

\bibitem[{{Pesce-Rollins} \& {for the Fermi-LAT
  Collaboration}(2009)}]{2009arXiv0912.3611P}
{Pesce-Rollins}, M. \& {for the Fermi-LAT Collaboration}. 2009, ArXiv e-prints

\bibitem[{{Pfeffermann} {et~al.}(1991){Pfeffermann}, {Aschenbach}, \&
  {Predehl}}]{1991A&A...246L..28P}
{Pfeffermann}, E., {Aschenbach}, B., \& {Predehl}, P. 1991, \aap, 246, L28

\bibitem[{{Pieri} {et~al.}(2009){Pieri}, {Lavalle}, {Bertone}, \&
  {Branchini}}]{2009arXiv0908.0195P}
{Pieri}, L., {Lavalle}, J., {Bertone}, G., \& {Branchini}, E. 2009, ArXiv
  e-prints

\bibitem[{{Pineault} {et~al.}(1993){Pineault}, {Landecker}, {Madore}, \&
  {Gaumont-Guay}}]{1993AJ....105.1060P}
{Pineault}, S., {Landecker}, T.~L., {Madore}, B., \& {Gaumont-Guay}, S. 1993,
  \aj, 105, 1060

\bibitem[{{Pineault} {et~al.}(1997){Pineault}, {Landecker}, {Swerdlyk}, \&
  {Reich}}]{1997A&A...324.1152P}
{Pineault}, S., {Landecker}, T.~L., {Swerdlyk}, C.~M., \& {Reich}, W. 1997,
  \aap, 324, 1152

\bibitem[{{Plucinsky} {et~al.}(1996){Plucinsky}, {Snowden}, {Aschenbach},
  {Egger}, {Edgar}, \& {McCammon}}]{1996ApJ...463..224P}
{Plucinsky}, P.~P., {Snowden}, S.~L., {Aschenbach}, B., {et~al.} 1996, \apj,
  463, 224

\bibitem[{{Porter} {et~al.}(2008){Porter}, {Moskalenko}, {Strong}, {Orlando},
  \& {Bouchet}}]{2008ApJ...682..400P}
{Porter}, T.~A., {Moskalenko}, I.~V., {Strong}, A.~W., {Orlando}, E., \&
  {Bouchet}, L. 2008, \apj, 682, 400

\bibitem[{{Profumo}(2008)}]{2008arXiv0812.4457P}
{Profumo}, S. 2008, ArXiv e-prints

\bibitem[{{Putze} {et~al.}(2010){Putze}, {Derome}, \&
  {Maurin}}]{2010A&A...516A..66P}
{Putze}, A., {Derome}, L., \& {Maurin}, D. 2010, \aap, 516, A66+

\bibitem[{{Redman} \& {Meaburn}(2005)}]{2005MNRAS.356..969R}
{Redman}, M.~P. \& {Meaburn}, J. 2005, \mnras, 356, 969

\bibitem[{{Reich} {et~al.}(1992){Reich}, {Fuerst}, \&
  {Arnal}}]{1992A&A...256..214R}
{Reich}, W., {Fuerst}, E., \& {Arnal}, E.~M. 1992, \aap, 256, 214

\bibitem[{{Reich} {et~al.}(2003){Reich}, {Zhang}, \&
  {F{\"u}rst}}]{2003A&A...408..961R}
{Reich}, W., {Zhang}, X., \& {F{\"u}rst}, E. 2003, \aap, 408, 961

\bibitem[{{Rosado} \& {Gonzalez}(1981)}]{1981RMxAA...5...93R}
{Rosado}, M. \& {Gonzalez}, J. 1981, Revista Mexicana de Astronomia y
  Astrofisica, 5, 93

\bibitem[{{Ruderman} \& {Sutherland}(1975)}]{1975ApJ...196...51R}
{Ruderman}, M.~A. \& {Sutherland}, P.~G. 1975, \apj, 196, 51

\bibitem[{{Salati}(2003)}]{2003PhLB..571..121S}
{Salati}, P. 2003, Physics Letters B, 571, 121

\bibitem[{{Schlickeiser} \& {Ruppel}(2009)}]{2009arXiv0908.2183S}
{Schlickeiser}, R. \& {Ruppel}, J. 2009, ArXiv e-prints

\bibitem[{{Shaviv} {et~al.}(2009){Shaviv}, {Nakar}, \&
  {Piran}}]{2009PhRvL.103k1302S}
{Shaviv}, N.~J., {Nakar}, E., \& {Piran}, T. 2009, Physical Review Letters,
  103, 111302

\bibitem[{{Shen}(1970)}]{1970ApJ...162L.181S}
{Shen}, C.~S. 1970, \apjl, 162, L181+

\bibitem[{{Shikaze} {et~al.}(2007){Shikaze}, {Haino}, {Abe}, {Fuke}, {Hams},
  {Kim}, {Makida}, {Matsuda}, {Mitchell}, {Moiseev}, {Nishimura}, {Nozaki},
  {Orito}, {Ormes}, {Sanuki}, {Sasaki}, {Seo}, {Streitmatter}, {Suzuki},
  {Tanaka}, {Yamagami}, {Yamamoto}, {Yoshida}, \&
  {Yoshimura}}]{2007APh....28..154S}
{Shikaze}, Y., {Haino}, S., {Abe}, K., {et~al.} 2007, Astroparticle Physics,
  28, 154

\bibitem[{{Shinn} {et~al.}(2006){Shinn}, {Min}, {Lee}, {Edelstein}, {Korpela},
  {Welsh}, {Han}, {Nam}, {Jin}, \& {Lee}}]{2006ApJ...644L.189S}
{Shinn}, J., {Min}, K.~W., {Lee}, C., {et~al.} 2006, \apjl, 644, L189

\bibitem[{{Shinn} {et~al.}(2007){Shinn}, {Min}, {Sankrit}, {Ryu}, {Kim}, {Han},
  {Nam}, {Park}, {Edelstein}, \& {Korpela}}]{2007ApJ...670.1132S}
{Shinn}, J.-H., {Min}, K.~W., {Sankrit}, R., {et~al.} 2007, \apj, 670, 1132

\bibitem[{{Silk} \& {Srednicki}(1984)}]{1984PhRvL..53..624S}
{Silk}, J. \& {Srednicki}, M. 1984, Physical Review Letters, 53, 624

\bibitem[{{Sofue} {et~al.}(1980){Sofue}, {Furst}, \&
  {Hirth}}]{1980PASJ...32....1S}
{Sofue}, Y., {Furst}, E., \& {Hirth}, W. 1980, \pasj, 32, 1

\bibitem[{{Strom} \& {Stappers}(2000)}]{2000ASPC..202..509S}
{Strom}, R.~G. \& {Stappers}, B.~W. 2000, in Astronomical Society of the
  Pacific Conference Series, Vol. 202, IAU Colloq. 177: Pulsar Astronomy - 2000
  and Beyond, ed. M.~{Kramer}, N.~{Wex}, \& R.~{Wielebinski}, 509--+

\bibitem[{{Strong} \& {Moskalenko}(1998)}]{1998ApJ...509..212S}
{Strong}, A.~W. \& {Moskalenko}, I.~V. 1998, \apj, 509, 212

\bibitem[{{Strong} {et~al.}(2007){Strong}, {Moskalenko}, \&
  {Ptuskin}}]{2007ARNPS..57..285S}
{Strong}, A.~W., {Moskalenko}, I.~V., \& {Ptuskin}, V.~S. 2007, Annual Review
  of Nuclear and Particle Science, 57, 285

\bibitem[{{Strong} {et~al.}(2000){Strong}, {Moskalenko}, \&
  {Reimer}}]{2000ApJ...537..763S}
{Strong}, A.~W., {Moskalenko}, I.~V., \& {Reimer}, O. 2000, \apj, 537, 763

\bibitem[{{Stupar} {et~al.}(2007){Stupar}, {Parker}, \&
  {Filipovi{\'c}}}]{2007MNRAS.374.1441S}
{Stupar}, M., {Parker}, Q.~A., \& {Filipovi{\'c}}, M.~D. 2007, \mnras, 374,
  1441

\bibitem[{{Sturrock}(1970)}]{1970Natur.227..465S}
{Sturrock}, P.~A. 1970, \nat, 227, 465

\bibitem[{{Sturrock}(1971)}]{1971ApJ...164..529S}
{Sturrock}, P.~A. 1971, \apj, 164, 529

\bibitem[{{Sun} {et~al.}(2006){Sun}, {Reich}, {Han}, {Reich}, \&
  {Wielebinski}}]{2006A&A...447..937S}
{Sun}, X.~H., {Reich}, W., {Han}, J.~L., {Reich}, P., \& {Wielebinski}, R.
  2006, \aap, 447, 937

\bibitem[{{Tan} \& {Ng}(1983)}]{1983JPhG....9.1289T}
{Tan}, L.~C. \& {Ng}, L.~K. 1983, Journal of Physics G Nuclear Physics, 9, 1289

\bibitem[{{Tatischeff}(2009)}]{2009A&A...499..191T}
{Tatischeff}, V. 2009, \aap, 499, 191

\bibitem[{{Taylor} {et~al.}(2003){Taylor}, {Gibson}, {Peracaula}, {Martin},
  {Landecker}, {Brunt}, {Dewdney}, {Dougherty}, {Gray}, {Higgs}, {Kerton},
  {Knee}, {Kothes}, {Purton}, {Uyaniker}, {Wallace}, {Willis}, \&
  {Durand}}]{2003AJ....125.3145T}
{Taylor}, A.~R., {Gibson}, S.~J., {Peracaula}, M., {et~al.} 2003, \aj, 125,
  3145

\bibitem[{{Taylor} {et~al.}(1993){Taylor}, {Manchester}, \&
  {Lyne}}]{1993ApJS...88..529T}
{Taylor}, J.~H., {Manchester}, R.~N., \& {Lyne}, A.~G. 1993, \apjs, 88, 529

\bibitem[{{Troja} {et~al.}(2008){Troja}, {Bocchino}, {Miceli}, \&
  {Reale}}]{2008A&A...485..777T}
{Troja}, E., {Bocchino}, F., {Miceli}, M., \& {Reale}, F. 2008, \aap, 485, 777

\bibitem[{{Uyaniker} {et~al.}(2002){Uyaniker}, {Kothes}, \&
  {Brunt}}]{2002ApJ...565.1022U}
{Uyaniker}, B., {Kothes}, R., \& {Brunt}, C.~M. 2002, \apj, 565, 1022

\bibitem[{{Valinia} \& {Marshall}(1998)}]{1998ApJ...505..134V}
{Valinia}, A. \& {Marshall}, F.~E. 1998, \apj, 505, 134

\bibitem[{{van den Bergh} \& {Tammann}(1991)}]{1991ARA&A..29..363V}
{van den Bergh}, S. \& {Tammann}, G.~A. 1991, \araa, 29, 363

\bibitem[{{Vinyaikin}(2007)}]{2007ARep...51..570V}
{Vinyaikin}, E.~N. 2007, Astronomy Reports, 51, 570

\bibitem[{{Welsh} \& {Sallmen}(2003)}]{2003A&A...408..545W}
{Welsh}, B.~Y. \& {Sallmen}, S. 2003, \aap, 408, 545

\bibitem[{{Welsh} {et~al.}(2001){Welsh}, {Sfeir}, {Sallmen}, \&
  {Lallement}}]{2001A&A...372..516W}
{Welsh}, B.~Y., {Sfeir}, D.~M., {Sallmen}, S., \& {Lallement}, R. 2001, \aap,
  372, 516

\bibitem[{{Woosley} \& {Janka}(2005)}]{2005NatPh...1..147W}
{Woosley}, S. \& {Janka}, T. 2005, Nature Physics, 1, 147

\bibitem[{{Xiao} {et~al.}(2008){Xiao}, {F{\"u}rst}, {Reich}, \&
  {Han}}]{2008A&A...482..783X}
{Xiao}, L., {F{\"u}rst}, E., {Reich}, W., \& {Han}, J.~L. 2008, \aap, 482, 783

\bibitem[{{Xiao} {et~al.}(2009){Xiao}, {Reich}, {F{\"u}rst}, \&
  {Han}}]{2009arXiv0904.3170X}
{Xiao}, L., {Reich}, W., {F{\"u}rst}, E., \& {Han}, J.~L. 2009, ArXiv e-prints

\bibitem[{{Yar-Uyaniker} {et~al.}(2004){Yar-Uyaniker}, {Uyaniker}, \&
  {Kothes}}]{2004ApJ...616..247Y}
{Yar-Uyaniker}, A., {Uyaniker}, B., \& {Kothes}, R. 2004, \apj, 616, 247

\bibitem[{{Y{\"u}ksel} {et~al.}(2009){Y{\"u}ksel}, {Kistler}, \&
  {Stanev}}]{2009PhRvL.103e1101Y}
{Y{\"u}ksel}, H., {Kistler}, M.~D., \& {Stanev}, T. 2009, Physical Review
  Letters, 103, 051101

\bibitem[{{Yusifov} \& {K{\"u}{\c c}{\"u}k}(2004)}]{2004A&A...422..545Y}
{Yusifov}, I. \& {K{\"u}{\c c}{\"u}k}, I. 2004, \aap, 422, 545

\bibitem[{{Zeiger} {et~al.}(2008){Zeiger}, {Brisken}, {Chatterjee}, \&
  {Goss}}]{2008ApJ...674..271Z}
{Zeiger}, B.~R., {Brisken}, W.~F., {Chatterjee}, S., \& {Goss}, W.~M. 2008,
  \apj, 674, 271

\bibitem[{{Zhang} \& {Cheng}(2001)}]{2001A&A...368.1063Z}
{Zhang}, L. \& {Cheng}, K.~S. 2001, \aap, 368, 1063

\end{thebibliography}

\begin{appendix}
\section{Analytical solutions of the spatial integral}
\label{app:analytic}

We briefly present the solutions of the spatial integral of the Green function convolved with a disk--like spatial distribution for sources. We distinguish two cases: (i) a homogeneous flat disk, which is relevant when dealing with secondary CR electrons and positrons (see Eq.~\ref{eq:flat_disk}), and (ii) a z-exponential disk (see Eq.~\ref{eq:rho}), relevant when dealing with primaries as long as the distribution of sources does not exhibit too strong gradients over 
a distance fixed by the half-thickness of the diffusion zone $\sim L$.

In both cases, disregarding the radial boundary conditions, the radial 
solution is simply given by:
\ben
{\cal I}_r &\equiv& \int_{r_{\rm min}}^{r_{\rm max}} dr_s \, r_s \, 
\frac{ \exp\left\{- \frac{r_s^2}{\lambda^2} \right\} }{\pi\,\lambda^2} \\
&=&\frac{\exp\left\{-\frac{r_{\rm min}^2}{\lambda^2}\right\}}{2\,\pi}\,
\left[ 1 - \exp\left\{-\frac{r_{\rm max}^2-r_{\rm min}^2}{\lambda^2}\right\} 
\right]\nn\\
&\overset{r_{\rm max}\rightarrow \infty}{\underset{r_{\rm min}\rightarrow 0}{\longrightarrow}}& \frac{1}{2\,\pi}\;.\nn
\een

If one wishes to take into account the radial boundary, it is enough to add a second propagator that will cancel out the first one at $r=R$. The 2D-propagator becomes:

\ben
\greenf_r (\vec{r}_\odot,E\leftarrow \vec{r}_s,E_s) &=& 
\frac{1}{\pi \,\lambda^2}\left(
\exp\left\{ -\frac{(\vec{r}_\odot-\vec{r}_s)^2}{\lambda^2} \right\} \right. \nonumber\\ & &- \left. \exp\left\{ -a(r_s)\frac{(\vec{r}_\odot-\vec{r}_{im})^2}{\lambda^2} \right\} \right)\;,
\een
where the image satisfies $\vec{r}_{im} = \vec{r}_{s}/a(r_s)$ with the scale parameter $a(r) = \frac{r^2}{R^2}$. This is equivalent with replacing $R_\odot$ by $\frac{R^2}{R_\odot}$ and $\lambda$ by $\lambda\frac{R}{R_\odot}$ (except for the first $\lambda^2$ which is in factor). In the case of secondary production, when the source term is homogeneous in the disk, the integration over the total disk is analytical and gives:

\begin{eqnarray*}
\lefteqn{{\cal I}_{r,\theta}^{\rm hom} = \int_0^{2\pi}\int_0^R r \greenf_r (\vec{r},E\leftarrow \vec{r}_s,E_s) dr d\theta =
 }\\
\lefteqn{  e^{-\frac{R_\odot^2}{\lambda^2}}\sum_{m=0}^{\infty}\left\lbrace \left(\frac{R_\odot^2}{\lambda^2}\right)^m \frac{1}{m!}\left(1-e^{-\frac{R^2}{\lambda^2}}\sum_{j=0}^{m}\left(\frac{R^2}{\lambda^2}\right)^{j} \frac{1}{j!} \right) \right\rbrace 
 }\\
 \lefteqn{ - \frac{e^{-\frac{R^2}{\lambda^2}}}{a(R_\odot)}\sum_{m=0}^{\infty}\left\lbrace \left(\frac{R^2}{\lambda^2}\right)^m \frac{1}{m!}\left(1-e^{-\frac{R_\odot^2}{\lambda^2}}\sum_{j=0}^{m}\left(\frac{R_\odot^2}{\lambda^2}\right)^{j} \frac{1}{j!} \right) \right\rbrace 
\;.}
\end{eqnarray*}

\subsection{Vertical solution for the homogeneous disk approximation}
\label{subapp:hom_disk}

We quote the result obtained by \citet{2009A&A...501..821D} for secondary 
positrons. We have:
\ben
{\cal I}_z^{\rm hom} &=& \int_{-z_{\rm max}}^{z_{\rm max}} dz_s \, 
\greenf_z(\lambda,z=0\leftarrow z_s)\\
&=& \begin{cases}
{\displaystyle
\frac{1}{2}\sum_{n=-\infty}^{\infty} \left\{ {\rm erf} \left( 
  \frac{z_n^{\rm max}}{\lambda}\right) -  
  {\rm erf} \left( \frac{z_n^{\rm min}}{\lambda}\right) \right\} \; ,}\\
{\displaystyle
  \frac{2}{L}\sum_{n=1}^{\infty}(-1)^{n+1}
  \frac{\cos \left( k_n(L-z_{\rm max}) \right) }{k_n}
  \times e^{-k_n^2\lambda^2/4 } \; .}
\end{cases}\nn
\een
The latter case corresponds to the Helmholtz solution, while the former is the image solution, for which we have $z_{n}^{\rm max}\equiv 2 n L + (-1)^n z_{\rm max}$ and $z_{n}^{\rm min}\equiv 2 n L - (-1)^n z_{\rm max}$. Throughout the paper, we have used a disk of half-thickness $z_{\rm max} = h = 0.1$ kpc.

\subsection{Vertical solution for the z-exponential disk approximation}
\label{subapp:exp_disk}

If we consider sources that exhibit exponential vertical profiles, the integral has to be performed over the complete diffusion zone:
\begin{align}
& {\cal I}_z^{\rm exp} = \int_{-L}^{L} dz_s \, 
\greenf_z(\lambda,z=0\leftarrow z_s) e^{\left(-\left|z\right|/z_0\right)} = \\ 
& \begin{cases}
{\displaystyle
\sum_{n=-\infty}^{\infty} (-1)^n e^{b_n^2-\left(\frac{2nL}{\lambda}\right)^2}\left\{ {\rm erf} \left( 
  \frac{L}{\lambda} + b_n\right) -  
  {\rm erf} \left( b_n\right) \right\} \; ,}\\
{\displaystyle
  \frac{2}{L}\sum_{n=1}^{\infty}\left((-1)^{n+1}z_0k_ne^{-L/z_0}
  +1 \right)
  \times \frac{z_0e^{-k_n^2\lambda^2/4 } }{1+z_0^2 k_n^2}\; .}
\end{cases} \nn
\end{align}
The first case corresponds to the image solution and $b_n$ represents $\frac{\lambda}{2z_0} + (-1)^n\frac{2nL}{\lambda}$. The second case corresponds to the Helmholtz solution.

\section{Fitting formulae for secondary electrons and positrons}
\label{app:fit}

Here, we provide useful fitting formulae for our template computations of the interstellar secondary flux, for both CR electrons and positrons, in the \propmed~propagation setup and with the M1 ISRF model for the full relativistic energy losses (see sections \ref{subsec:prop_par} and \ref{sec:secondaries} for more details). These formulae do not account for the solar modulation that one has to implement oneself, and are relevant from $\sim 1$ GeV up to a few TeV. We used the following parameterization for the flux, expressed in units of $({\rm GeV\,cm^2\,s\, sr})^{-1}$:
\ben
\phi_{\sec}(E) = \exp\left\{ \sum_{i=0}^{N}\alpha_i \,
\left[\ln\left( \frac{E}{1\,{\rm GeV}} \right) \right]^i \right\}\;.
\label{eq:fitsec}
\een
For secondary positrons, we find
\ben
\alpha_i^{e^+} &=& \Big\{ -5.46298, -3.52896, -0.0887432, 0.0319396,\nn \\
&& -0.00339393, 0.000107393 \Big\}\;,
\een
while for electrons, we have
\ben
\alpha_i^{e^-} &=& \Big\{  -6.00407, -3.40715, -0.0977409, 0.0346854, \nn \\
&& -0.00441225, 0.00019 \Big\} \;.
\een
Note that rescaling these formulae by a global factor mostly means rescaling the averaged local value of the gas density in the disk.

\section{Local supernova remnants}
\label{app:SNR}

SNRs are very numerous but not always easy objects to detect. The most complete available catalog is that of \citet{2005MmSAI..76..534G}, which was later updated \citep{2009BASI...37...45G}. Moreover the Canadian Galactic Plane Survey \citep[][CGPS hereafter]{2003AJ....125.3145T} has focused on many of these objects. In the following, we discuss the SNRs that 
lie less than $\sim$ 2~kpc away from the Earth. We classified them by 
distance to the Earth. All the important data we have used are summed up in 
\citetab{tab:SNRs}. Considering the number of recent discoveries in this 
region, it is more than probable that other remnants will be discovered in 
the future.

\begin{center}
\begin{table*}[ht] 
{\small
\hfill{}
\begin{tabular}{|c|c|c|c|c|c|c|c|} 
\hline
\# & SNR  & other name & distance  & radio index & Brightness  & age  & Pulsar\\
 & G+long+lat & & [kpc] & & [Jy] & [kyr] & \\
\hline
\hline
1&18.95-1.1 & & 2. \textbf{$\pm$ 0.1} & 0.28 & 40 & 11.75~$\pm$~0.85
& ? \\
\hline
2&65.3+5.7 & & 0.9~$\pm$~0.1 & 0.58~$\pm$~0.07& 52 & 26~\textbf{$\pm
$~1} & - \\
\hline
3&65.7+1.2 & DA 495 & 1.0~$\pm$~0.4 & 0.45~$\pm$~0.1 & 5 & 16.75~$\pm
$~3.25 & unknown\\
\hline
4&69.0+2.7 & CTB 80 & 2.0 \textbf{$\pm$ 0.1} & 0.20~$\pm$~0.10 & 60~
$\pm$~10 & 20~\textbf{$\pm$~1} & J1952+3252 \\
\hline 
5&74.0-8.5 & Cygnus Loop & 0.54$^{+0.10}_{-0.08}$ & 0.4~$\pm$~0.06 &
175~$\pm$~30 & 10~\textbf{$\pm$~1} &- \\
\hline 
6&78.2+2.1 & $\gamma$ Cygni & 1.5 \textbf{$\pm$ 0.1} & 0.75~$\pm
$~0.03 & 275~$\pm$~25 & 7~\textbf{$\pm$~1} & -\\
\hline 
7&82.2+5.3 & W63 & 2.3 $\pm$ 1.0 & 0.36~$\pm$~0.08 & 105~$\pm$~10 &
20.1~$\pm$~6.6 &- \\
\hline
8&89.0+4.7 & HB 21 & 1.7~$\pm$~0.5 & 0.27~$\pm$~0.07 & 200~$\pm$~15 &
5.60~$\pm$~0.28 &- \\
\hline
9&93.7-0.2 & CTB 104A or DA 551 & 1.5~$\pm$~0.2 & 0.52~$\pm$~0.12 & 42~
$\pm$~7 & 50~$\pm$~20& -\\
\hline
10&114.3+0.3 & & 0.7 & 0.49~$\pm$~0.25 & 6.4~$\pm$~1.4 & 7.7 ~
\textbf{$\pm$~0.1}& - \\
\hline
11&116.5+1.1 & & 1.6 & 0.16~$\pm$~0.11 & 10.9~$\pm$~1.2 & 20~$\pm$~5
&B2334+61 ? \\
\hline
12&116.9+0.2 & CTB 1 & 1.6 & 0.33~$\pm$~0.13 & 6.4~$\pm$~1.4 &20~$\pm$~5
& B2334+61 ? \\
\hline
13&119.5+10.2 & CTA 1 & 1.4~$\pm$~0.3 & 0.57~$\pm$~0.06 & 42.5~$\pm$~2.5
& 10~$\pm$~5 & J0010+7309 \\
\hline
14&127.1+0.5 & R5 & 1.$\pm$0.1 & 0.43~$\pm$~0.1 & 12~$\pm$~1 & 25~$\pm
$~5 & -\\
\hline 
15&156.2+5.7 & & 0.8~$\pm$~0.5 & 2.0$^{+1.1}_{-0.7}$ & 4.2~$\pm$~0.1 &
10~\textbf{$\pm$~1} & B0450+55 ?\\
\hline 
16&160.9+2.6 & HB 9 & 0.8~$\pm$~0.4 & 0.48~$\pm$~0.03 & $\sim$ 75 & 5.5~
$\pm$~1.5 & B0458+46 \\
\hline
17&180.0-1.7 & S147 & 1.2~$\pm$~0.4 & 0.75 & 74~$\pm$~12 & 600~\textbf{$\pm$~10} & J0538+2817 \\
\hline
\multirow{2}{*}{18}&\multirow{2}{*}{184.6-5.8} & Crab nebula &
\multirow{2}{*}{2.0~$\pm$~0.5} & \multirow{2}{*}{0.3} &
\multirow{2}{*}{1,040} & \multirow{2}{*}{7.5 $\star$} &
\multirow{2}{*}{B0521+31} \\
& & or 3C144 or SN1054 & & & & &\\
\hline 
19&189.1+3.0 & IC 443 & 1.5 \textbf{$\pm$ 0.1} & 0.36~$\pm$~0.04 &
160~$\pm$~5 & 30 or 4 & - \\
\hline 
20&203.0+12.0 & Monogem ring & 0.288$^{+0.033}_{-0.027}$ & & & 86~
\textbf{$\pm$~1} & B0656+14\\
\hline 
21&205.5+0.5 & Monoceros Nebula & 1.63~$\pm$~0.25 & 0.66~$\pm$~0.2 &
156.1~$\pm$~19.9 & 29~\textbf{$\pm$~1} & -\\
\hline 
22&263.9-3.3 & Vela(XYZ) & 0.295~$\pm$~0.075 & variable & 2,000~$\pm
$~700 & 11.2~\textbf{$\pm$~0.1} & B0833-45\\
\hline
\multirow{2}{*}{23}&\multirow{2}{*}{266.2-1.2} & RX J0852.0-4622 or Vela
Jr & \multirow{2}{*}{0.75 \textbf{$\pm$ 0.01}}& & &
\multirow{2}{*}{3.5~$\pm$~0.8~$\star$?} & \multirow{2}{*}{J0855-4644 ?} \\
& & or SN1300 & & & & & \\
\hline 
24&276.5+19.0 & Antlia & 0.2 $\pm$ 0.14 & & & $\geq$~1,000 &B0950+08 \\
\hline 
25&315.1+2.7& & 1.7~$\pm$~0.8 & 0.7 & & \textbf{50~$\pm$~10} &
J1423-56\\
\hline 
26&330.0+15.0 & Lupus Loop & 1.2~$\pm$~0.3 & & & 50~\textbf{$\pm
$~10} &B1507-44 ? \\
\hline
27&347.3-0.5 & SN393 & 1.$\pm$ 0.3 & & & 4.9~$\star$ & - \\
\hline 
\end{tabular}}
\hfill{}
\caption{\small Characteristics of nearby SNRs. Spectral index and brightness are inferred from measurements made at 1~GHz. Uncertainties in bold are not taken from bibliographic references, but just correspond to a rough uncertainty in the last relevant digit; hence they can be underestimated. An age is flagged with a $\star$ for an historical remnant; in this case, the age uncertainty is set from the distance uncertainty. Note that these ages are the {\em observed} ages, which differ from the {\em actual} ages by $d/c$.}
\label{tab:SNRs}
\end{table*}
\end{center}

From Fig.~\ref{fig:local_sources}, it clearly appears  that all sources do not contribute the same way. We first list here only the most important ones:

\paragraph{G203.0+12.0-3.3}
Also called Monogem ring, this object is considered as a "probable remnant" by 
the Green catalog. Results based on observations from 
\citet{1996ApJ...463..224P} derived a distance of $\sim$ 300 pc and an age of 86 kyr. Parallax measurements by \citet{2003ApJ...593L..89B} for the associated pulsar gave a more precise distance of 288$^{+33}_{-27}$ pc. Being 25$^\circ$ in diameter in the sky, it is possible to infer neither a correct power spectrum nor a brightness.

\paragraph{G263.9-3.3}
This SNR is one of the brightest radio object in the sky. Its large apparent size was the reason it was long considered as three different objects of the Vela constellation, and it is often called Vela (XYZ). \citet{1999ApJ...515L..25C} estimated its distance to be 250~$\pm$~30 pc from absorption measurements. This is consistent with the parallax measurement of the associated pulsar (PSR) B0833--45 from \citet{2001ApJ...561..930C} which gave 294$_{-50}^{+76}$ pc. Pulsar period derivative and nebula evolution implied the same age of $\sim$ 11,200 yr \citep{1993ApJS...88..529T,2008ApJ...676.1064M}. Radio observations reviewed by \citet{2001A&A...372..636A} gave the three spectral indices of the structure: 
$\alpha_X = 0.39 \pm 0.03$, $\alpha_Y = 0.70 \pm 0.10$, and $\alpha_Z = 0.81 
\pm 0.16$. They also gave brightnesses at 1~GHz of each parts: $S_X = 1,160 
\pm 200$~Jy, $S_Y = 440 \pm 240$~Jy, and  $S_Z = 400 \pm 260$~Jy, which imply that $S_{XYZ} = 2,000 \pm 700$~Jy consistent with the value of 
1,750~Jy given by \citet{2009BASI...37...45G}.

\paragraph{Antlia}
Although not yet in the Green catalog, this remnant was discovered by \citet{2002ApJ...576L..41M} and confirmed by \citet{2007ApJ...670.1132S}. It 
lies at coordinates 276.5+19.0 and at a distance of only 60-340 pc from the Sun 
(Sedov estimation). Other considerations (maximal size, column density and 
$^{26}$Ti abundance) favors the lower bound of the distance. Its age is probably
more than 1 Myr. However its angular diameter is so large (24$^\circ$) that a 
correct spectral index cannot be inferred.

Hereafter we list other nearby SNRs ($\leq$2.~kpc). Some authors did not 
correct their age estimations by the light travel length, sometimes leading to 
results outside our lightcone. Therefore, a supplemental uncertainty of 
$\Delta d/c$ in the age should be added in some cases.

\paragraph{G18.95-1.1}
Using HI absorption observations, \citet{1989A&A...209..361F} estimated its 
distance to the Earth to be $\sim$ 2~kpc, but they could not exclude the 
possibility that it could be an extremely and unusually bright object located 
15~kpc away. The study from \citet{2004ApJ...603..152H} estimated 
the age of the object at 4,400 to 6,100 yr (without taking into account the 
distance uncertainty). However, this value is inconsistent with causality 
constraints, so we added the value corresponding to the distance 
6.5~kyr. Using ROSAT a observation, \citet{1997A&A...319..655F} found a spectral index of 0.28 but the remnant is hosts a great deal of substructures that makes this value difficult to appreciate. They also infer a luminosity of order 40~Jy at 1~GHz.

\paragraph{G65.3+5.7}
Sometimes this remnant is also called G65.2+5.7. 
The radio power spectrum  is  0.58~$\pm$~0.07 from 83 MHz up to 4.8 GHz 
\citep{2009arXiv0904.3170X}, and the estimated distance varies from 800~pc to 
1~kpc. In earlier studies, in the context of a pulsar survey in several 
SNRs \citep{1996ApJ...458..257G}, among which G65.2+5.7, the flux is estimated to 
52~Jy at 1~GHz, but this last quantity is rough estimated. Optical observation 
from \citet{2002A&A...388..355M} leads to an age of $\sim$~26~kyr.

\paragraph{G67.5+1.2}
Also called DA 495, this object is highly polarized. 
\citet{2006A&A...457.1081K} suggested that point-like sources may alter the 
measurement of the power spectrum, leading in that case to a lower limit of 
0.38~$\pm$~0.08. The brightness at 1~GHz they obtained is 4.6~$\pm$~0.2~Jy, 
which should be considered as an upper limit since the remnant seems to sit on 
a diffuse emission plateau. However, \citet{2008ApJ...687..516K} proposed that it is not a shell-type 
SNR but a PWN. Indeed the spectrum has a power break at $\sim$~1.3~GHz and no 
shell. The power spectrum is 0.45~$\pm$~0.1 and the spectral break is of 
0.42~$\pm$~0.22. The corresponding brightness is of order $\sim$~5~Jy at 1~GHz. 
They also give a distance of 1.0~$\pm$~0.4~kpc and an age of 20,000~yr. 
However, if there is a pulsar in this remnant, it has not yet been detected.

\paragraph{G69.0+2.7}
More often called CTB 80, this remnant is estimated by 
\citet{2003AJ....126.2114C} to be $\sim$~20,000 yr old. Multi-frequency 
review from \citet{2005A&A...440..171C} gave a spectral index of 
0.36~$\pm$~0.02 and a brightness of $\sim$~65~Jy at 1~GHz. However, the Canadian Galactic plane survey \citep{2006A&A...457.1081K} gave a 
spectrum of 0.2~$\pm$~0.1 and a brightness of 60~$\pm$~10~Jy. Many authors cite 
\cite{2000ASPC..202..509S} as a reference for a distance of $\sim$~2~kpc based 
on HI absorption observations. The study by \citet{2008ApJ...674..271Z} seems 
to reveal an association with the pulsar B1951+32. However it might be much 
older (51~kyr) than the remnant.

\paragraph{G74.0-8.5}
Also known as Cygnus Loop this $\sim$~10,000 yr old remnant has been shown 
by \citet{2009ApJ...692..335B} to be farther than 576~$\pm$~61~pc away. This is 
consistent with Hubble Space Telescope observations from 
\citet{2005AJ....129.2268B}, which found 540$^{+100}_{-80}$~pc. 
\citet{2006A&A...447..937S} found a spectral index of 0.40~$\pm$~0.06 and an 
integrated flux of 175~$\pm$~30~Jy at 1~GHz.

\paragraph{G78.2+2.1}
This remnant is also called $\gamma$ Cygni. The catalog based on the CGPS 
gives a brightness of 
275~$\pm$~25~Jy at 1~GHz taking into account the work of 
\citet{2008A&A...490..197L}, which yields a very precise spectral index of 
0.75~$\pm$~0.03 correctly taking into account the thermal subtraction. A deep 
optical survey by \citet{2003A&A...408..237M} revealed an age of $\sim$ 7,000 
yr. However, it is not possible from HI observation to infer a distance, 
depending on the method they find either 1 or 4~kpc with large error bars. The value of 1.5~kpc seems to preferred by most authors \citep[see][and references therein]{2004A&A...427L..21B}.

\paragraph{G82.2+5.3}
Also called W63, this remnant is estimated by \citet{2004A&A...415.1051M} to be at a 
distance between 1.6 and 3.3~kpc and an age between 13.5 and 26.7 kyr. However, 
it is not clear whether the Sedov analysis they have performed is licit in this 
case. Previous review from \citet{1981RMxAA...5...93R} suggests a distance of 
1.6~$\pm$~0.3~kpc. The catalog based on the CGPS gives a brightness of 
105~$\pm$~10~Jy at 1~GHz and a spectral index of 0.36~$\pm$~0.08.

\paragraph{G89.0+4.7}
According to \citet{2006ApJ...637..283B},  this remnant, also known as HB 21, 
is at a distance of 1.7~$\pm$~0.5~kpc from the Sun. The review by 
\citet{2003A&A...408..961R} suggests a spectral index of 0.41~$\pm$~0.02 and a 
flux density of 228~$\pm$~5~Jy at 1~GHz. However, the more recent CGPS gives an 
index of 0.27~$\pm$~0.07 and a brightness of 200~$\pm$~15~Jy. A study from 
\citet{2006ApJ...647..350L} estimates the age to be 5,600~$\pm$~280~yr but they 
used the former distance estimation of 0.8 kpc by \citet{1978ApJS...38..309H} ; 
it is not clear how this age would change for a larger distance of 1.7~kpc.

\paragraph{G93.7-0.2}
Also referred to as CTB 104A or DA 551, this highly polarized and large object 
was studied by \citet{2006A&A...457.1081K}. They measured a spectral 
index of 0.52~$\pm$~0.12 and a brightness at 1~GHz of 42~$\pm$~7~Jy. 
\citet{2002ApJ...565.1022U} estimated its distance to be 1.5~$\pm$~0.2~kpc but in 
\citet{2006A&A...457.1081K}, citing the same reference, the authors give 
1.4~kpc. The only age estimation we could find is the one from 
\citet{1982A&A...105..176M}, which suggests 29,000 to 74,000 yr. However most 
of the other results of these observations have been corrected by later works 
and therefore it is not clear to us how trustworthy this age estimation is.

\paragraph{G114.3+0.3}
Although some authors \citep[see \eg][]{2006ApJS..163..344K} still use the 
large value of 3.5~kpc, the study of a HI association by \citet{2004ApJ...616..247Y} 
gives 0.7~kpc. Same authors also estimated an age of 7,700 yr. Because of its 
proximity to Cas A, \citet{2006A&A...457.1081K} were not able to measure the 
flux and the spectral index. However, they give, as an average of previous 
works, a spectral index of 0.49~$\pm$~0.25 and a flux of 6.4~$\pm$~1.4~Jy at 
1~GHz.

\paragraph{G116.5+1.1}
While some authors \citep[see \eg][]{2006ApJS..163..344K} still use the large 
value of 4~kpc, the study of a HI association by \citet{2004ApJ...616..247Y} gives 
1.6~kpc. Same authors also estimated its age to be between 15,000 and 50,000 
yr, they wrote that they tend to trust more the younger age. We decided to assume 20~$\pm$~5~kyr. 
\citet{2006A&A...457.1081K} measured a spectral index of 0.16~$\pm$~0.11 and a 
flux of 10.9~$\pm$~1.2~Jy at 1~GHz. It is possible that the pulsar B2334+61 is 
associated with this remnant (or to the next one), although the age and distance 
estimations given by the ATNF catalog \citep{2005AJ....129.1993M} are not in close agreement.

\paragraph{G116.9+0.2}
Also called CTB 1, this remnant is considered by some authors 
\citep[see \eg][]{2006ApJS..163..344K} to lie 3.1~kpc away. However the study of HI 
association by \citet{2004ApJ...616..247Y} gives 1.6~kpc. It seems that it 
is very close to G116.5+1.1. The same authors also estimated its age to be between 
15,000 and 50,000 yr favoring a younger age. \citet{2006A&A...457.1081K} 
measured a spectral index of 0.33~$\pm$~0.13 and a flux of 7.9~$\pm$~1.3~Jy at 
1~GHz.

\paragraph{G119.5+10.2}
Also known as CTA1, this remnant is the first object seen by Fermi 
\citep{2008Sci...322.1218A}. The most recent estimations of its age and distance 
are those of \citet{1993AJ....105.1060P}, who estimated the distance to be 
1.4~$\pm$~0.3~kpc and the age to be between 5,000 and 15,000 yr. The later 
study of \citet{1997A&A...324.1152P} revealed a spectral index of 
0.57~$\pm$~0.06 and a brightness of 40 $\sim$ 45~Jy at 1~GHz.

\paragraph{G127.1+0.5}
Also known as R 5, this remnant was studied by \citet{2006A&A...451..251L}, who used radio observations from the CGPS. They 
determined an age of 2--3 10$^4$~yr, a spectral index of 0.43~$\pm$~0.10, and a 
brightness of 12~$\pm$~1~Jy at 1~GHz. For the distance, thanks to HI line 
survey, they found 1.15~kpc which is consistent with an association with NGC 
559, located 0.9--1.13~kpc away from the Sun.

\paragraph{G156.2+5.7}
Discovered during ROSAT survey by \citet{1991A&A...246L..28P}, this remnant is 
one of the faintest in the Galaxy with a brightness of 4.2~$\pm$~0.1~Jy at 
1~GHz \citep[see \eg][]{1992A&A...256..214R}. In their study of the X-ray 
emission, \citet{2004AdSpR..33..434P} found that the photon emission is 
in agreement with an electron density with a power law of 2.0~$^{+1.1}_{-0.7}$ 
and a cut-off at about 10 TeV. Its distance from the Sun was first estimated to be a few pc, although it seems from optical study by \citet{2007MNRAS.376..929G} 
that the remnant may be as close as 300 pc and younger than 10,000 yr. In their 
latest paper, \citet{2009PASJ...61..155K} use 1~kpc.

\paragraph{G160.9+2.6}
Also called HB9, \citet{2007A&A...461.1013L} using the 
CGPS and HI observations estimated a spectral index of 0.48~$\pm$~0.03, and a distance 
of 0.8~$\pm$~0.4~kpc. Concerning the age, they conclude that the Sedov age for 
HB9 is 6,600 yr and the evaporative cloud model yields ages of 4,000 -- 7,000 
yr. However, the age of the corresponding pulsar B0458+46 seems to be 7~kyr. 
The brightness is of the order of 75~Jy at 1~GHz.

\paragraph{G180.0-1.7}
This bright radio remnant is also called S147. The 143 ms pulsar PSRJ0538+2817 
\citep[][]{1996ApJ...468L..55A} is located within S147 and believed to be 
associated as the distance estimates for both the remnant (0.8--1.6~kpc) and 
the pulsar (1.2~kpc) agree. The characteristic age of the pulsar of 600 kyr is 
much older than the estimated 100 kyr based on radio data from 
\citet{1980PASJ...32....1S}. However the complexity of the environment of this 
remnant makes any age analysis extremely difficult. We therefore adopt the 
age of the pulsar. \citet{2003A&A...408..961R} measured the brightness of the 
source at 863 MHz. Using the spectral index of  $0.30 \pm 0.15$ from 
\citet{2008A&A...482..783X}, one finds a brightness of 74~$\pm$~12~Jy at 1~GHz. 
\citet{2008A&A...482..783X} also showed that the spectrum is broken at $\sim$ 
1.5 GHz above which the index increases to $1.35 \pm 0.20$; this is probably 
due to the diffuse component.

\paragraph{G184.6-5.8}
One of the most famous, more known as Crab nebula or 3C144, this is the remnant of the historical Supernova SN1054. As explained in 
\citet{2008ApJ...677.1201K}, even though this object is used to calibrate many 
instruments and hence very well studied, its exact distance cannot be measured 
precisely because its extreme brightness prevents from a parallax study. The 
nominal distance in the literature is 2.0~$\pm$~0.5~kpc. The corresponding age 
is therefore 6,000 to 9,000 yr. According to the Green catalog, its spectral 
index is $\sim$ 0.3 and it brightness is 1,040~Jy at 1~GHz. However, 
\citet{2007ARep...51..570V} showed that this value is decreasing with time.

\paragraph{G189.1+3.0}
Also named IC 443, this 30,000 yr old remnant is well-known for its rich 
chemical composition \cite[see][and references therein]{2007ApJ...664..890N}. 
\citet{2003A&A...408..545W} have calculated a distance of $\sim$ 1.5~kpc. The 
review by \citet{1986AJ.....92.1349M} gives a power spectrum of 0.36~$\pm$~0.04 
and \citet{2003A&A...408..961R} provide an intensity of 160~$\pm$~5~Jy. An  XMM-Newton observation \citep{2008A&A...485..777T} implies a much 
younger age of 4,000 yr.

\paragraph{G205.5+0.5}
Also known as Monoceros Nebula or the Monoceros Loop, this SNR is believed to 
be associated with the Rosette Nebula \citep[see][]{1986ApJ...301..813O} which 
lies at a distance of 1.6~$\pm$~0.2~kpc from the Sun 
\citep{2009MNRAS.tmp..278B}. This special position makes it a very interesting 
probe for cosmic ray acceleration study \citep[see][]{2008ICRC....2..719F}. 
\citet{1986MNRAS.220..501L} have estimated the age of the object to $\sim$ 
29,000 yr. However, this value is very model-dependent, and ultraviolet 
spectroscopy analysis by \citet{2001A&A...372..516W} suggests that it is even 
older (up to 150,000 yr). The review by \citet{1982A&A...109..145G} suggests 
a spectral index of 0.47~$\pm$~0.06 for the radio emission and a flux density 
of 156.1~$\pm$~19.9~Jy at 1~GHz. The results of \citet{2009arXiv0904.2261B} which have higher statistical significance prefer an index of 
0.66~$\pm$~0.20 with a brightness at 1~GHz in agreement with previous ones and 
a distance of 1,630~$\pm$~250~pc.

\paragraph{G266.2-1.2}
Often referred to as RX J0852.0-4622 or Vela Junior, the detection of the 
radioactive decay line of $^{44}$Ti seems to prove that it is a very young 
object. However its proximity to Vela makes any estimation of it brightness and 
spectral index quite difficult. Based on an evolution study,
\citet{2008ApJ...678L..35K} estimated its age to be between 
2,700 and 4,300 yr and its distance to be $\sim$ 750 pc. It is a little 
puzzling that no historical record of this supernova explosion, that happened 
around year 1,300, has been found. It was proposed in 
\citet{2005MNRAS.356..969R} an association with pulsar PSR J0855-4644, based 
on a reestimate of the pulsar distance.

\paragraph{G315.1+2.7}
The discovery by \citet{2007MNRAS.374.1441S} of this extremely large remnant 
revealed a distance of $\sim$1.7~kpc. Reanalyzing the former radio observation
by \citet{1997MNRAS.287..722D} they found a spectral index of 0.7. However, it 
seems that some ambiguity remains concerning the brightness. The age is not 
given precisely but it is considered as old. Without anymore precision we have 
taken 50~$\pm$~10~kyr.

\paragraph{G330.0+15.0}
Also called Lupus Loop, this remnant is 800 pc away from the Sun 
\citep{2004ApJS..153..269K}. However, a review by 
\citet{2006ApJ...644L.189S} gives 1.2~$\pm$~0.4~kpc. According to 
\citet{2006ApJS..163..344K} (and references therein) it is 50 kyr old. Very 
few data are available for Lupus Loop in radio wavelength. To our knowledge, 
the latest are from \citet{1991ApJ...374..218L}, who concluded that its 
spectrum could not be reproduced by a single power law.

\paragraph{G347.3-0.5}
Associated with the gamma source RX J1713.7-3946, this SNR lies about 1~kpc 
away. The age of the first light on Earth is estimated to be $\sim$ 1,600 yr, 
consistent with the historical Chinese record of an SN exploded in AD 393 
\citep[see \eg][]{2009MNRAS.392..240M}. For a distance of $\sim 1$kpc away, 
this means an age of $\sim 4.9$~kyr. \citet{2005ApJ...632..920E} explained that 
because its environment is extremely complex, it is very difficult to subtract the
background and to fit both a power spectrum and the brightness.

\end{appendix}

\end{document}